\documentclass[useAMS,usenatbib,fleqn]{mnras}

\usepackage[utf8]{inputenc}
\usepackage{graphicx,color}
\usepackage{mathtools}	%needed for dcase environment, loads amsmath package
\usepackage{amsmath}
\usepackage{amssymb}
\usepackage{bm}
\usepackage{fixltx2e} %to display {figure*}'s in right order
\usepackage{pstricks} % PSTricks package
\usepackage{arydshln} % for \hdashline
\usepackage[normalem]{ulem}

   \usepackage[british]{babel}             % British English hyphenation
   \usepackage{newtxtext}                  % Good fonts
   \usepackage[slantedGreek]{newtxmath}    %   "    "   (slanted Greek)
   \usepackage[T1]{fontenc}                % Good font encoding

\def\apj{ApJ}

\def\mnras{MNRAS}
\def\aap{A\&A}
\def\apjl{ApJ}

\def\apjs{ApJS}

\def\na{New\ Astron.}
\def\aapr{ARA\&A}

\newcommand{\f}{_\textrm{0}}

\newcommand{\1}{_\mathrm{1}}					   	%forcing
\newcommand{\core}{_\mathrm{1,c}}					   	%forcing
\newcommand{\2}{_\mathrm{2}}					   	%forcing
\newcommand{\diff}{_\mathrm{d}}			   		%diffusion, subscript
\newcommand{\gas}{_\mathrm{gas}}

\newcommand{\oneenv}{_\mathrm{1,e}}

\newcommand{\soft}{_\mathrm{soft}}

\newcommand{\Gn}{\mathrm{G}}
\newcommand{\init}{_\mathrm{i}}
\newcommand{\final}{_\mathrm{f}}

\newcommand{\Rsol}{\,\mathrm{R_\odot}}

\newcommand{\Rsun}{\,\mathrm{R_\odot}}
\newcommand{\Msun}{\,\mathrm{M_\odot}}
\newcommand{\Msunyr}{\,\mathrm{M_\odot\,yr^{-1}}}

\newcommand{\amb}{_\mathrm{amb}}
\newcommand{\bo}{_\mathrm{box}}

\newcommand{\rmD}{\mathrm{\Delta}}

\newcommand{\CE}{_\mathrm{CE}}

\newcommand{\refine}{_\mathrm{ref}}

\newcommand{\unb}{_\mathrm{unb}}

\newcommand{\sound}{_\mathrm{s}}

\newcommand{\abar}{\overline{a}}

  \newcommand{\erg}{\,{\rm erg}}
  
  \newcommand{\gcmcmcm}{\,{\rm g\,cm^{-3}}}

  \newcommand{\kms}{\,{\rm km\,s^{-1}}}

  \newcommand{\Gyr}{\,{\rm Gyr}}

  \newcommand{\yr}{\,{\rm yr}}     
  \newcommand{\da}{\,{\rm d}}     
  \newcommand{\dynecmcm}{\,{\rm dyn\,cm^{-2}}}

\title[Common Envelope Evolution on the AGB] 
{Common Envelope Evolution on the Asymptotic Giant Branch: Unbinding within a Decade?}
\author[L.~Chamandy et al.]{Luke Chamandy,\thanks{lchamandy@pas.rochester.edu}
Eric G.~Blackman,\thanks{blackman@pas.rochester.edu}
Adam Frank,\thanks{afrank@pas.rochester.edu}
Jonathan Carroll-Nellenback\thanks{jcarrol5@ur.rochester.edu}
\newauthor % starts a new line in the author environment 
and Yisheng Tu\thanks{ytu7@ur.rochester.edu}
\\
Department of Physics and Astronomy, University of Rochester, Rochester NY 14627, USA
}

\begin{document}

%\pagerange{\pageref{firstpage}--\pageref{lastpage}} \pubyear{2018}

\maketitle

\begin{abstract}
  Common envelope (CE) evolution is a critical but still poorly understood progenitor phase of many high-energy astrophysical phenomena.  Although 3D global hydrodynamic CE simulations have become more common in recent years, 
  those involving an asymptotic giant branch (AGB) primary
  are scarce, due to the high computational cost from the larger dynamical range compared to red giant branch (RGB) primaries.  But CE evolution with AGB progenitors is desirable to simulate because such events 
  are the likely progenitors of most bi-polar planetary nebulae (PNe), and prominent observational testing grounds for CE physics.
  Here we present a high resolution global simulation of CE evolution involving an AGB primary and  $1\Msun$ secondary,  
  evolved for  $20$ orbital revolutions.
  During the last $16$ of these orbits, the envelope unbinds at an almost constant rate of about $0.1$--$0.2\Msunyr$.  
  If this rate were maintained, the envelope would be unbound in less than $10\yr$.
  The dominant source of this unbinding is consistent with inspiral; we assess the influence of the ambient medium to be subdominant.
  We compare this run with a previous run that used an RGB phase primary evolved from the same $2\Msun$ main sequence star
  to assess the influence of the evolutionary state of the primary.
  When scaled appropriately, the two runs are quite similar, but with some important differences.
\end{abstract}
\begin{keywords}
binaries: close -- stars: AGB and post-AGB -- stars: kinematics and dynamics -- stars: mass loss -- stars: winds, outflows -- hydrodynamics
\end{keywords}

\defcitealias{Chamandy+18}{Paper~I}
\defcitealias{Chamandy+19a}{Paper~II}
\defcitealias{Chamandy+19b}{Paper~III}
%----------------------------------------------------------------------------

\section{Introduction}
\label{sec:intro}
Common envelope (CE) evolution is a brief but strongly interacting phase of binary stellar evolution 
whose consequences are fundamental to understanding many phenomena including planetary nebulae (PNe), 
the progenitors of supernovae type Ia,  and the progenitors of compact binaries that become observable gravitational wave sources.  
The CE phase occurs when a binary orbit decays to the point that the secondary plunges into the envelope of the primary,
and dissipative losses drive a fast inspiral of the secondary (\citealt{Paczynski76}; 
see \citet{Ivanova+13a} and \citet{Jones20a} for recent reviews).
Two possible outcomes are expected: either ejection of the envelope or a merger.
In this way CE evolution is thought to be the principal mechanism of forming short period binaries. 

Simulations have not yet produced unbound envelopes without invoking recombination energy \citep{Nandez+15,Nandez+Ivanova16,Ohlmann16,Prust+Chang19,Reichardt+20} in addition to the released orbital energy;
see \citet{Iaconi+17} and \citet{Iaconi+Demarco19} for compilations of global CE simulations from the literature.
However, both the importance and universality of the recombination energy in assisting unbinding remain unclear.
This is mainly because convection and radiative losses could change the estimates
but are as yet unaccounted for in simulations \citep{Sabach+17,Grichener+18,Wilson+Nordhaus18,Ivanova18}.
Energy liberated to the envelope as gas accretes onto the secondary
\citep{Soker04,Macleod+17,Morenomendez+17,Soker17b,Chamandy+18,Lopez-camara+19,Shiber+19} could also assist unbinding,
but  how far into the CE  this could be sustained, and at what rate,
remain to be determined.
Processes that redistribute energy,
such as convection, radiation pressure exerted on dust \citep{Glanz+Perets18,Iaconi+20}, 
excitation of pressure waves by the inspiralling secondary \citep{Soker92a},
or interaction of the stellar cores with envelope material that has fallen back \citep{Kashi+Soker11},
could also help to unbind the envelope.

The inter-particle separation at the end of existing simulations is generally still too large
to \textit{expect} the envelope to be unbound using the standard CE energy formalism, 
so simulations and theory are  consistent at this basic level \citep{Chamandy+19a,Iaconi+Demarco19}.%
\footnote{There are a few exceptions for which the envelope should be unbound according to the energy formalism if $\alpha\CE=1$ (see Section~\ref{sec:theory}),
but is not, which can be used to obtain an upper limit for $\alpha\CE$ \citep{Iaconi+Demarco19}.
However, at the ends of the highest resolution simulations \citep[e.g.][]{Ohlmann+16a}, 
the separation is still too large to set an upper limit on $\alpha\CE$ \citep{Chamandy+19a}.}
Possible, not necessarily mutually exclusive,  reasons that simulations do not succeed in unbinding the envelope are:
(i)~insufficient duration (as orbital energy is still being liberated at the end of simulations, albeit very slowly in some cases);
(ii)~insufficient resolution \citep{Ohlmann16,Iaconi+18,Chamandy+19a}; and
(iii)~non-inclusion of relevant physical processes (affecting total energy budget and energy redistribution).

In addition, 
global 3D simulations have so far focused on systems involving red giant branch (RGB) primaries,
whose envelopes are more strongly bound compared to the asymptotic giant branch (AGB) counterparts 
into which they would have evolved, absent binary interaction. 
The larger spatial and temporal dynamic ranges of  AGB stars, 
which have comparably dense cores but more distended envelopes, make them 
more challenging to simulate. However, 
this extra computational cost might be compensated by a smaller envelope binding energy.

\citet{Sandquist+98} performed five CE simulations with AGB primaries of $3\Msun$ or $5\Msun$, 
and companions of $0.4\Msun$ or $0.6\Msun$.
They used a nested grid with smallest
resolution element $\delta=2.2\Rsun$ and a \citet{Ruffert+93} potential with smoothing length $1.5\delta$.
They found  final separations between $4\Rsun$ and $9\Rsun$, 
but deemed them upper limits due to sensitivities to resolution and smoothing length.  Smaller smoothing lengths and higher resolution produced smaller final separations.  Nevertheless, \citet{Iaconi+17} estimate that $\sim21$--$46\%$ of the envelope mass
unbinds by the end of the \citet{Sandquist+98} simulations.
More recently, \citet{Staff+16a}, performed AGB CE simulations  primarily 
to explain a particular observed system,  using a high initial orbital eccentricity.  Most of their simulations consisted of a $3.05\Msun$ $473\Rsun$ AGB primary (zero-age main sequence (ZAMS) $3.5\Msun$) with a secondary of mass $1.7\Msun$.   Comparing their simulations ``4'' and ``4hr'',
with resolutions  $\delta=25\Rsun$ and $12\Rsun$, respectively,
and smoothing length of $39\Rsun$ \citep{Ruffert+93} for both runs,
they obtain final separations of $86\Rsun$ and $43\Rsun$, showing a lack of convergence with resolution. 
They therefore report the $\sim10\%$ fraction of mass unbound 
at the end of their simulations to be a lower limit.

Both \citet{Sandquist+98} and \citet{Staff+16a} 
find \textit{multiple}  mass loss events between periods of little  unbinding.
The initial event is nearly contemporaneous with first periastron passage,
analogous to what is seen in most RGB CE simulations.
A  longer  quiescent phase passes until the second unbinding event, followed  by another quiescent phase.
In \citet{Staff+16a} the second event occurs around the time of second periastron passage, but in \citet{Sandquist+98} it happens much later.
In \citet{Ohlmann16}, a second unbinding phase is also seen at the end of a simulation involving an RGB primary, using an ideal gas equation of state without recombination. 
Here we explore the outcome of a high resolution CE simulation involving an AGB primary, focusing on energy transfer and mass unbinding.
We also compare this simulation with our extensively studied earlier fiducial RGB CE simulation \citep{Chamandy+18,Chamandy+19a,Chamandy+19b}, 
whose setup was very similar\footnote{See also \cite{Ohlmann+16a} and \cite{Prust+Chang19} for simulations with very similar initial conditions.} 
apart from the nature of the primary.
In Section~\ref{sec:setup} we summarize the numerical setup.
Then, in Section~\ref{sec:theory}, we use the CE energy formalism to predict the final separation for our system.
Simulation results can be found in Section~\ref{sec:results}.
We summarize and conclude in Section~\ref{sec:conclusions}.

\begin{table}
  \begin{center}
  \caption{Physical parameters for the two runs discussed in this work.
           Both runs have zero initial orbital eccentricity 
           and in both cases the primary is initialized with zero rotation.
          \label{tab:params}
          }
  \begin{tabular}{@{}lccc@{}}
  %computed with original data from full resolution run 143 and with Paper_energy/en.pro
    \hline
    Quantity		&Symbol		&AGB Run			&RGB Run			\\
    \hline
    Primary age  	&--		&$1.175\Gyr$			&$1.041\Gyr$			\\
    Primary mass	&$M_1$		&$1.78\Msun$			&$1.96\Msun$			\\
    Core particle mass	&$M\core$	&$0.53\Msun$			&$0.37\Msun$			\\
    Envelope mass       &$M\oneenv$	&$1.25\Msun$			&$1.59\Msun$			\\
    Secondary mass	&$M_2$		&$0.98\Msun$			&$0.98\Msun$			\\
    Primary radius	&$R_1$		&$122.2\Rsun$			&$48.1\Rsun$			\\
    Initial separation	&$a\init$	&$124.0\Rsun$			&$49.0\Rsun$			\\
    Ambient density	&$\rho\amb$	&$1.0\times10^{-9}\gcmcmcm$	&$6.7\times10^{-9}\gcmcmcm$	\\
    Ambient pressure	&$P\amb$	&$1.1\times10^{4}\dynecmcm$	&$1.0\times10^{5}\dynecmcm$	\\
    \hline                  	 
  \end{tabular}
  \end{center}
\end{table}

\section{Simulation Setup}
\label{sec:setup}
The setup for our new run, which we refer to as the AGB run, 
is similar to that of the RGB run,
i.e.~Model~A of \citet{Chamandy+18}, which was also studied in \citet{Chamandy+19a} and \citet{Chamandy+19b}.
Both simulations are performed in the inertial frame of reference for which the system centre of mass is initially at rest, but can shift slightly owing to transport through the domain boundaries.

The initial stellar and orbital parameters for both the AGB and RGB runs are presented in Table~\ref{tab:params},
as well as the stellar age of the primary (with zero corresponding to the ZAMS).
The 1D stellar profile is obtained by running a Modules for Experiments in Stellar Astrophysics (MESA; \citealt{Paxton+11,Paxton+13,Paxton+15}) simulation.
To obtain the initial mass density and pressure profiles of the AGB star, 
we used a later snapshot of the same 1D simulation used for the RGB run.
Specifically, we evolved a ZAMS star of mass $2\Msun$ with metallicity $Z=0.02$, 
and chose snapshots corresponding as closely as possible to the ``RG'' and ``AGB'' models of \citet{Ohlmann+17}, 
for easy comparison with results of that work.%
\footnote{Our RGB star has surface luminosity $\log_{10} L_\mathrm{surf}=2.73$ and effective temperature $\log_{10} T_\mathrm{eff}=3.60$,
while our AGB star has $\log_{10} L_\mathrm{surf}=3.31$ and $\log_{10} T_\mathrm{eff}=3.55$.}

Both stellar cores cannot be resolved on the 3D mesh, 
so the core was expunged and replaced by  a gravitation-only point particle and $n=3$ polytrope 
that matches smoothly to the MESA profile at stellar radius $r$ equal to the spline softening radius of the particle $r\soft$,
but retaining the original core mass \citep{Ohlmann+17,Chamandy+18}.
Furthermore, a uniform ambient medium with pressure  slightly larger than that at the surface of the primary was included 
in order to truncate the pressure profile near the surface and hence prevent scale heights that would be too small to resolve.
No additional damping of velocities was performed as this was  found to be unnecessary in the RGB case \citep{Chamandy+18}.
We used an ideal gas equation of state with adiabatic index $5/3$.
Appendix~\ref{sec:ic} shows our initial density profiles of mass, internal energy, and potential energy  in comparison with the MESA profiles.
The secondary is modeled as a point particle with the same spline softening radius as the primary core particle $r\soft=2.4\Rsun$.
Both particles have fixed mass and no subgrid accretion model is included in the runs presented.

Our 3D hydrodynamic simulations use the adaptive mesh refinement (AMR) multi-physics code AstroBEAR \citep{Cunningham+09,Carroll-Nellenback+13}.
AstroBEAR fully accounts for all gravitational interactions (particle-gas, particle-particle and gas self-gravity),
and uses the \textit{hypre}\footnote{\textit{hypre}: High Performers Preconditioners (see http://www.llnl.gov/CASC/hypre/).}
library to solve for the gas gravitational potential on each AMR level. 
The hydrodynamics are solved using the corner transport upwind (CTU) method \citep{COLELLA1990171} 
with  piecewise linear reconstruction, along with the necessary modifications 
to include self-gravitational forces in a momentum conserving manner.  
Particle-gas interactions are treated as a separate source, 
but  conserve momentum  between the particles and the gas.

For both runs, the simulation domain has dimension $L\bo=1150\Rsun$ and extrapolating hydrodynamic boundary conditions.
The boundary conditions for the Poisson solver are calculated using a multipole expansion of the gas distribution. 
The base and highest resolutions are $\delta_0=2.25\Rsun$ and $0.07\Rsun$ respectively. 
See \citet{Chamandy+18} for a discussion of the numerics for the RGB run.
For the AGB run, the mesh was refined at AMR level 3 with resolution $\delta_3\approx0.28\Rsun$
everywhere inside a spherical region of radius $r\refine$, 
taken initially to be somewhat larger than the initial separation $a\init$ 
and gradually decreased as the binary separation decreased. 
This $r\refine$ is centred on the AGB core particle and, after $t=65\da$, on the particles' centre of mass.
Additionally, a roughly spherical region of radius $\approx12\Rsun$ 
was resolved at AMR level 5 or $\delta_5\approx0.07\Rsun$ around the primary core particle,
and the same extra refinement was added around the secondary after $t=44.9\da$.
Thus, $r\soft=2.4\Rsun\approx34\delta_5$ 
and the softening radius was kept constant during the run.
A buffer zone of $8$ cells per level allowed the resolution to transition gradually between the lowest and highest refinement levels.
As shown in Appendix~\ref{sec:energy_conservation}, the total energy $E$ in the simulation, 
accounting for fluxes through the domain boundaries, 
gradually increases, and both simulations were stopped when the energy gain reached $\rmD E/|E\init|\approx0.05$.

\section{Theoretical Expectations}
\label{sec:theory}
The energy formalism is a statement of energy conservation,
expressed by equating the initial binding energy of the envelope
with the change in orbital energy of the system multiplied by an efficiency, $\alpha\CE$, where $\alpha\CE\le1$ \citep{Webbink84,Iben+Tutukov84,Ivanova+13a}:
\begin{equation}
  \label{alpha}
  \frac{\Gn M\1 M\oneenv}{\lambda R\1} = \alpha\CE \frac{\Gn M\2}{2}\left(\frac{M\core}{a\final} -\frac{M\1}{a\init}\right).
\end{equation}
Here $\Gn$ is Newton's constant and $M\oneenv= M\1 -M\core$ is the mass of the primary's envelope. 
The parameter $\lambda$ can be computed directly from the envelope binding energy, 
and evaluates to $0.91$ ($1.31$) for the AGB (RGB) star simulated.
If the initial and final orbits are assumed to be circular, 
$a\init$ and $a\final$ are equal to the initial and final orbital separations, respectively.
The initial (final) state entails a completely bound (unbound) envelope.

Even without  sinks like radiation, $\alpha\CE<1$ is ensured
because unbound gas generally contains more than the threshold energy it needs to  unbind \citep{Ivanova+13a,Chamandy+19a} (regardless of the precise energy condition for unboundedness adopted) and this excess  is not otherwise accounted for in equation~\eqref{alpha}.
Population synthesis studies obtain  $0.1\le \alpha\CE\le 0.3$ 
\citep{Davis+10,Zorotovic+10,Cojocaru+17,Briggs+18}; $\alpha\CE$ likely varies from one binary system to another.

\begin{table}
  \begin{center}
  \caption{Initial energy components, in units of $10^{47}\erg$, for the AGB run, 
           with initial separation $a\init=124\Rsol$. 
           Values for $a\init=284\Rsol$ (Roche limit separation) are also shown for reference.
           A Newtonian potential is assumed for $|\bm{r}-\bm{r}\1|<r\soft$;
           using instead the spline potential employed in the simulation 
           results in a positive change of $<0.02\times10^{47}\erg$ in $E_\mathrm{pot,e-1,i}$.
           Particle~1 refers to the AGB core particle, and particle~2 to the secondary.
          \label{tab:initial_energy_terms}
          }
  \begin{tabular}{@{}llrr@{}}
  %computed with original data from full resolution run 143 and with Paper_energy/en.pro
    \hline                  	 
    Energy component at $t=0$		&Symbol			&$a\init=124\Rsol$	&$a\init=284\Rsol$	\\
    \hline                  		 
    Particle~1 kinetic			&$E_\mathrm{bulk,1,i}$	&$0.03$			&$0.01$	\\
    Particle~2 kinetic			&$E_\mathrm{bulk,2,i}$	&$0.17$			&$0.07$	\\
    Particle-particle potential 	&$E_\mathrm{pot,1-2,i}$	&$-0.16$		&$-0.07$\\
    \hline
    Envelope bulk kinetic		&$E_\mathrm{bulk,e,i}$	&$0.07$			&$0.03$	\\
    Envelope internal			&$E_\mathrm{int,e,i}$	&$0.71$			&$0.71$	\\
    Envelope-envelope potential		&$E_\mathrm{pot,e-e,i}$	&$-0.57$		&$-0.57$\\
    Envelope-particle~1 potential	&$E_\mathrm{pot,e-1,i}$	&$-0.88$		&$-0.88$\\
    Envelope-particle~2 potential	&$E_\mathrm{pot,e-2,i}$	&$-0.37$		&$-0.16$\\
    \hline
    Particle total			&$E_\mathrm{1-2,i}$	&$0.04$			&$0.02$	\\
    Envelope total			&$E_\mathrm{e,i}$	&$-1.05$		&$-0.88$\\
    \hline
    Total particle and envelope		&$E_\mathrm{e-1-2,i}$	&$-1.01$		&$-0.86$\\
    \hline                  	 
  \end{tabular}
  \end{center}
\end{table}

\begin{table}
  \begin{center}
  \caption{Final inter-particle separations $a\final$ predicted 
           by equation~\eqref{alpha} for various assumed values of $\alpha\CE$.
           The smaller of the two initial separations $a\init$ shown is that used in the given simulation,
           while the larger of the two is the Roche limit separation.
           Larger initial separation means larger initial orbital energy,
           so more orbital energy is released down to a given final separation $a\final$.
           \label{tab:alpha}
          }
  \begin{tabular}{lcrrrr}
  %computed with original data from full resolution run 143 and with Paper_energy/en.pro
    \hline                  	 
    &$\alpha\CE$:		&$0.1$	&$0.25$	&$0.5$	&$1$	\\
    \hline
    &$a\init$ ($\!\Rsun$)	&\multicolumn{4}{c}{$a\final$ ($\!\Rsun$)}	\\
    AGB 		&124	&1.3	&3.0	&5.6	&9.8			\\
    $\lambda=0.91$  	&284	&1.3	&3.2	&6.2	&11.5			\\
    RGB		        &49	&0.3	&0.8	&1.5	&2.6			\\
    $\lambda=1.31$	&109	&0.4	&0.9	&1.7	&3.1			\\
    \hline                  	 
  \end{tabular}
  \end{center}
\end{table}

\begin{figure}
  %produced using psep_multipleruns_183_143.py
  \includegraphics[width=\columnwidth,clip=true,trim= 8 7 0 5]{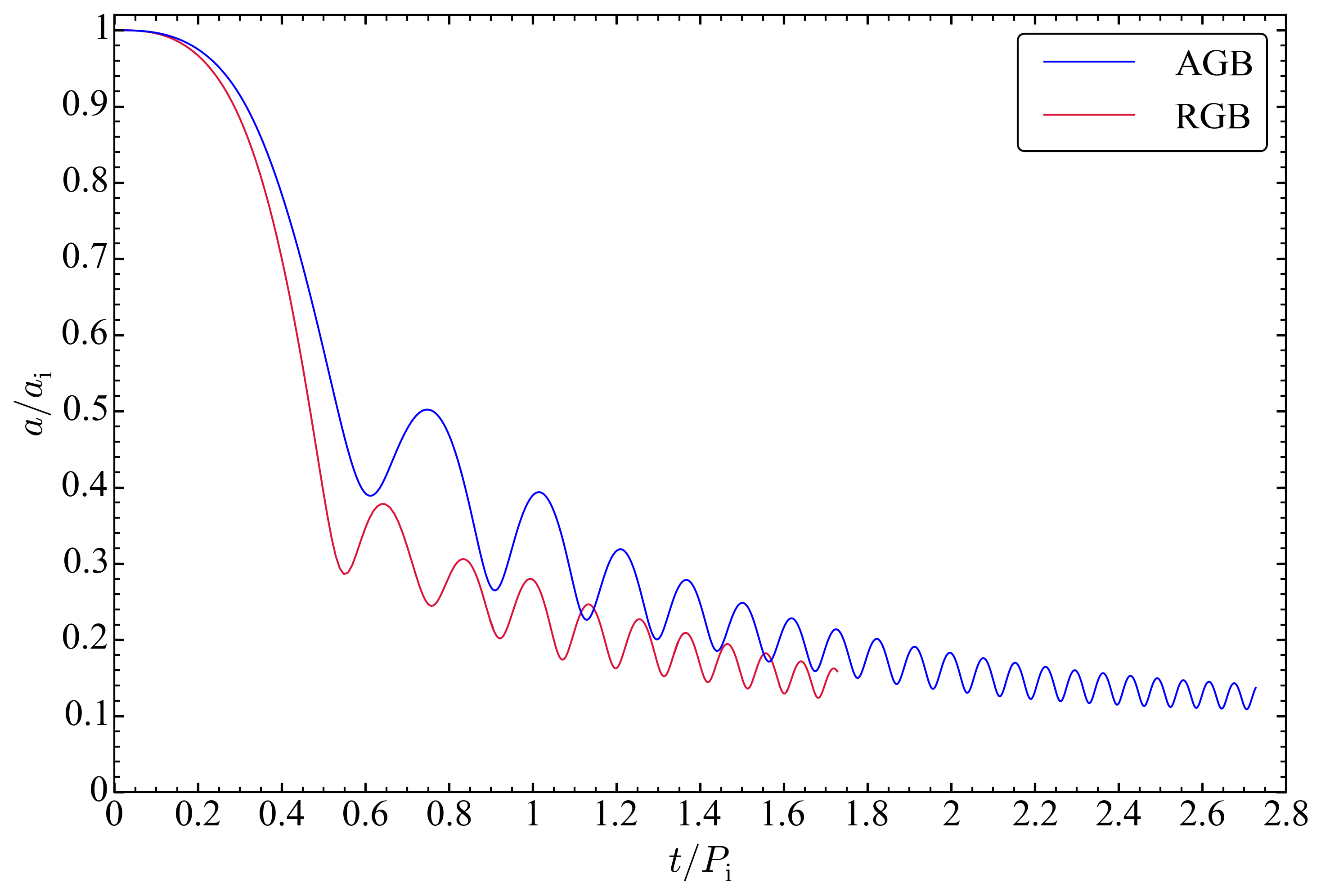}
  \vspace{-0.3cm}
  \caption{Evolution of inter-particle separation for the AGB and RGB runs. 
           Time is normalized by the respective initial orbital period: 
           $P_\mathrm{i}=96.5$ for the AGB run and $P_\mathrm{i}=23.2$ days for the RGB run.
           Separation is normalized with respect to the initial orbital separation, 
           $a_\mathrm{i}=124\Rsun$ and $a_\mathrm{i}=49\Rsun$, respectively.
           (See Figure~\ref{fig:mass_183_143} for the separation evolution with days as the unit of time.)
           \label{fig:separation}
          }
\end{figure}

\begin{figure*}
  %produced using force_comparison_183_143.py
  \vspace{-0.3cm}
  \includegraphics[width=\textwidth,clip=true,trim= 0 0 0 5]{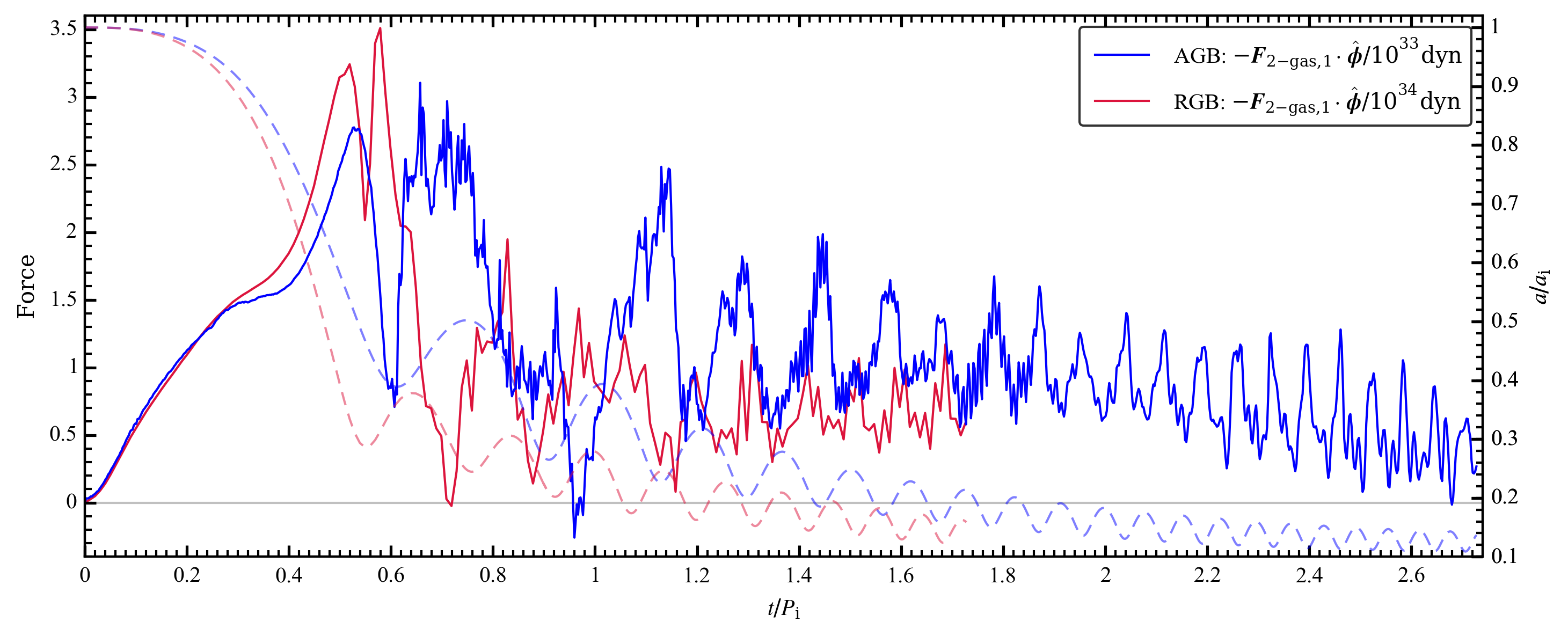}
  \vspace{-0.3cm}
  \caption{Azimuthal component of the drag force on the secondary (labeled `2') 
           in the reference frame of the primary core particle (labeled `1'),
           for the AGB run (blue) and the RGB run (red).
           \label{fig:force}
          }
\end{figure*}

Initial values of the various energy components associated with the particles, or integrated over the envelope gas
(not including ambient gas) are listed in Table~\ref{tab:initial_energy_terms} for the initial orbital separation
of $124\Rsol$ as well as the Roche limit separation \citep{Eggleton83} of $284\Rsol$ 
(see \citet{Chamandy+19a} for details and  the RGB simulation).
These values can be used to estimate $a\final$ from equation~\eqref{alpha}, given a choice for $\alpha\CE$.
The results are shown in Table~\ref{tab:alpha} for both the AGB and RGB runs.
The envelope would thus be expected to be ejected with greater $a\final$ for the AGB run than for the RGB run, 
assuming similar values of $\alpha\CE$, although this assumption may not be justified \citep{Iaconi+Demarco19}.

\section{Simulation Results}
\label{sec:results}

\subsection{Orbital Evolution}\label{sec:orbit}
In Figure~\ref{fig:separation} we show the inter-particle separation $a$, normalized by the initial separation $a\init$, 
plotted against time in initial orbital periods $P\init$, for the AGB (blue) and RGB (red) runs.
In these units, the separation evolution for the two runs is fairly similar,
but the plunge of the secondary (here defined to be down to the first periastron) 
is slightly slower and  shallower in  $a/a\init$ by $\sim 4/3$ for
the AGB run.   However, by the tenth apastron passage, 
which is just prior to the end of the RGB simulation and at $t/P\init\approx2$ in the AGB simulation, 
this factor has reduced to about $9/8$, 
implying that the AGB run tightens faster when time is measured in orbits and distance in $a/a\init$.

The particles have not reached a stationary orbit by the end of either simulation,  
since $\abar$, the time-averaged value of $a$ over one orbital revolution, continues to decrease.  
Moreover, the envelope is not fully unbound (Section~\ref{sec:mass}).
Hence, if  $\abar$ at the end of the simulation 
were less than $a\final$ predicted for $\alpha\CE=1$ in Table~\ref{tab:alpha}, 
then $\abar$ could have been used to place an upper limit on $\alpha\CE$.%
\footnote{A careful comparison between theory and simulation might try to account for non-circularity of the orbit,
but this detail is not necessary for present purposes.}
At the end of the AGB run $\abar\approx15.5\Rsun$,
or about $1.6$ times larger than the threshold value of $9.8\Rsun$ needed to 
constrain $\alpha\CE$ in this way.
At ten orbits, corresponding to the end of the RGB run, 
$\abar\approx19.5\Rsun$ for the AGB run and $\abar\approx7.0\Rsun$ for the RGB run,
so we are slightly closer to the value of $\abar$ needed to place an upper limit on $\alpha\CE$ in the AGB run -- a ratio of $2.0$, 
as compared to $2.7$ for the RGB run.

\subsection{Drag Force Evolution}\label{sec:force}
The azimuthal component of the gas dynamical friction force on the secondary, 
computed in the non-inertial rest frame of the primary core particle, 
$\bm{F}_\mathrm{2-gas,1}\cdot\bm{\hat{\phi}}$,
is shown in Figure~\ref{fig:force} for the AGB and RGB runs.
This  frame is chosen to facilitate comparison with theory and local ``wind tunnel'' simulations;
see \citet{Chamandy+19b} for an extensive discussion of drag force for  RGB CE simulations.

The force in the AGB case evolves similarly  to that of the RGB case.
At late times the force varies with the same periodicity as $a$, but with a half-period phase difference.
At early times the evolution is also similar to the RGB case, and in both cases the force momentarily declines to $\approx0$ 
around the time of the second periastron passage.
However, the overall magnitude of the force is about an order of magnitude lower in the AGB case.

A simple estimate based on Bondi-Hoyle-Lyttleton theory \citep{Hoyle+Lyttleton39,Bondi+Hoyle44} 
gives $F\f= 4\uppi\Gn^2M\2^2\rho\f v\f/(c\f^2+v\f^2)^{3/2}$,  
where $\rho\f$, $v\f$, and $c\f$ are, respectively, the original gas density, secondary orbital speed and sound speed,
computed using the unperturbed primary at radius $r=a(t)$
(the orbital speed is computed using the mass interior to the orbit).
In both runs, this formula correctly predicts the drag force to within a factor of $\sim2$ 
just prior to the first periastron passage,
and at late times correctly predicts the periodicity and phase, but greatly overestimates the magnitude.
At ten orbits (at apastron: $t\approx193\da$ for the AGB run and $t\approx40\da$ for the RGB run)
we find $\bm{F}_\mathrm{2-gas,1}\cdot\bm{\hat{\phi}}/F\f\approx0.05$ for both runs.
At $t\sim20$ orbits, around the end of the AGB run, the drag force, averaged over a few periods, is about $0.03F\f$.
While this discrepancy is slightly reduced in the RGB case  when  density stratification in the surrounding medium is accounted for
\citep{Dodd+Mccrea52},  the discrepancy in the AGB case remains about the same.
The more refined treatment of \citet{Ostriker99} was also found to be inadequate in general \citep{Chamandy+19b}. 
Discrepancies arise because the assumptions of these models are not always justified in the CE context.

Recent studies have progressed our understanding of how the dynamical friction force on the perturbers in a gaseous medium 
behaves under  complicating conditions that are present in the CE context.  
These include curvilinear motion, stratification, binarity, nonlinear perturbations owing to large perturber masses, 
and motion of the perturber centre of mass \citep{Sanchez-salcedo+Brandenburg01,Escala+04,Kim+Kim07,Kim+08,Kim10,Sanchez-salcedo+Chametla14}.
Further understanding the drag force evolution in CE simulations will include applying and extending the theory from those studies.
For example, a study exploring the dependence of drag force on  orbital eccentricity could be helpful to understand their mutual feedback and evolution.  Orbital eccentricity might be driven resonantly by the toroidal circumbinary envelope gas \citep{Kashi+Soker11},
or perhaps by the spiral wakes trailing the cores.
 
\subsection{Secondary Accretion and Primary Core Stripping}\label{sec:accretion}
\begin{figure}
  %produced using maccsphere_comparison.py
  \vspace{-0.3cm}
  \includegraphics[width=\columnwidth,clip=true,trim= 0 0 70  5]{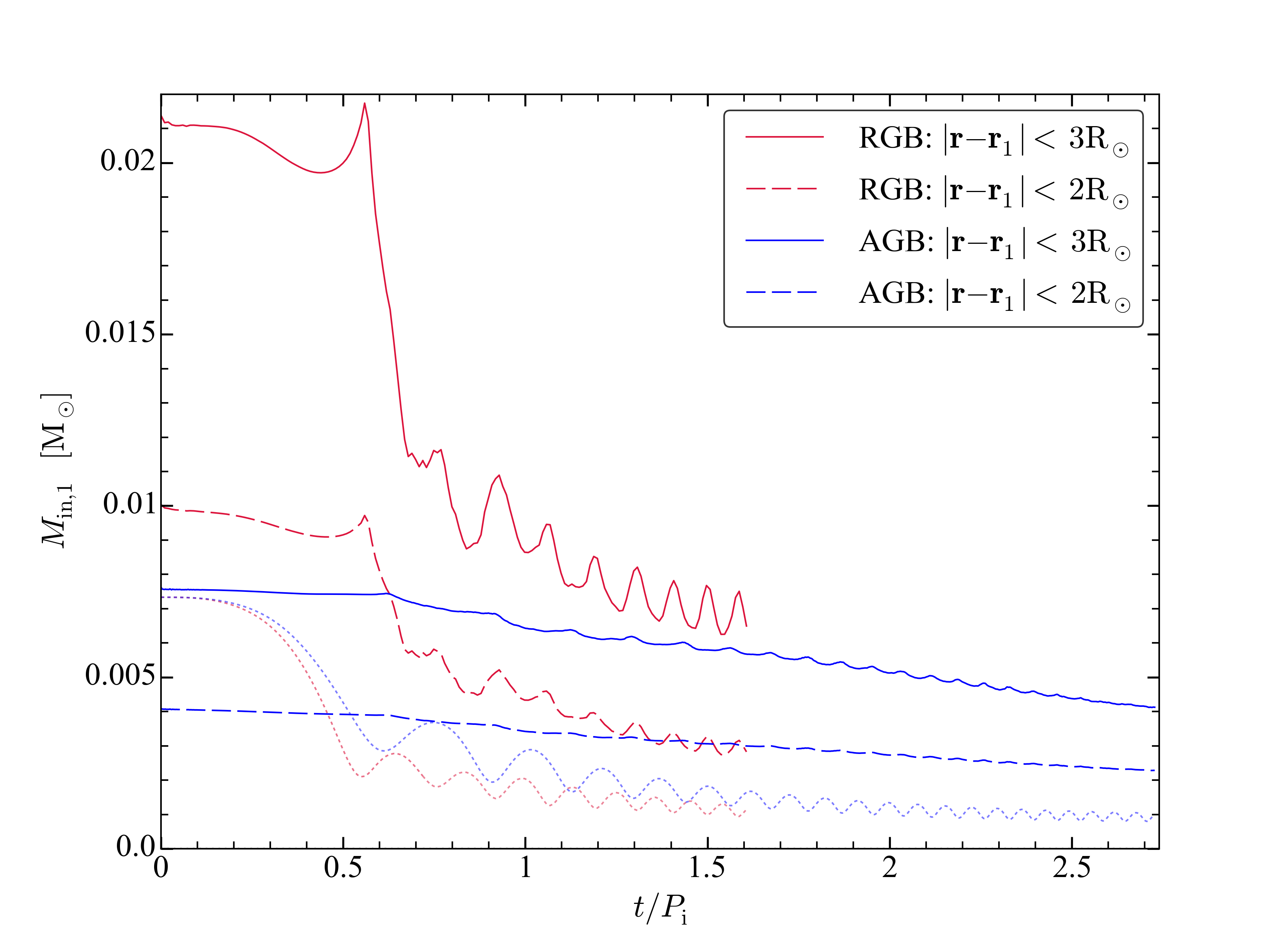}\\
  \includegraphics[width=\columnwidth,clip=true,trim= 0 0 70 35]{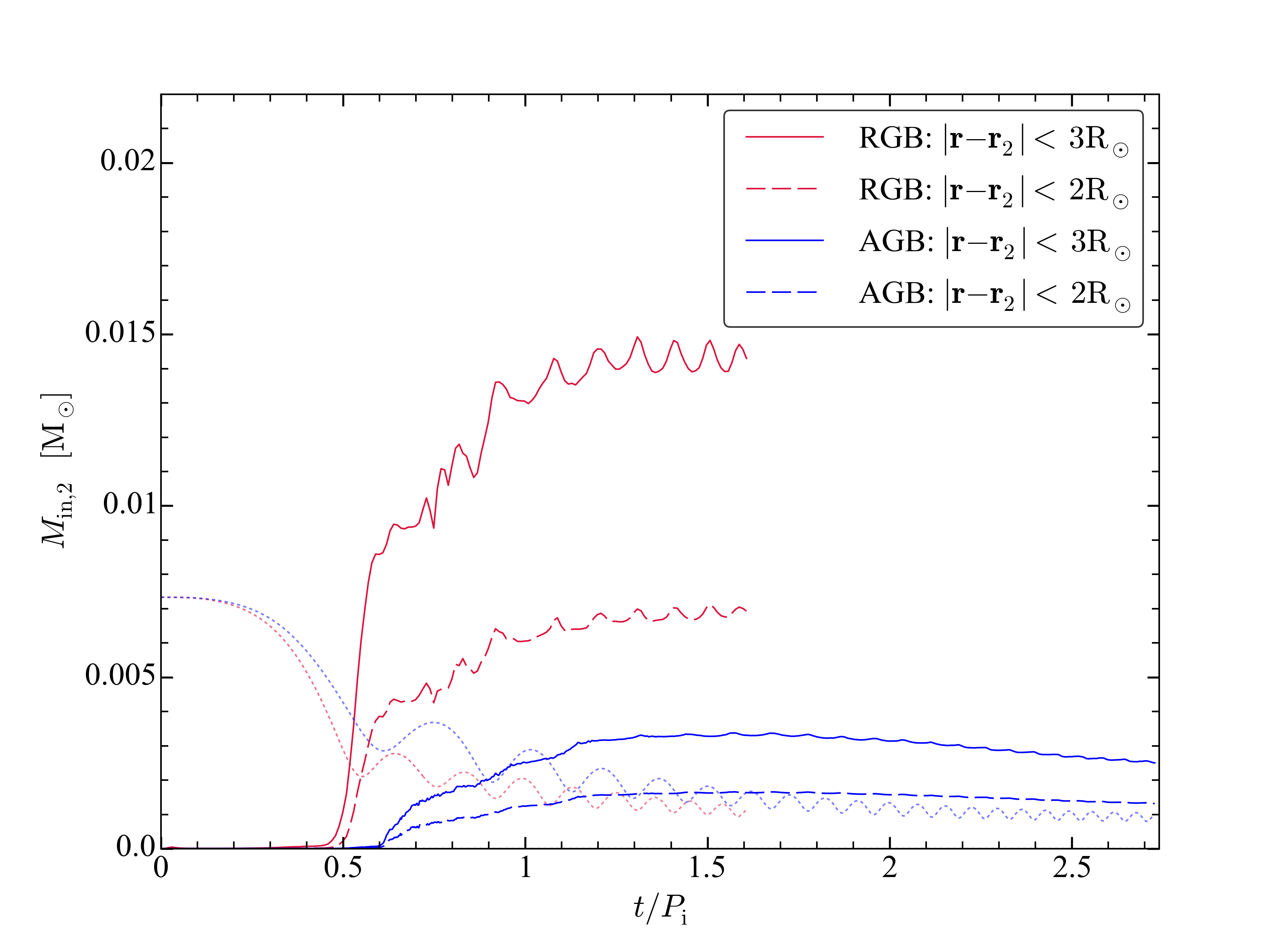}
  \vspace{-0.3cm}
  \caption{Evolution of mass within control spheres around the primary core particle, labeled `1' (top), 
           and the secondary, labeled `2' (bottom), 
           for the AGB run (blue) and the RGB run (red; but see footnote~\ref{foot:rgb}).
           The dotted curves show the separation $a$ in arbitrary units, for reference.
           \label{fig:accretion}
          }
\end{figure}

\begin{figure*}
\begin{tabbing}
\=\includegraphics[scale=0.158,clip=true,trim=  50 202 238 180]{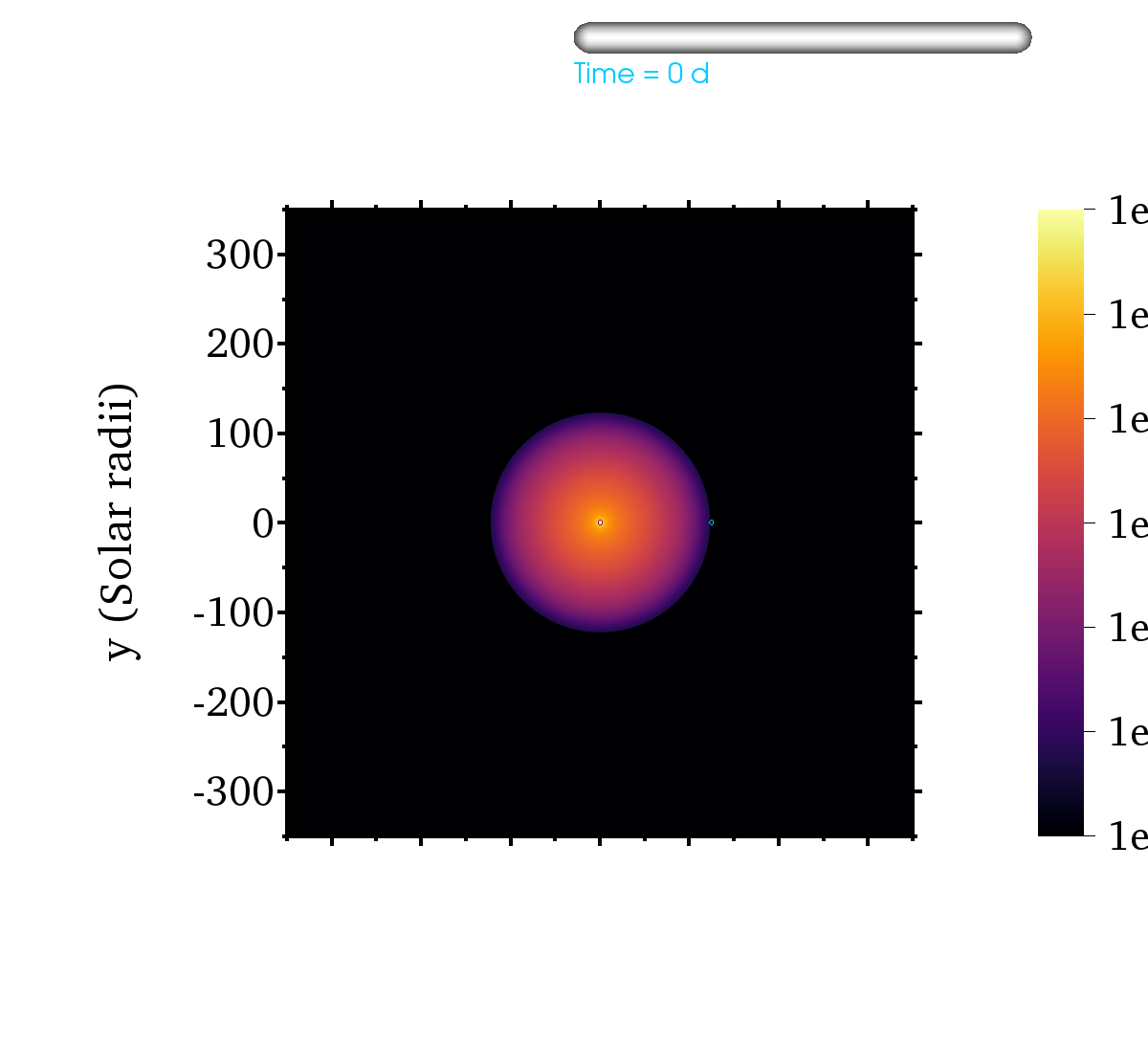}
  \includegraphics[scale=0.158,clip=true,trim= 292 202 238 180]{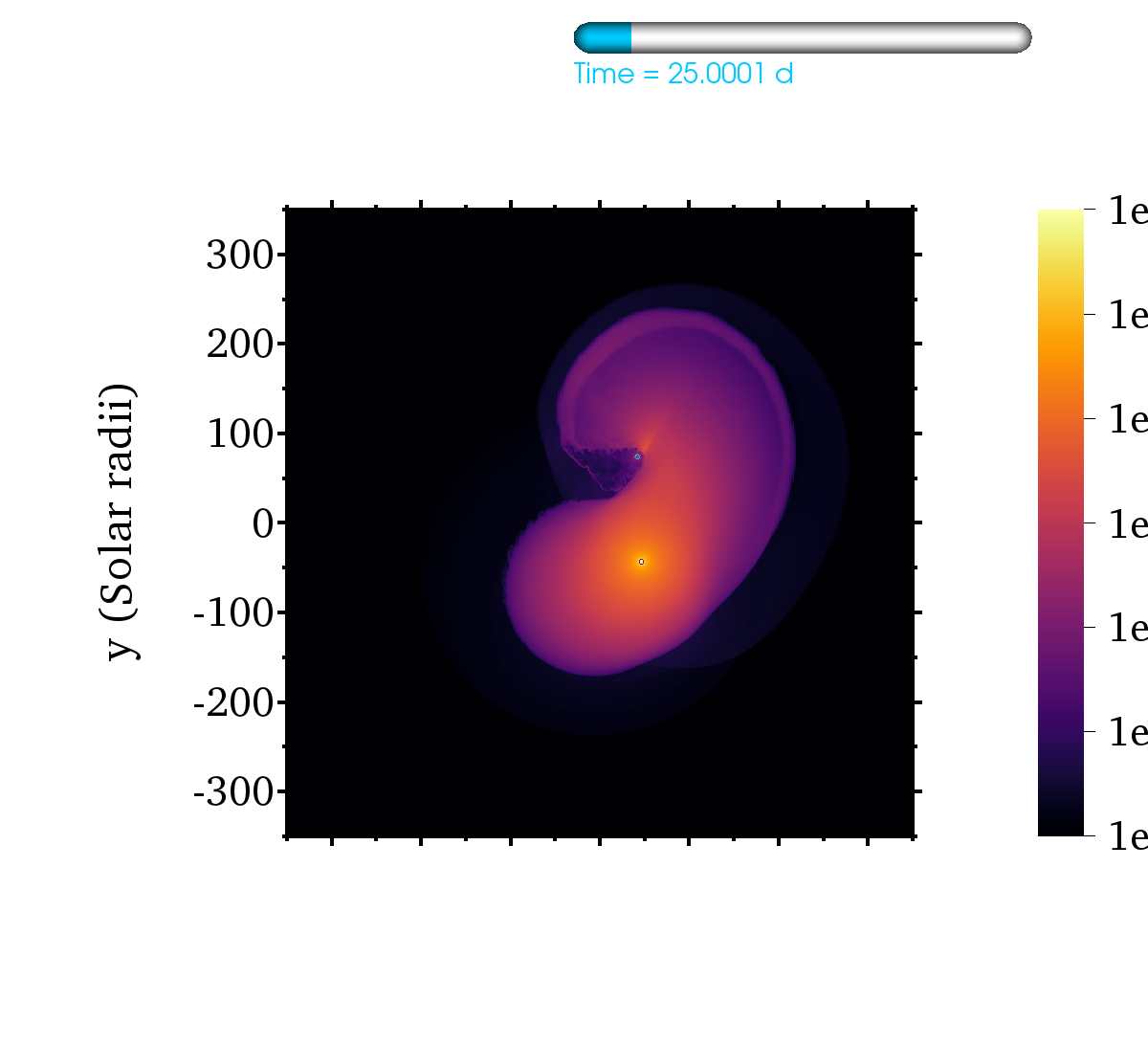}
  \includegraphics[scale=0.158,clip=true,trim= 292 202 238 180]{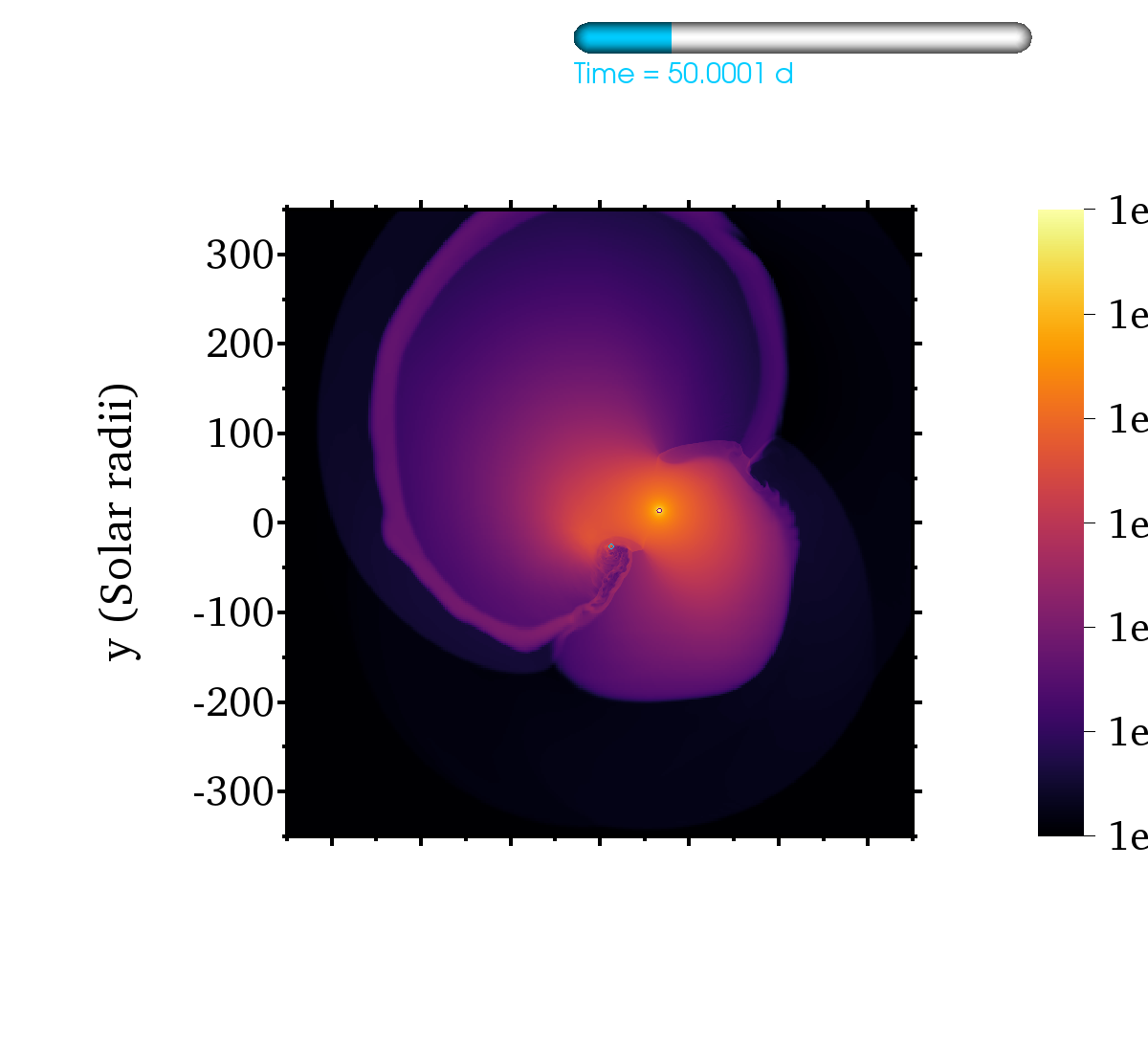}
  \includegraphics[scale=0.158,clip=true,trim= 292 202   0 180]{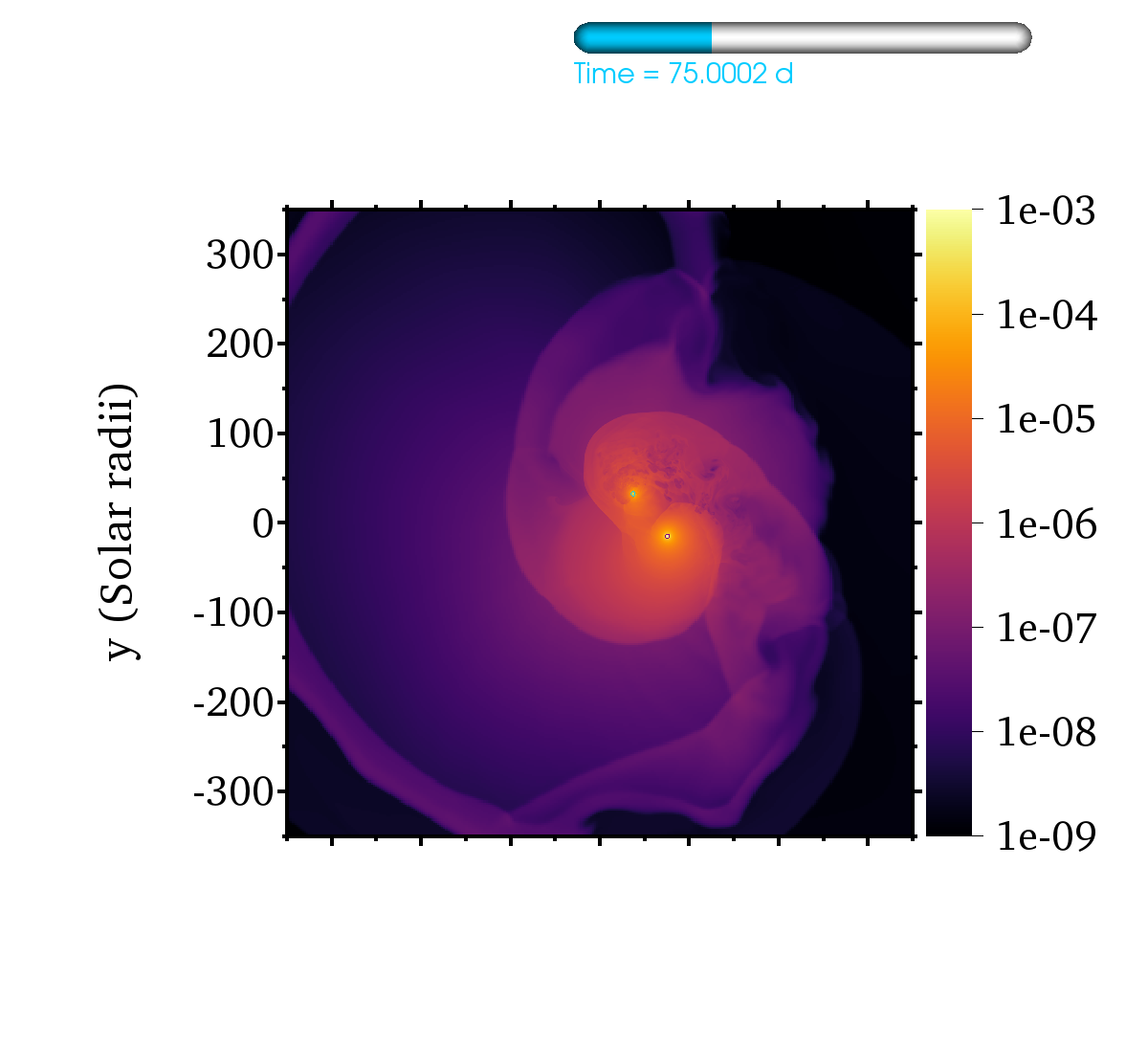}\\
\>\includegraphics[scale=0.158,clip=true,trim=  50 120 238 211]{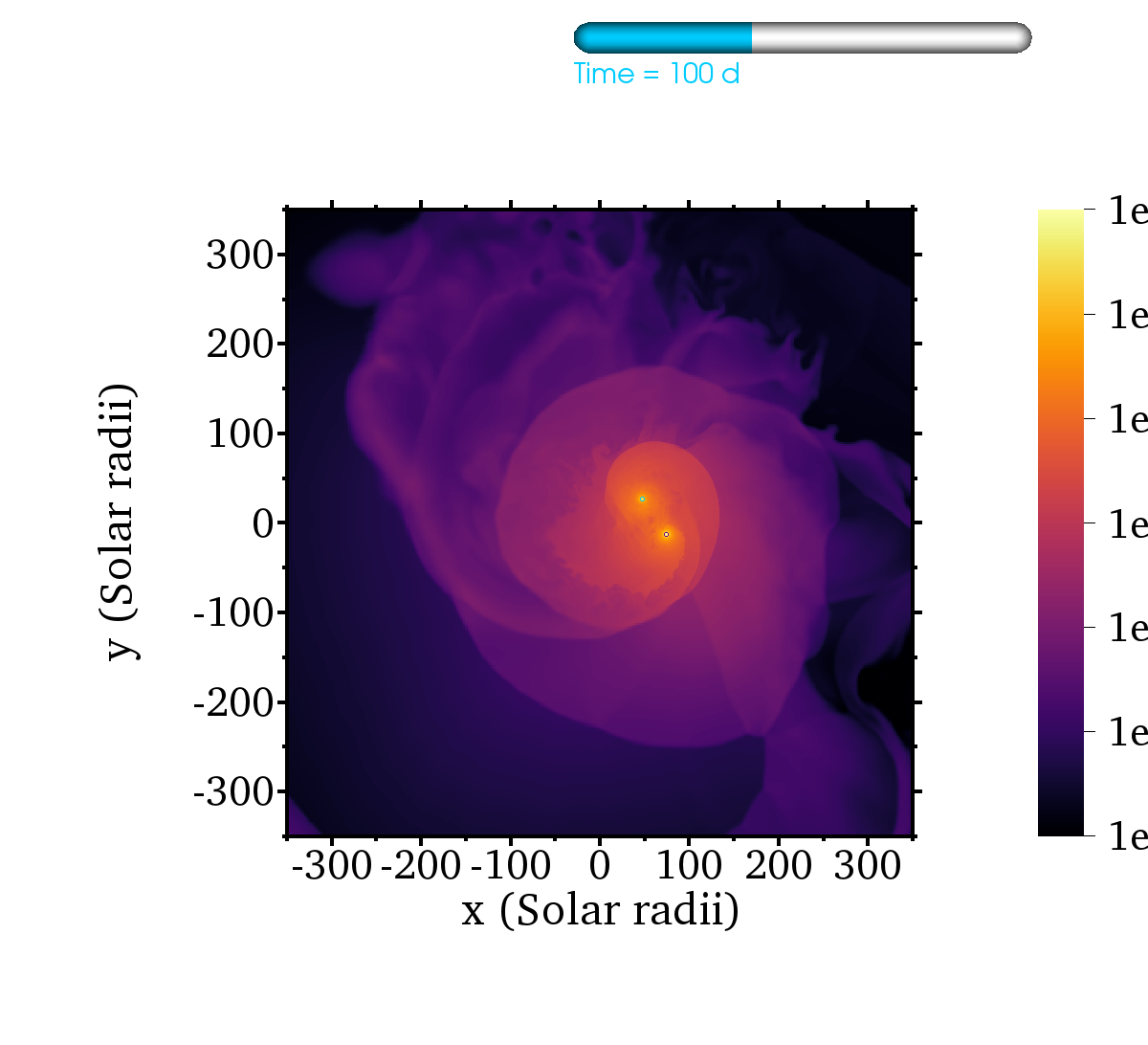}
  \includegraphics[scale=0.158,clip=true,trim= 292 120 238 211]{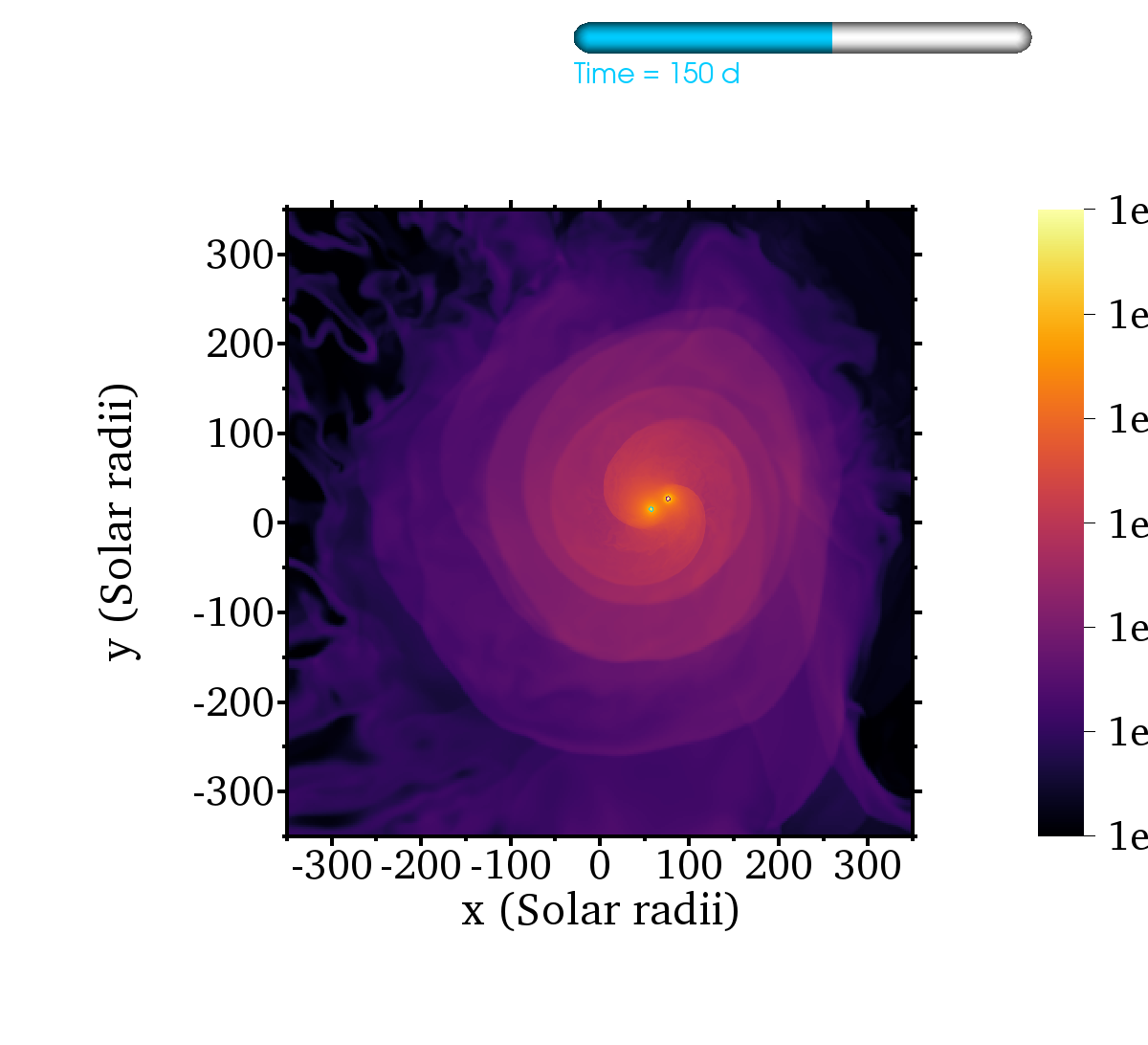}
  \includegraphics[scale=0.158,clip=true,trim= 292 120 238 211]{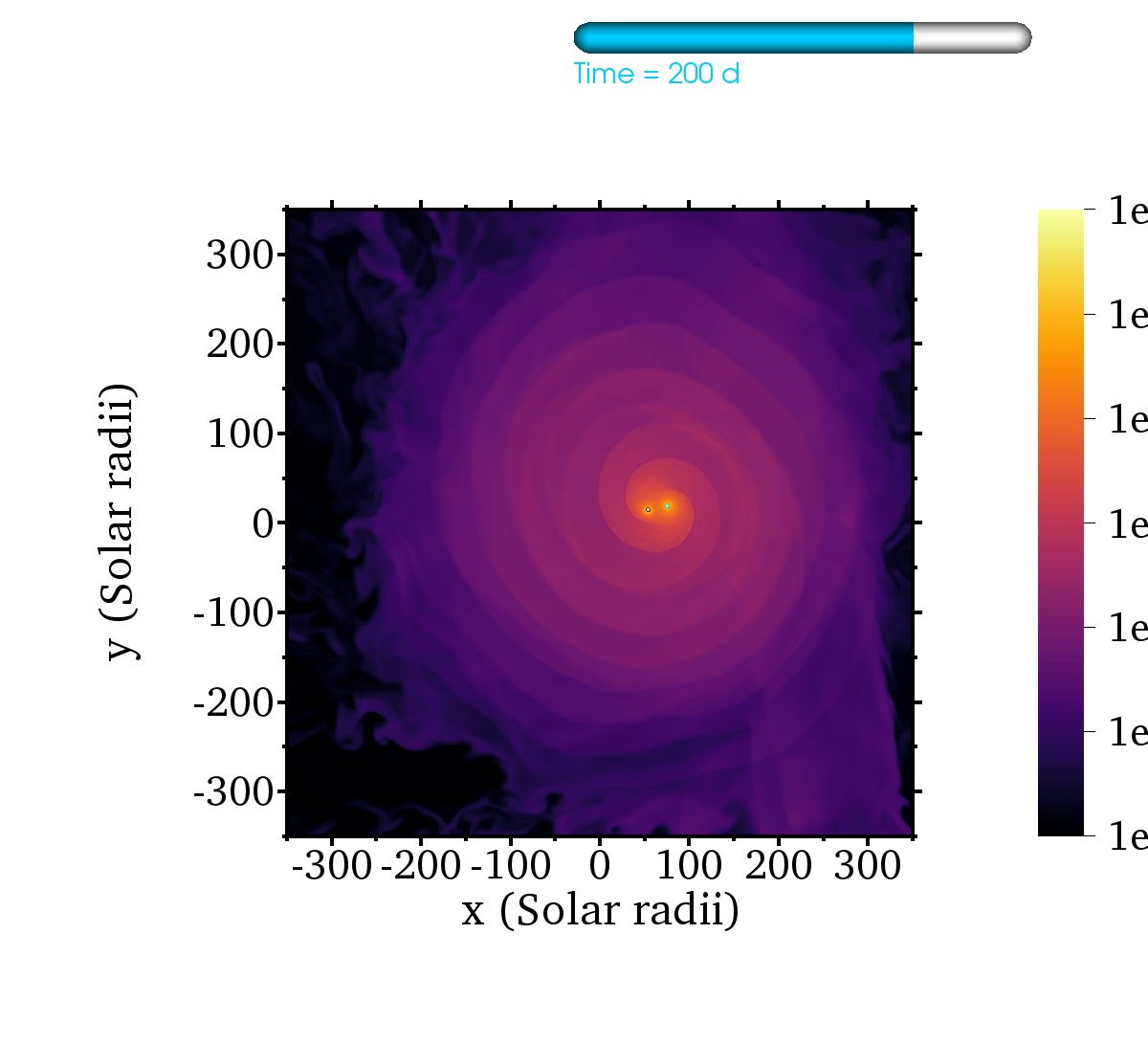}
  \includegraphics[scale=0.158,clip=true,trim= 292 120 238 211]{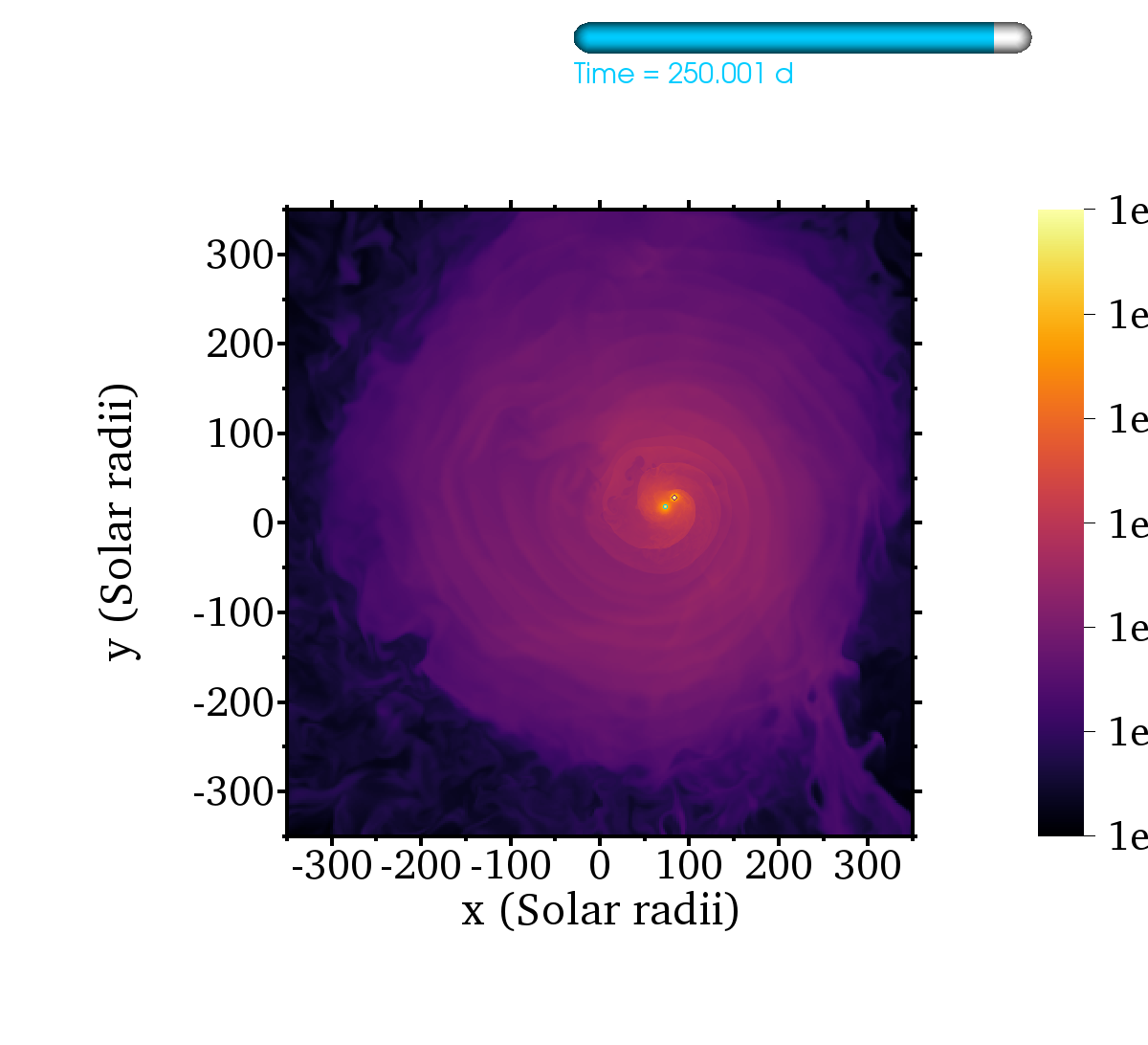}
\end{tabbing}
\caption{Slice through the orbital plane showing the gas density $\rho$ in $\gcmcmcm$. 
         Particle softening spheres (radius $2.4\Rsun$) are shown with purple and blue circles, 
         for the primary core particle and the secondary, respectively.
         Snapshots (row by row from left to right)  show $t=0$, $25$, $50$, $75$, $100$, $150$, $200$, and $250\da$,
         in the simulation rest frame, with the origin of the simulation domain at $(0,0)$.
         \label{fig:density_faceon_183}
        }            
\end{figure*}

\begin{figure*}
\begin{tabbing}
\includegraphics[scale=0.158,clip=true,trim=  50  120 238 180]{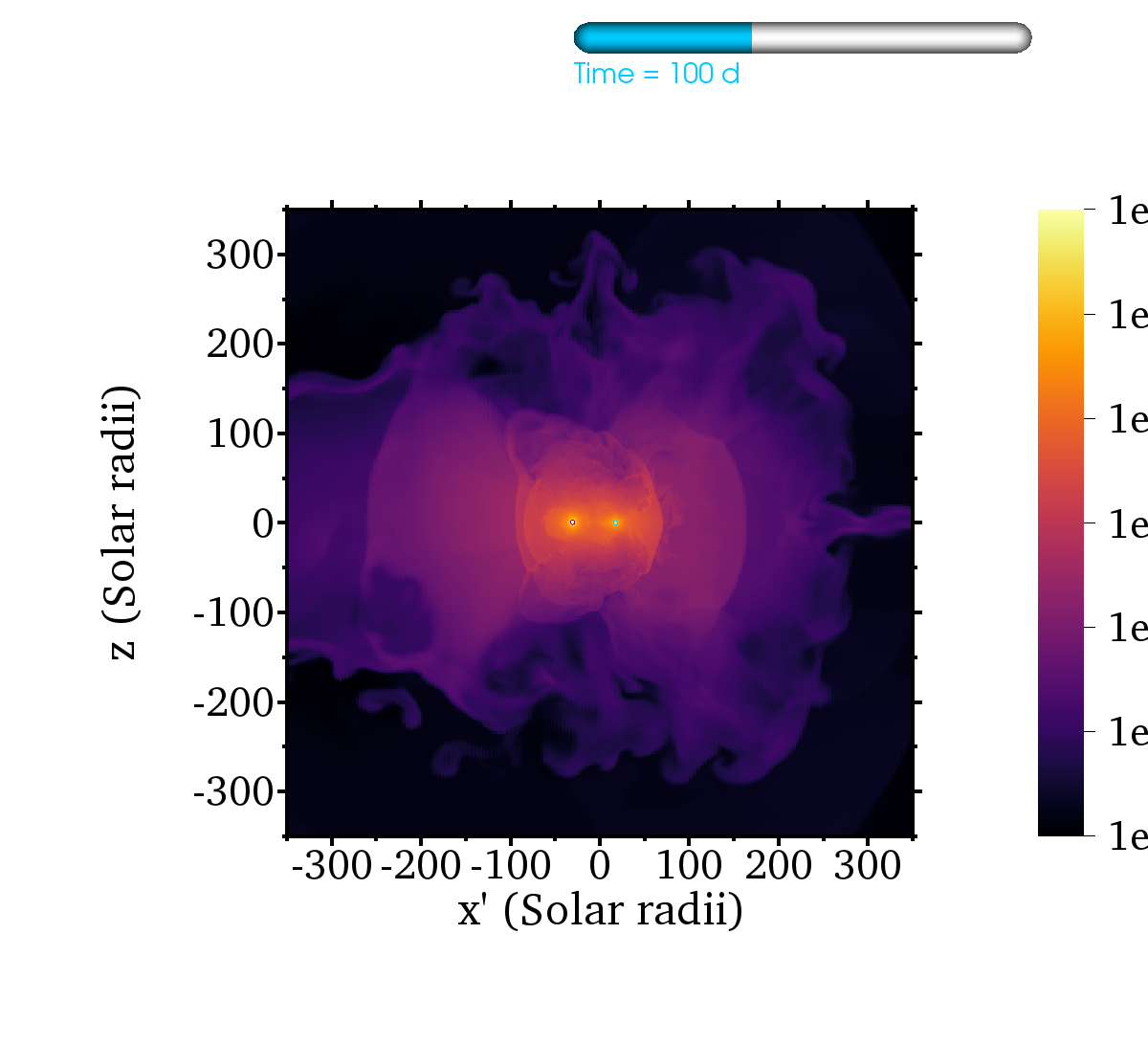}
\includegraphics[scale=0.158,clip=true,trim= 292  120 238 180]{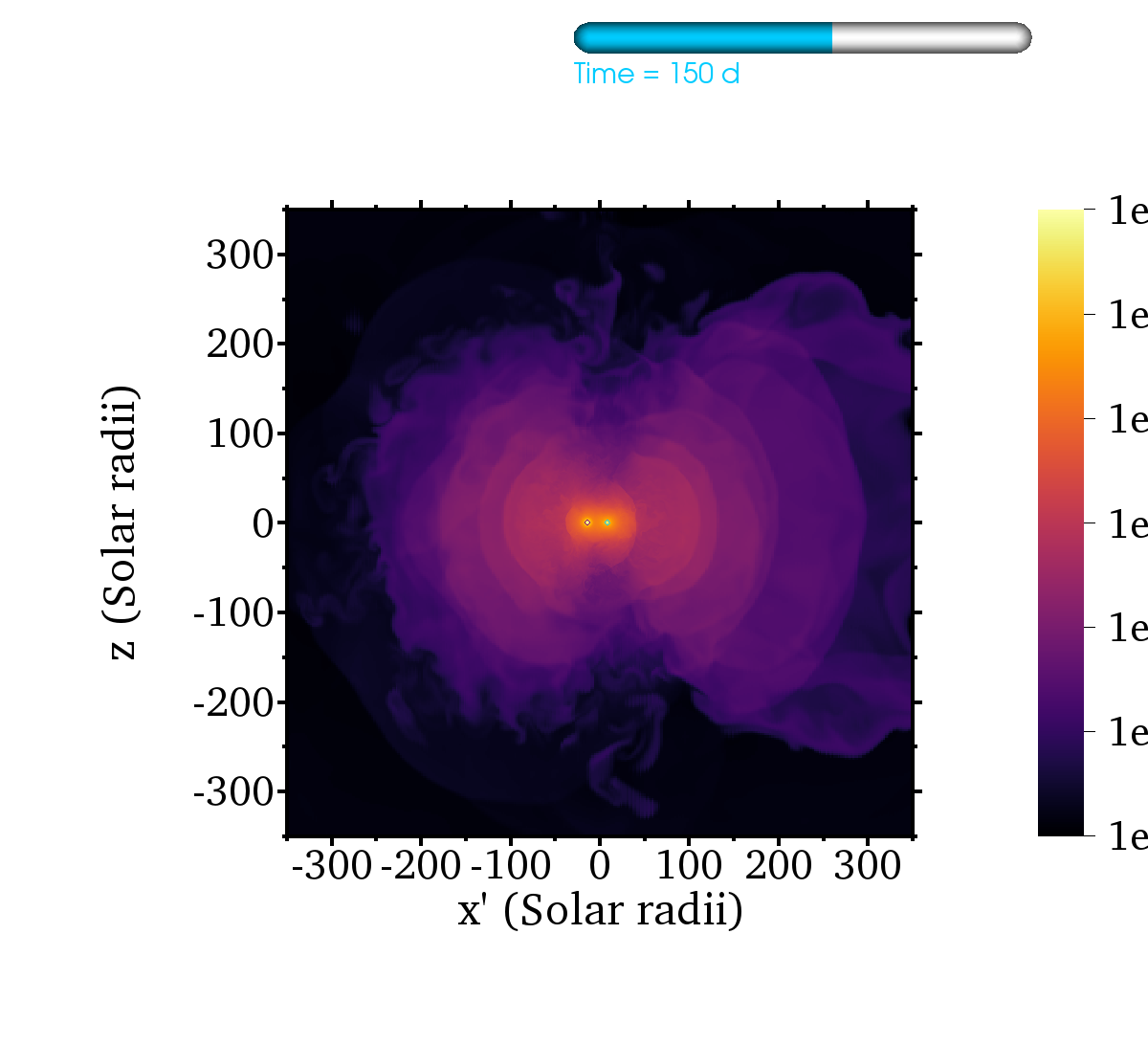}
\includegraphics[scale=0.158,clip=true,trim= 292  120 238 180]{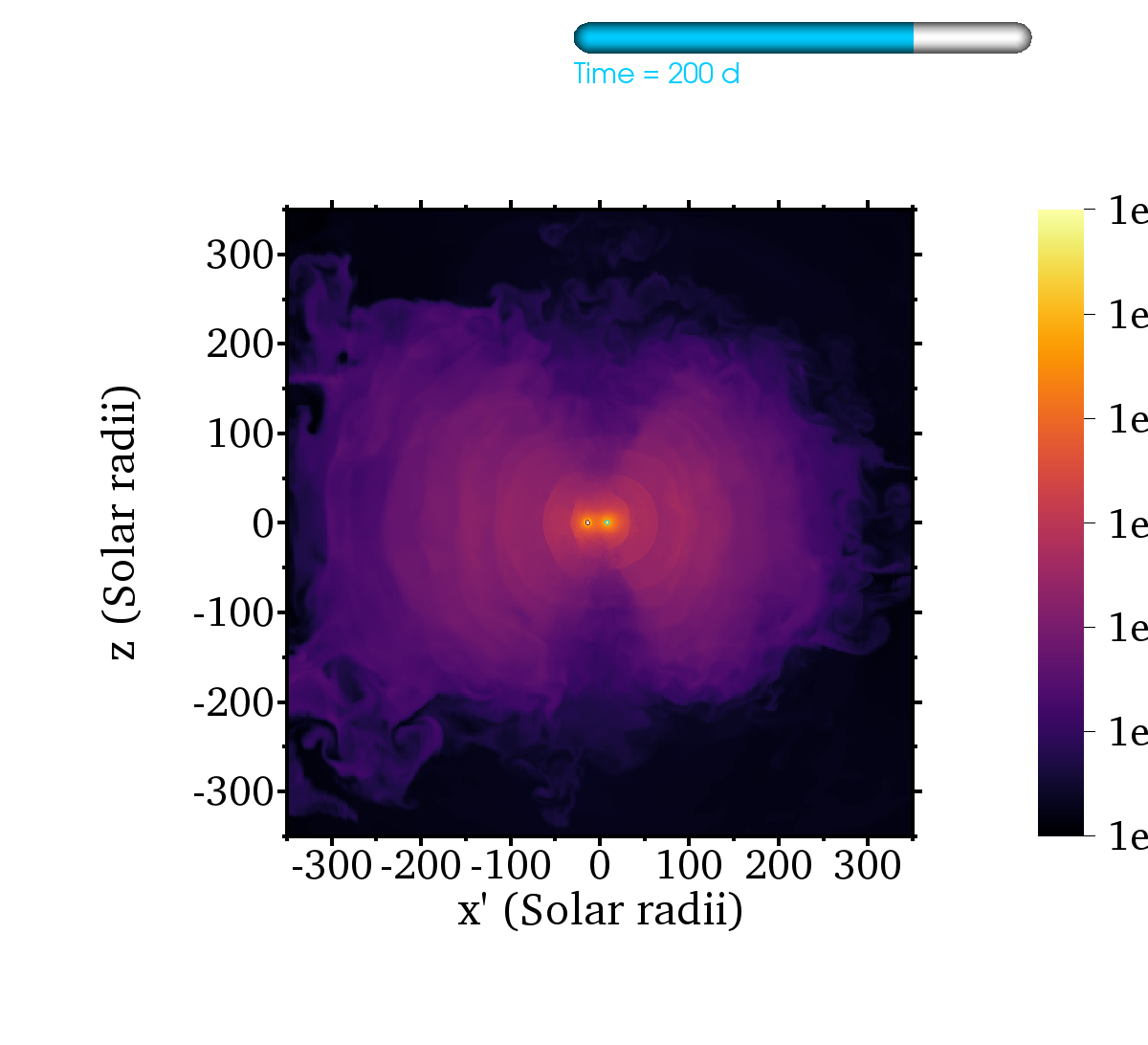}
\includegraphics[scale=0.158,clip=true,trim= 292  120   0 180]{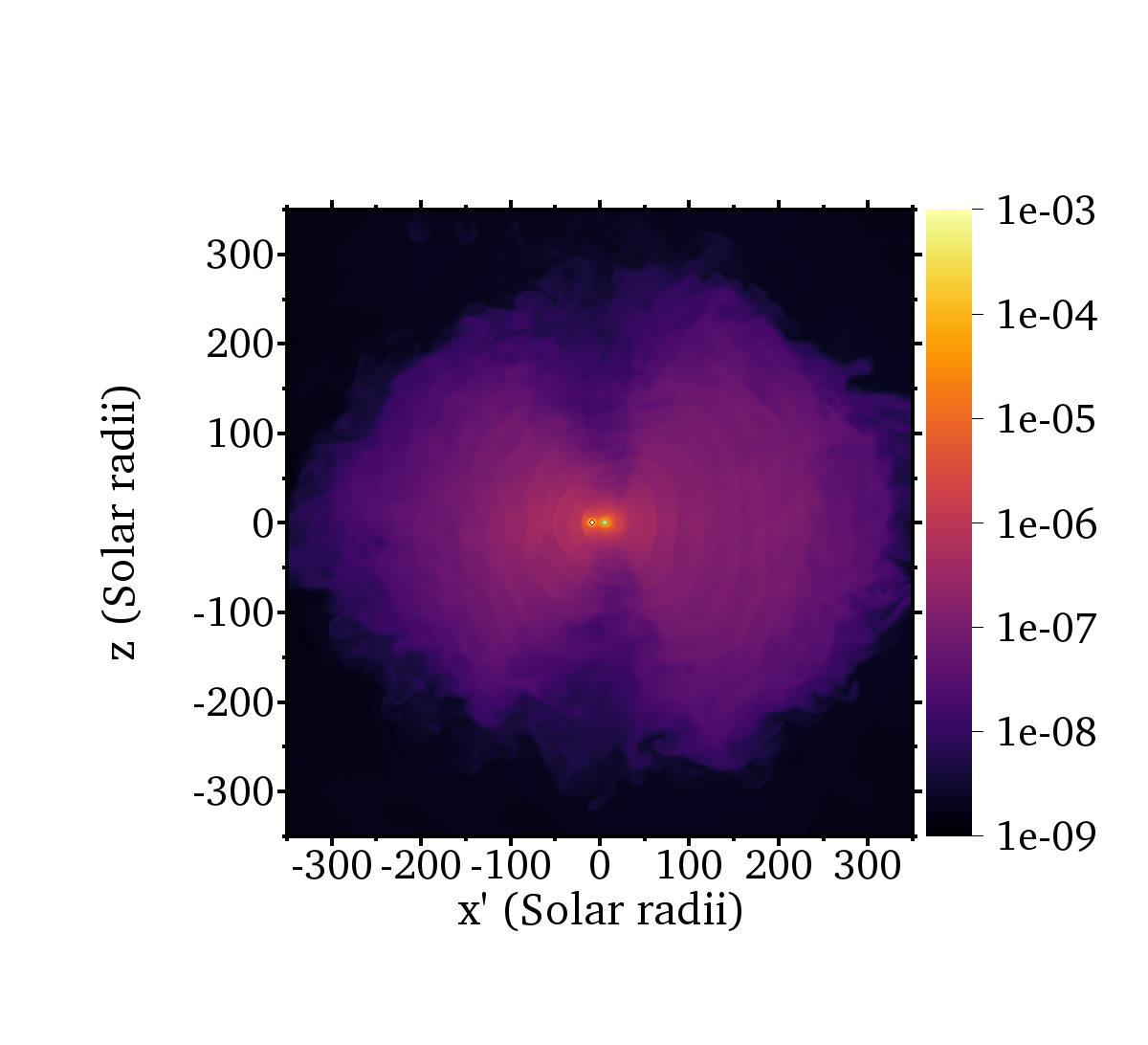}
\end{tabbing}
\caption{Snapshots of gas density in $\gcmcmcm$, at $t=100$, $150$, $200$, and $250\da$,
         showing vertical slices containing both particles.
         The primary core particle is on the left, the secondary is on the right, 
         and the particle centre of mass is placed at the origin.
         \label{fig:density_edgeonthroughparticles_183}
        }            
\end{figure*}

\begin{figure*}
\begin{tabbing}
\=\includegraphics[scale=0.140,clip=true,trim= 50  120  40 200]{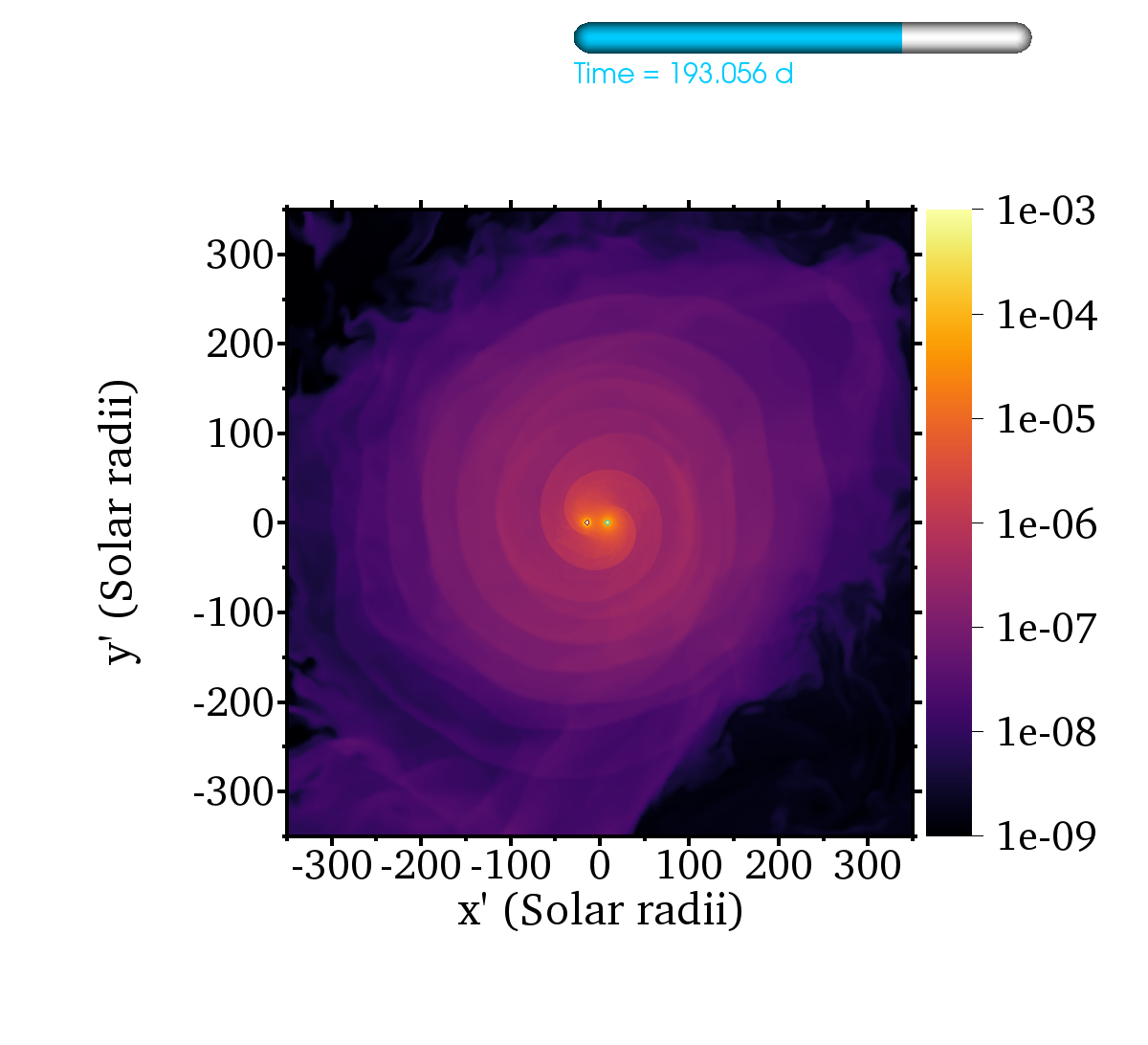}
\=\includegraphics[scale=0.140,clip=true,trim=292  120  80 200]{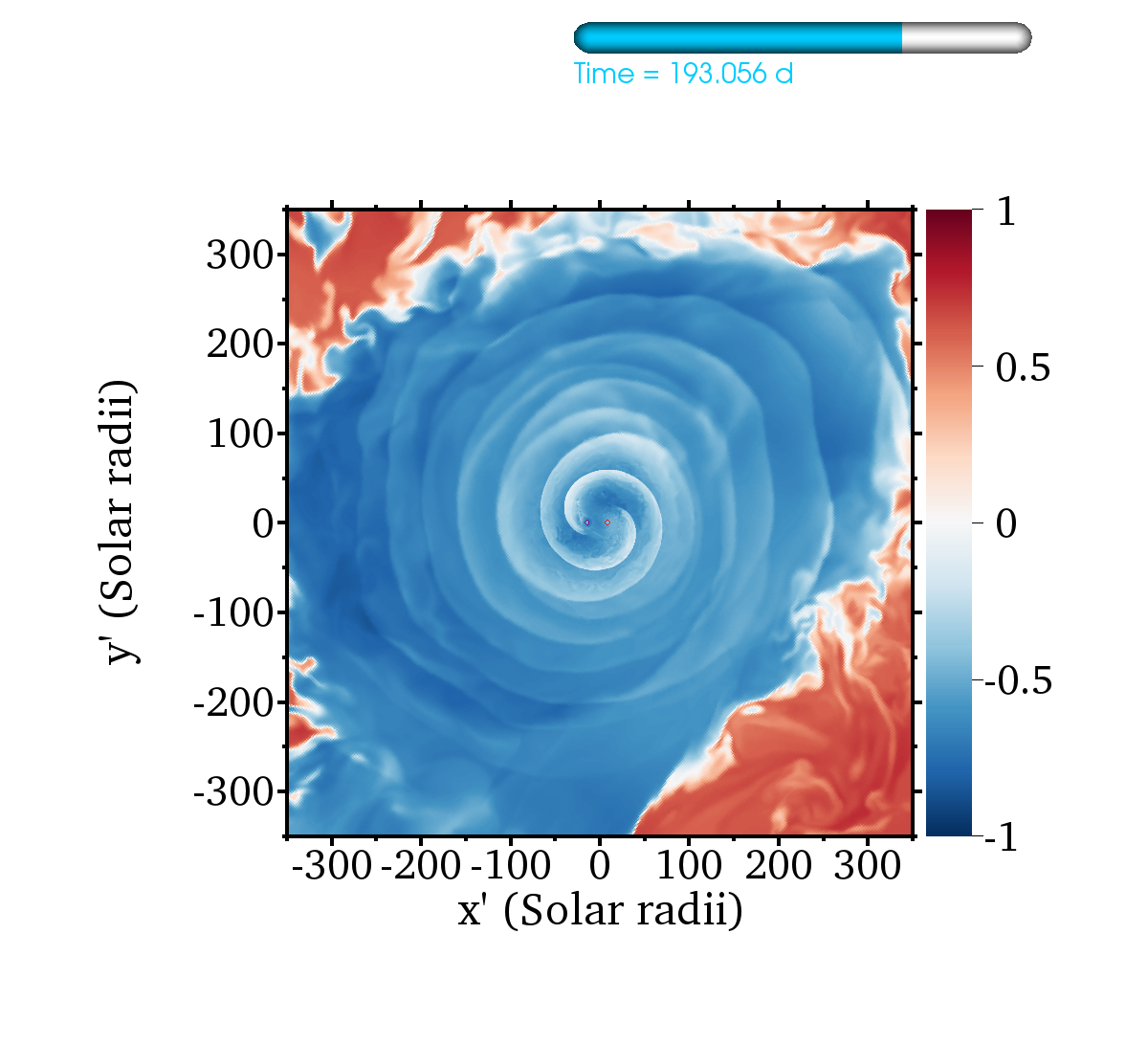}
\=\includegraphics[scale=0.140,clip=true,trim=100  120 238 200]{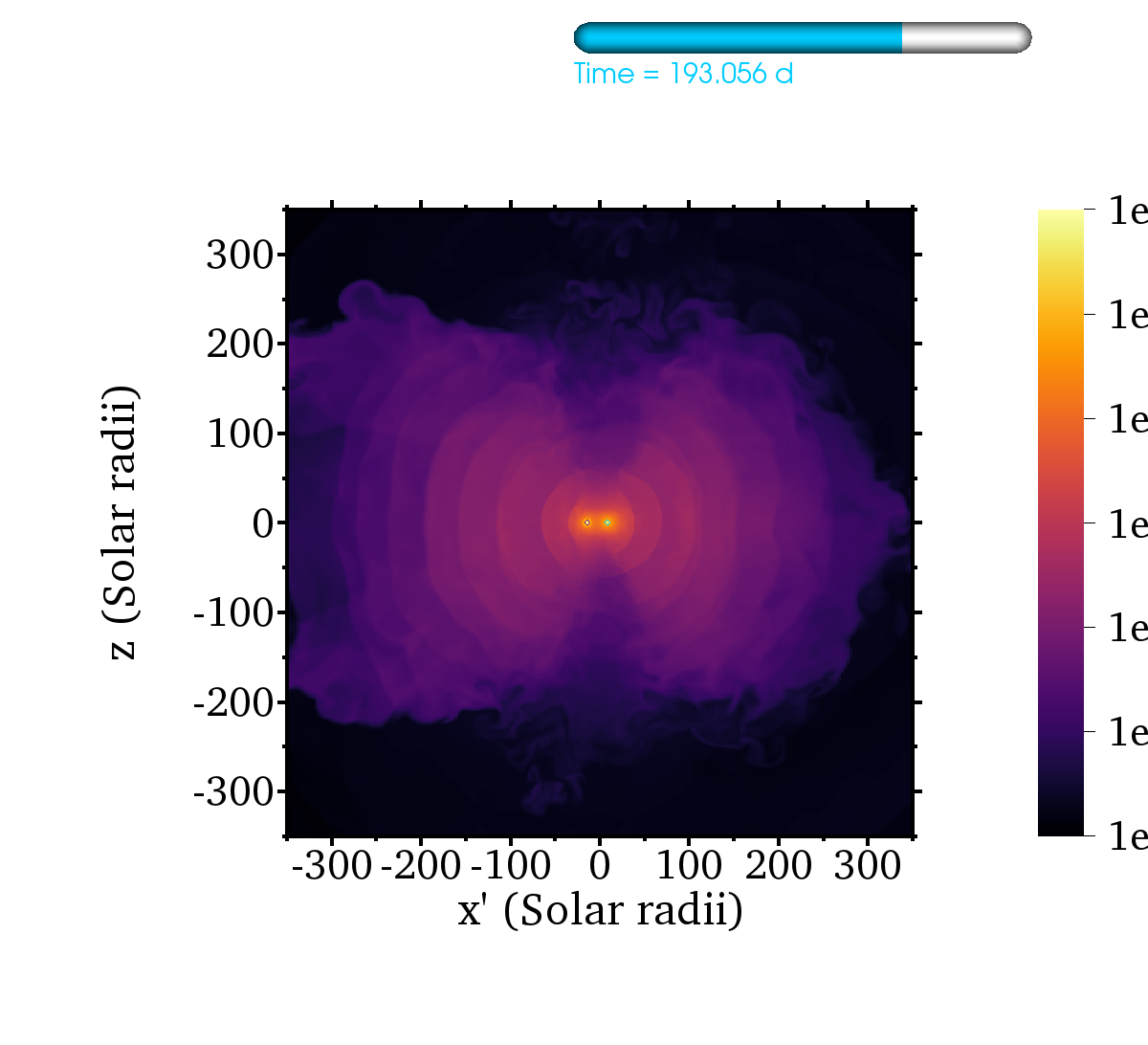}
\=\includegraphics[scale=0.140,clip=true,trim=292  120 238 200]{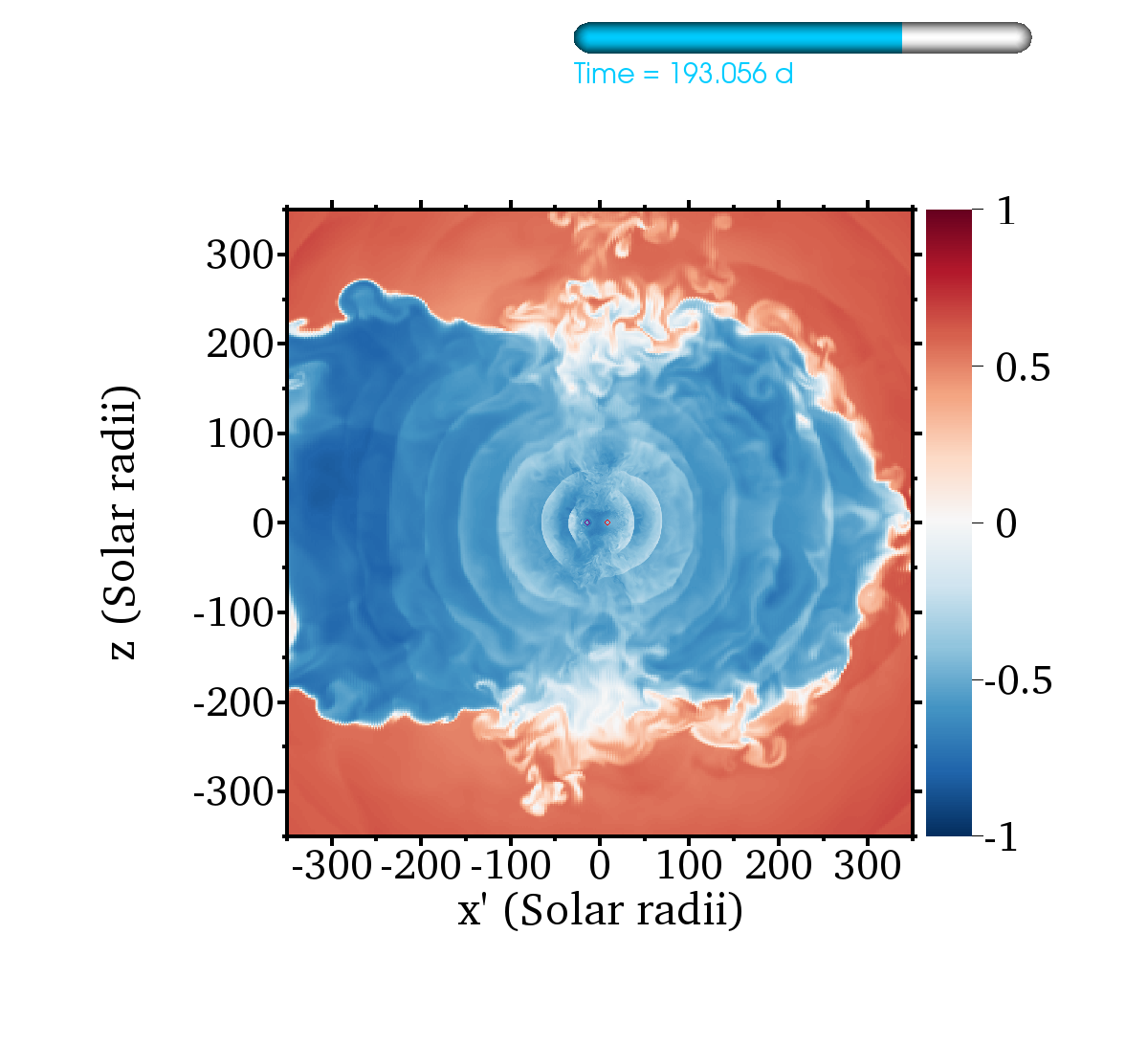}\\
\>\includegraphics[scale=0.140,clip=true,trim= 50  120  40 200]{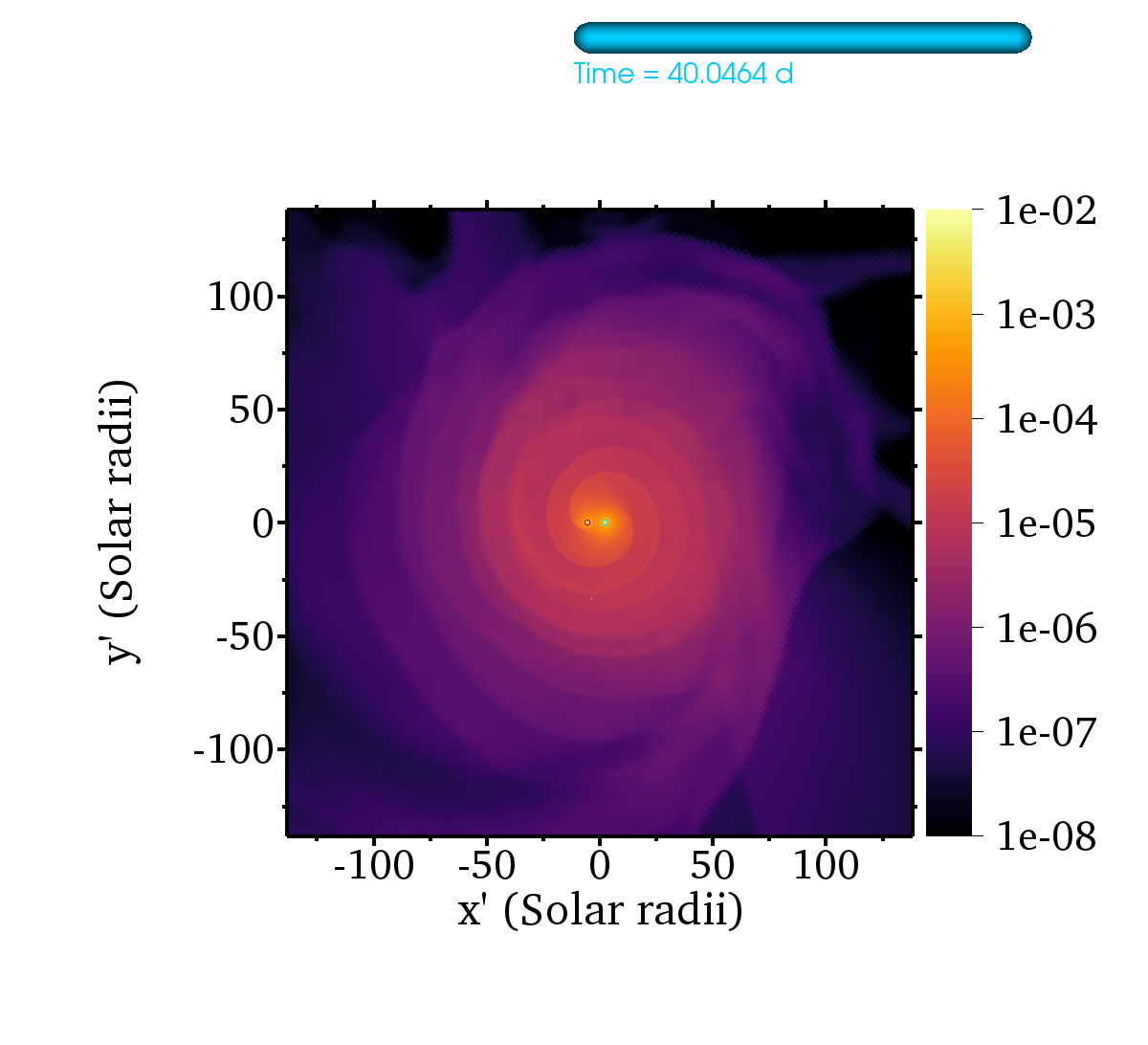}
\>\includegraphics[scale=0.140,clip=true,trim=292  120  80 200]{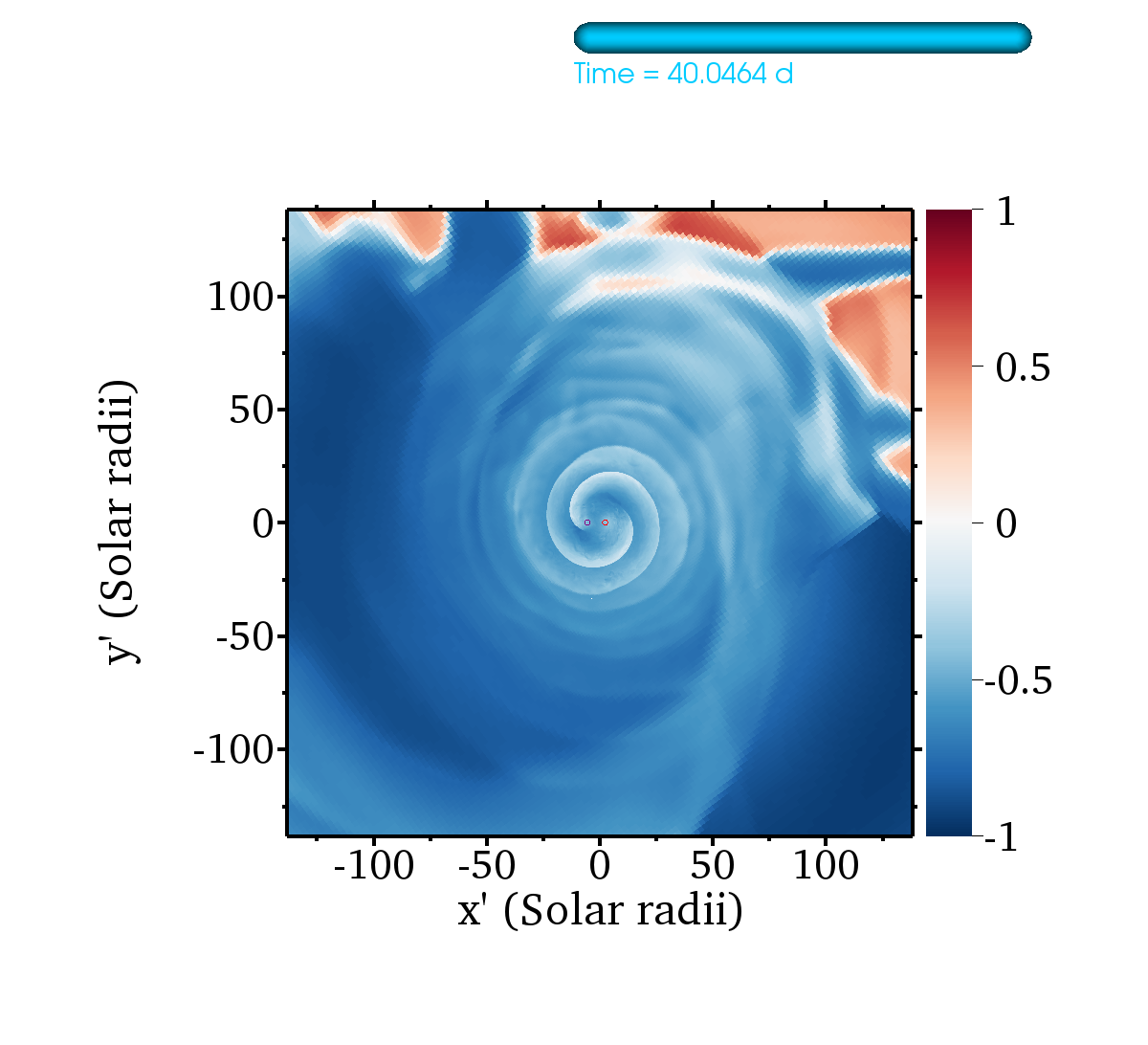}
\>\includegraphics[scale=0.140,clip=true,trim=100  120 238 200]{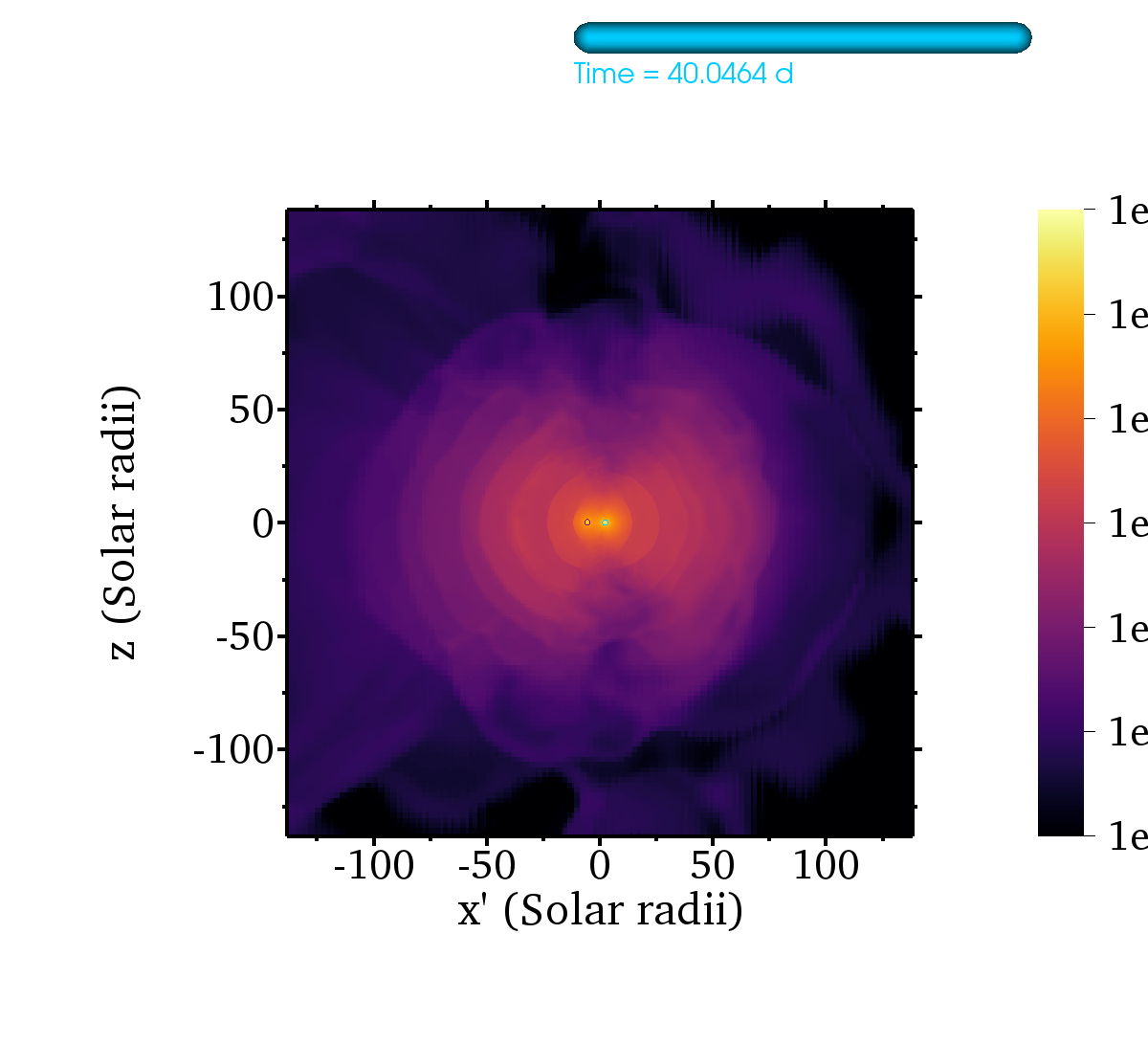}
\>\includegraphics[scale=0.140,clip=true,trim=292  120 238 200]{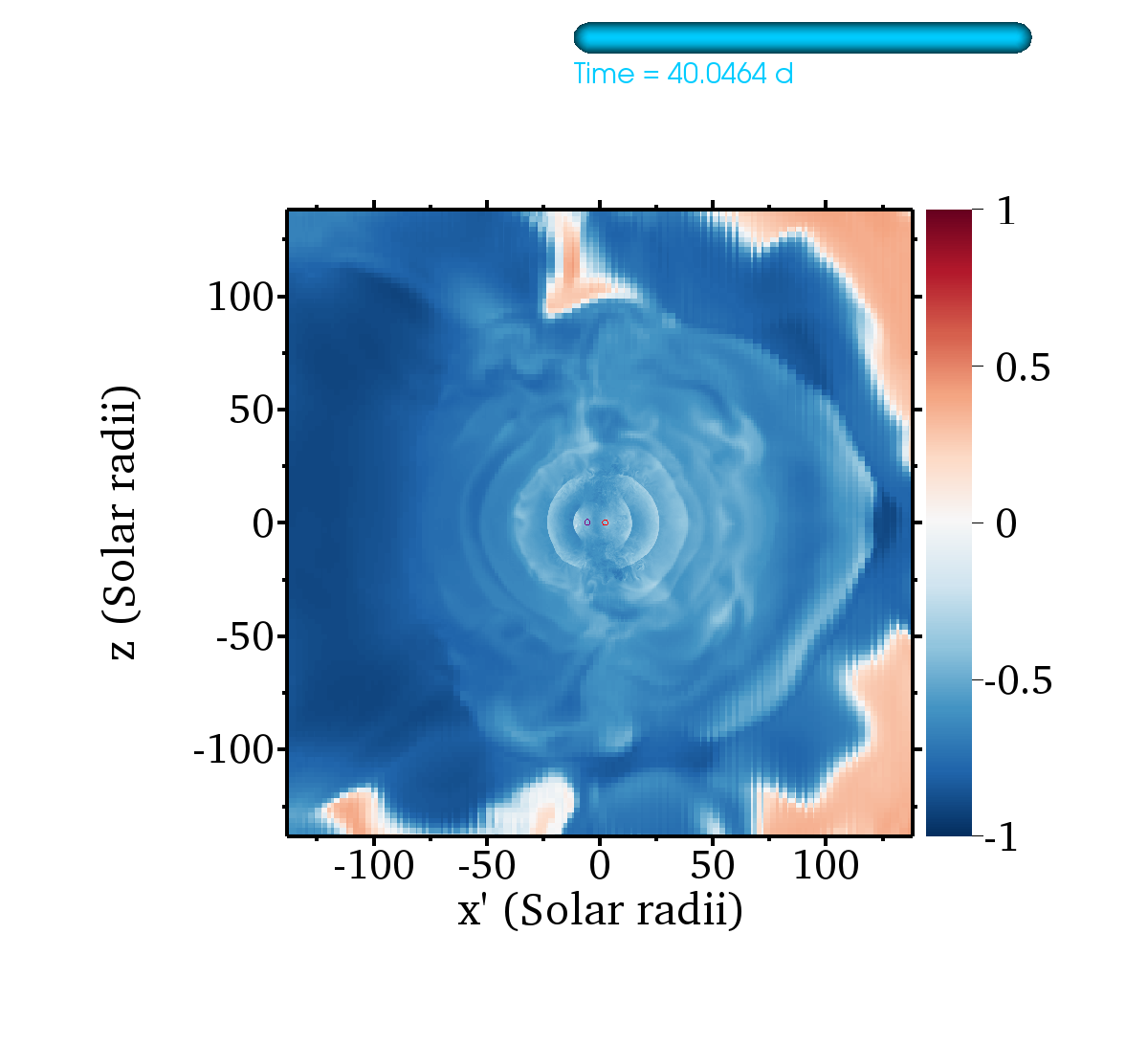}
\end{tabbing}
\caption{Comparison of AGB run (top row), and RGB run (bottom row),
         at 10 orbits, corresponding to $t=193\da$ for the AGB run and to $t=40\da$ for the RGB run.
         The left column shows snapshots of gas density $\rho$ in the orbital plane. 
         The primary core particle is located on the left, the secondary on the right, 
         and the particle centre of mass at the origin. 
         The second column shows the same slice but for the quantity $\widetilde{\mathcal{E}}_\mathrm{gas}$.
         Here blue denotes bound gas and red unbound.
         The third and fourth columns are similar but now for the orthogonal slice, also through the particles.
         The size of the field of view is proportional to the initial orbital separation $a_i$.
         \label{fig:comparison_143_183}
        }            
\end{figure*}

In Figure~\ref{fig:accretion} we plot the integrated mass within control spheres centred on the primary core particle (top panel)
and the secondary (bottom panel) for the AGB run (blue) and RGB run (red).
In each case, results using control spheres of radius $2\Rsun$ (dashed) and $3\Rsun$ (solid) are shown \citep[see also][]{Chamandy+18}.
Solid and dashed curves are separated by about a factor of two in mass but otherwise look similar.
The inter-particle separation for each run is also plotted in arbitrary units, using dotted lines in faint blue (AGB) and faint red (RGB).%
\footnote{\label{foot:rgb} In Figure~\ref{fig:accretion} we have actually used a slightly shorter RGB run (Model~F of \citealt{Chamandy+19a}),
identical to the fiducial RGB run used elsewhere except that the softening length is not halved at $t=16.7\da$ 
but stays constant at $r\soft=2.4\Rsun$, as in the AGB model.
The reason for this choice is that this arbitrary reduction in $r\soft$ has a small 
but significant effect on the mass distribution near the particles,
so this provides a fairer comparison with the AGB run, for which the softening length also remains constant.
For other aspects of our analysis this does not make a significant difference and we use the fiducial RGB run.}

In the RGB run, the mass around the primary core particle, 
as shown by the solid and dashed red lines in the top panel of Figure~\ref{fig:accretion},  
peaks sharply at the first periastron passage, 
and then decreases suddenly until about halfway between the first apastron passage and second periastron passage.
The average mass then decreases secularly, modulated by oscillations such that it peaks at each periastron passage.
These oscillations are in phase with oscillations of the mass around the secondary,
and \citet{Chamandy+18} suggested that the individual mass distributions around the particles 
overlap more as the particles approach, leading to a larger mass within each control sphere.
The mass around the secondary in the RGB run (bottom panel, red) increases dramatically just before the first periastron passage,
before increasing more slowly, and then still more slowly after about the third periastron passage, before leveling off. 

While the overall behaviour in the AGB run is similar, there are differences. 
Most strikingly, the decrease in mass around the primary core particle
after the first periastron passage is much smaller than in the RGB case,
and there is a much smaller amplitude of oscillations.
These differences can be explained qualitatively. 
Firstly, the secondary does not come nearly as close to the primary core particle in the AGB case, 
and is thus less able to tidally disrupt and draw matter away from the AGB core at early times. 
Also,  the individual ``envelopes'' around the particles do not overlap as much. 
This may explain the smaller oscillations,
though a more detailed explanation is warranted.
Secondly, the AGB core represented by the primary core particle is about $1.5$ times more massive than the RGB core, 
so it retains the more gas within the control sphere in spite of strong tidal perturbations.  
Thirdly, the gas is more centrally condensed around the core in the AGB case compared to the RGB case 
(Section~\ref{sec:morphology} and Appendix~\ref{sec:ic}).
It would be interesting to use theoretic models and local hydrodynamic wind tunnel simulations 
to further study the evolution of the mass distributions around the particles during CE evolution.
\begin{figure}
  %produced using anal_mass_183_143.py
  \vspace{-0.3cm}
  \includegraphics[width=\columnwidth,clip=true,trim= 0 0 0 5]{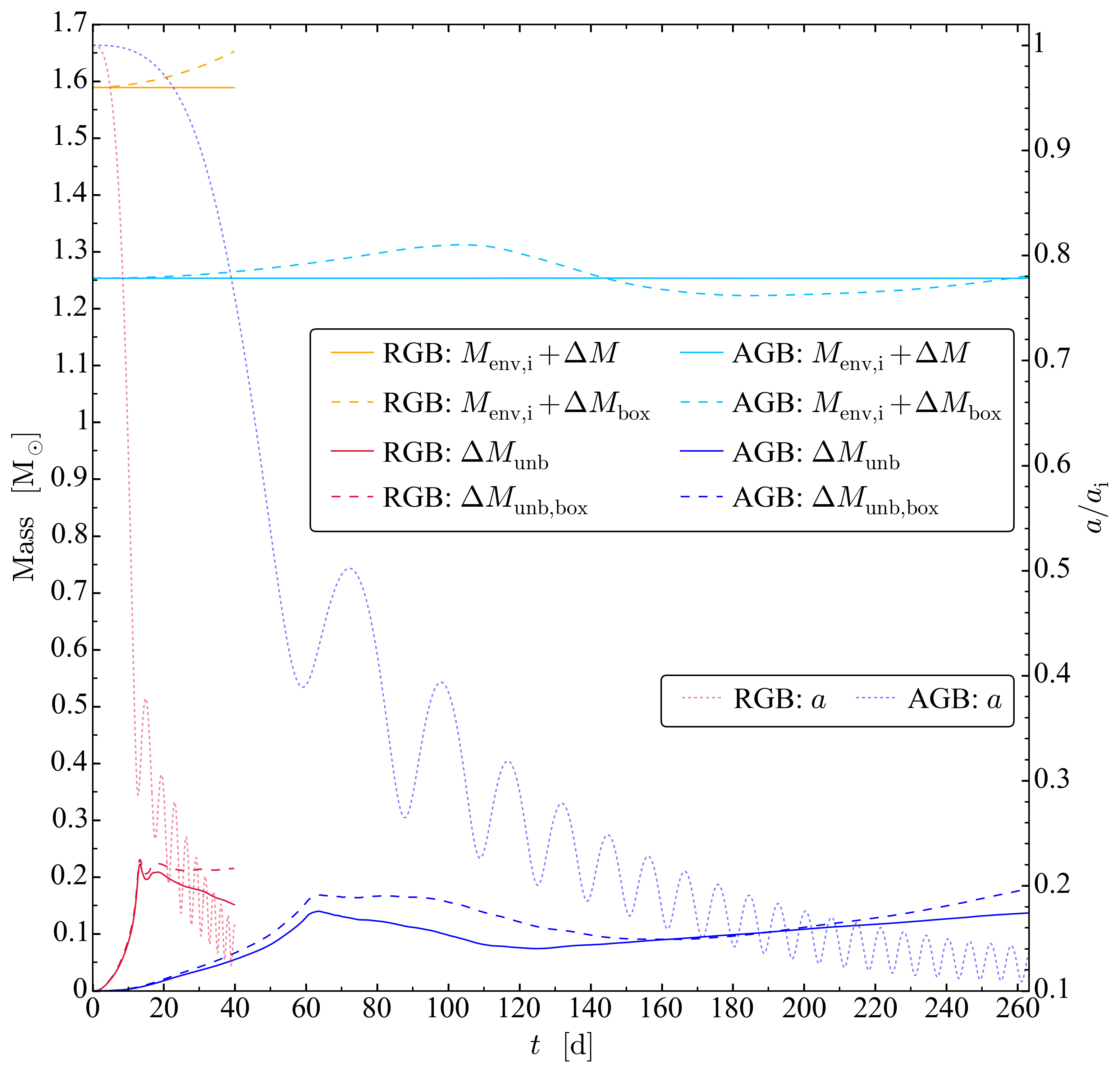}
  \vspace{-0.3cm}
  \caption{Mass evolution in the AGB and RGB runs.
           Upper solid lines (light blue for the AGB run and orange for the RGB run) 
           show the sum of the initial envelope mass $M_\mathrm{env,i}$
           and change in total mass $\rmD M$, 
           accounting for mass that has entered or exited through the domain boundaries; 
           lines are horizontal because mass is conserved ($\rmD M\approx0$).
           For the upper dashed lines, the change in mass $\rmD M\bo$ includes
            only mass inside the simulation box, 
           and hence does not account for mass flux through the boundaries.
           Lower solid lines (dark blue for the AGB run and red for the RGB run) 
           show the change in the unbound mass $\rmD M\unb$, 
           accounting for the flux through the domain boundaries. 
           Gas is called ``unbound'' if $\widetilde{\mathcal{E}}\gas>0$.
           Dashed lines do not account for flux through the boundaries.
           The inter-particle separation (dotted lines, right vertical axis) is also shown for comparison.
           \label{fig:mass_183_143}
          }
\end{figure}

\begin{figure*}
\begin{tabbing}
\includegraphics[scale=0.158,clip=true,trim=  50 120 238 180]{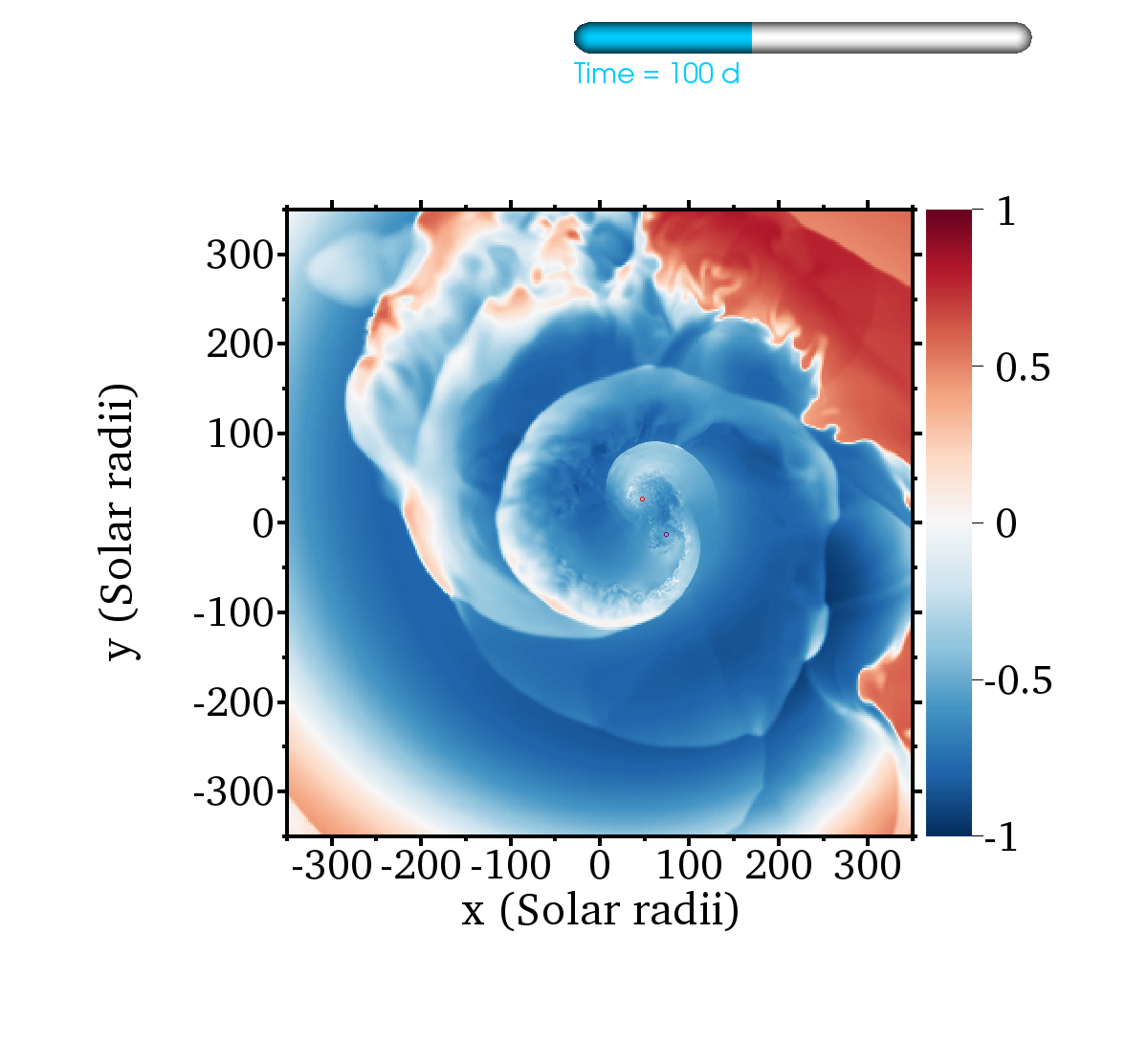}
\includegraphics[scale=0.158,clip=true,trim= 292 120 238 180]{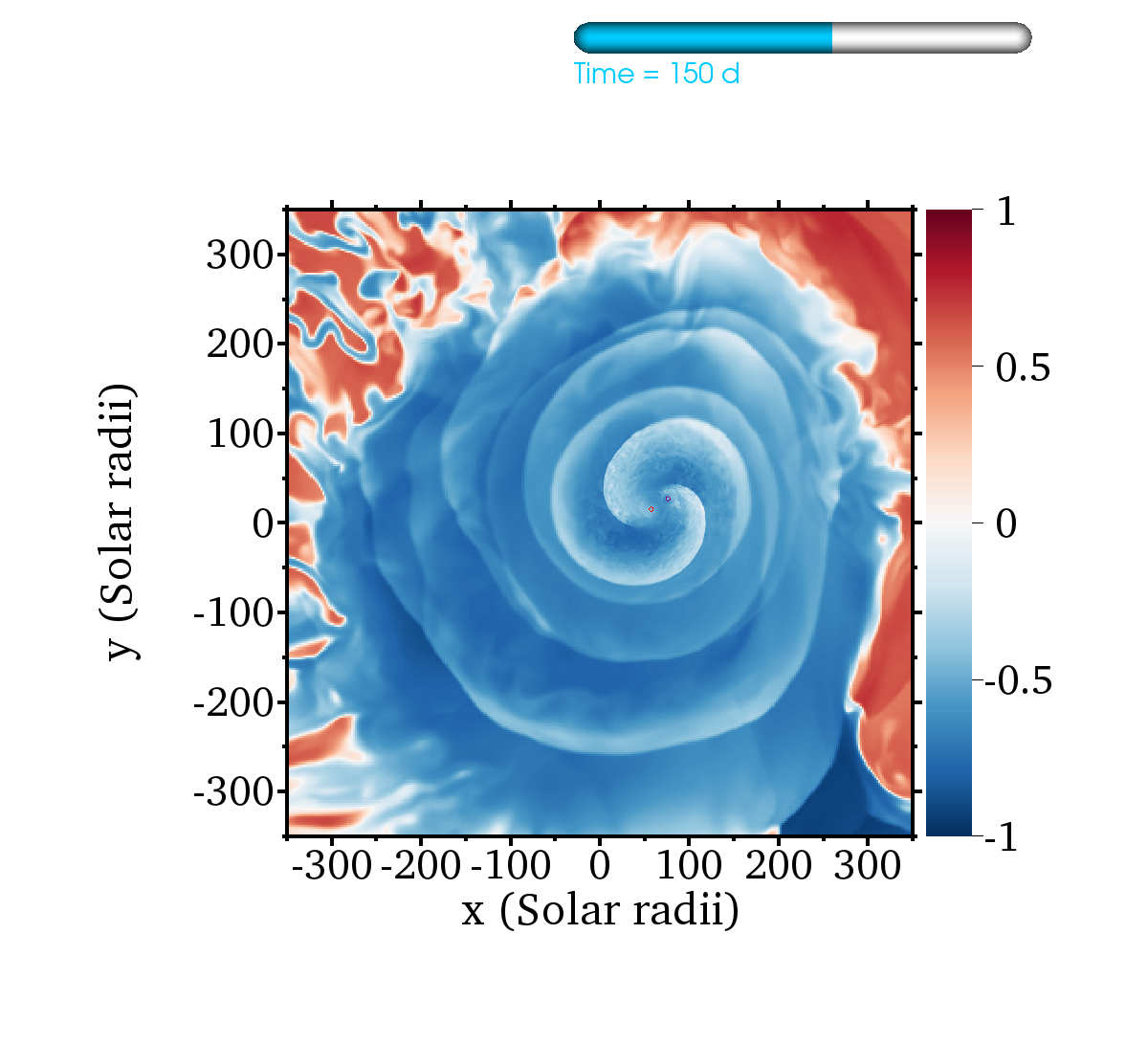}
\includegraphics[scale=0.158,clip=true,trim= 292 120 238 180]{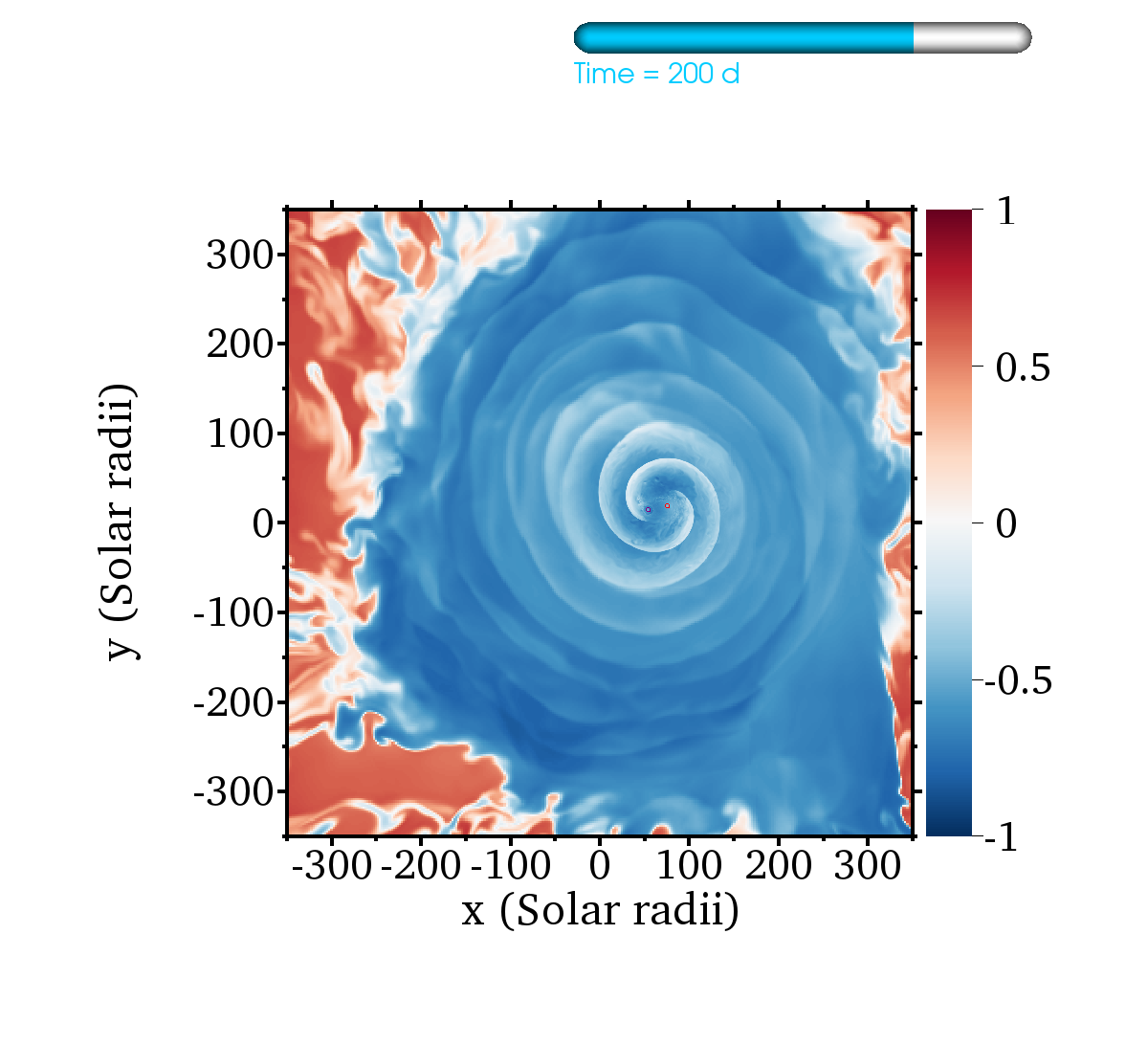}
\includegraphics[scale=0.158,clip=true,trim= 292 120   0 180]{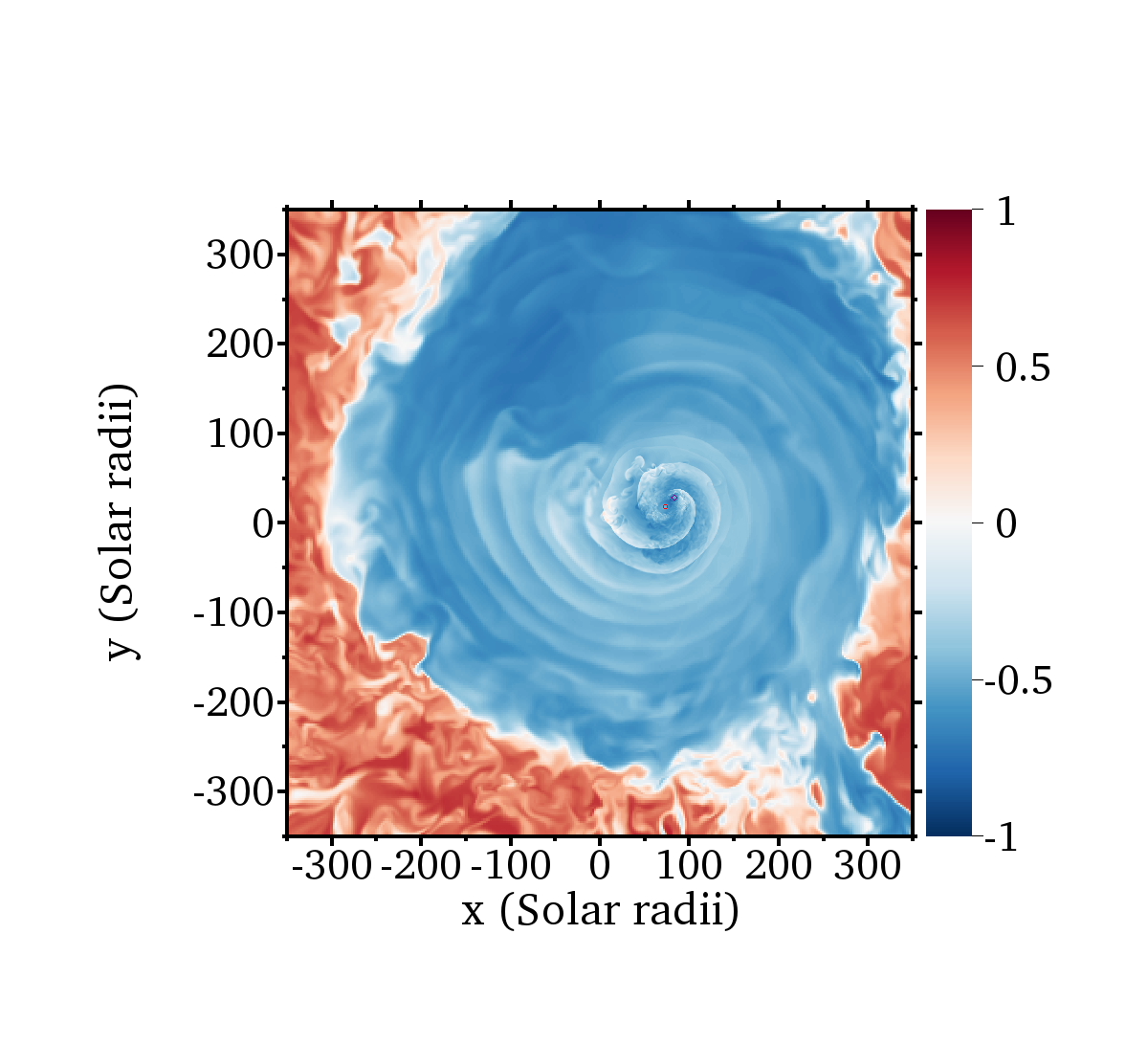}
\end{tabbing}
\caption{Same slices as the bottom row of Fig.~\ref{fig:density_faceon_183}, at $t=100$, $150$, $200$, and $250\da$,
         but now showing the quantity $\widetilde{\mathcal{E}}\gas$.
         Positive red values show unbound gas, negative blue values show bound gas, 
         and white shows marginally bound or unbound gas.
         \label{fig:energy_faceon_183}
        }            
\end{figure*}

\subsection{Morphological Evolution}\label{sec:morphology}
Density slices through the orbital plane at various evolutionary times are shown in Figure~\ref{fig:density_faceon_183},
with primary core particle and secondary softening spheres labeled by small purple and blue circles, respectively.
Figure~\ref{fig:density_edgeonthroughparticles_183} shows vertical slices through the particles for the last four times shown in Figure~\ref{fig:density_faceon_183}.
Morphologies are broadly consistent with those found in other CE simulations, and so we do not describe them in detail here.
However, at $t=250\da$ or about $18$ orbits (final snapshot), we note evidence for mixing between spiral layers, 
particularly to the left of the particles in the bottom-right panel of Figure~\ref{fig:density_faceon_183}. 
Similar mixing was also noted by \citet{Ohlmann+16a}, who attributed it to Kelvin-Helmholtz instabilities between adjacent layers.
Whether such mixing in our simulation is physical or caused by numerical effects should be explored in future work.

We now compare morphologies obtained for the AGB and RGB runs.
In Figure~\ref{fig:comparison_143_183}, the top row shows results for the AGB run and the bottom row the RGB run.
We plot the final frame of the RGB run, which ended after $10$ orbits ($t=40\da$), 
and the AGB run is plotted after the same number of orbits ($t=193\da$) for comparison.
The first and third columns respectively show horizontal and vertical slices of gas density, sliced through the particles.
The field of view in the bottom row is equal to that in the top row if lengths are normalized by the value of $a\init$ in each run.
Note that the colour bar for the RGB run is shifted up by one order of magnitude to account for the larger densities in that run.

The morphology for the two runs is strikingly similar. 
However, the mass in the AGB run is more centrally concentrated as compared to the RGB run. 
This is true even at $t=0$, as seen by comparing the density profile in Appendix~\ref{sec:ic} with that in \citealt{Chamandy+19a}.
For the AGB run, the density is largest at the location of the primary core particle.  
Outside of the high-density region of diameter $\sim80\Rsun$ around the particles, 
there is a gradual, approximately exponential decline with radius.
For the RGB run, the density is largest at the secondary.
There is a comparable high-density region surrounding the particles,
but  surrounded by a region of diameter $\sim110\Rsun$ where the density decreases  weakly with radius, 
outside of which it decreases  more steeply.

The edge-on view after $10$ orbits, presented in the third column of Figure~\ref{fig:comparison_143_183},
is also very similar for the two runs.
A partially evacuated conical region has developed with axis roughly coincident with the vertical axis passing through the particle centre of mass.
The maximum density contrast between the walls of this cavity and its interior,
along lines parallel to the orbital plane, is typically in the range $2$--$4$ 
(measured using the slices shown and orthogonal vertical slices through the particle centre of mass),
with the contrast marginally higher in the AGB case than the RGB case.
The full opening angle is of order $50^\circ$--$70^\circ$, with values in the AGB case being slightly smaller than for the RGB case.

\subsection{Envelope Unbinding and Mass Budget}\label{sec:mass}
\begin{figure}
  %produced using anal_energy_terms_183_143.py
  \vspace{-0.3cm}
  \includegraphics[width=\columnwidth,clip=true,trim= 0 0 0 5]{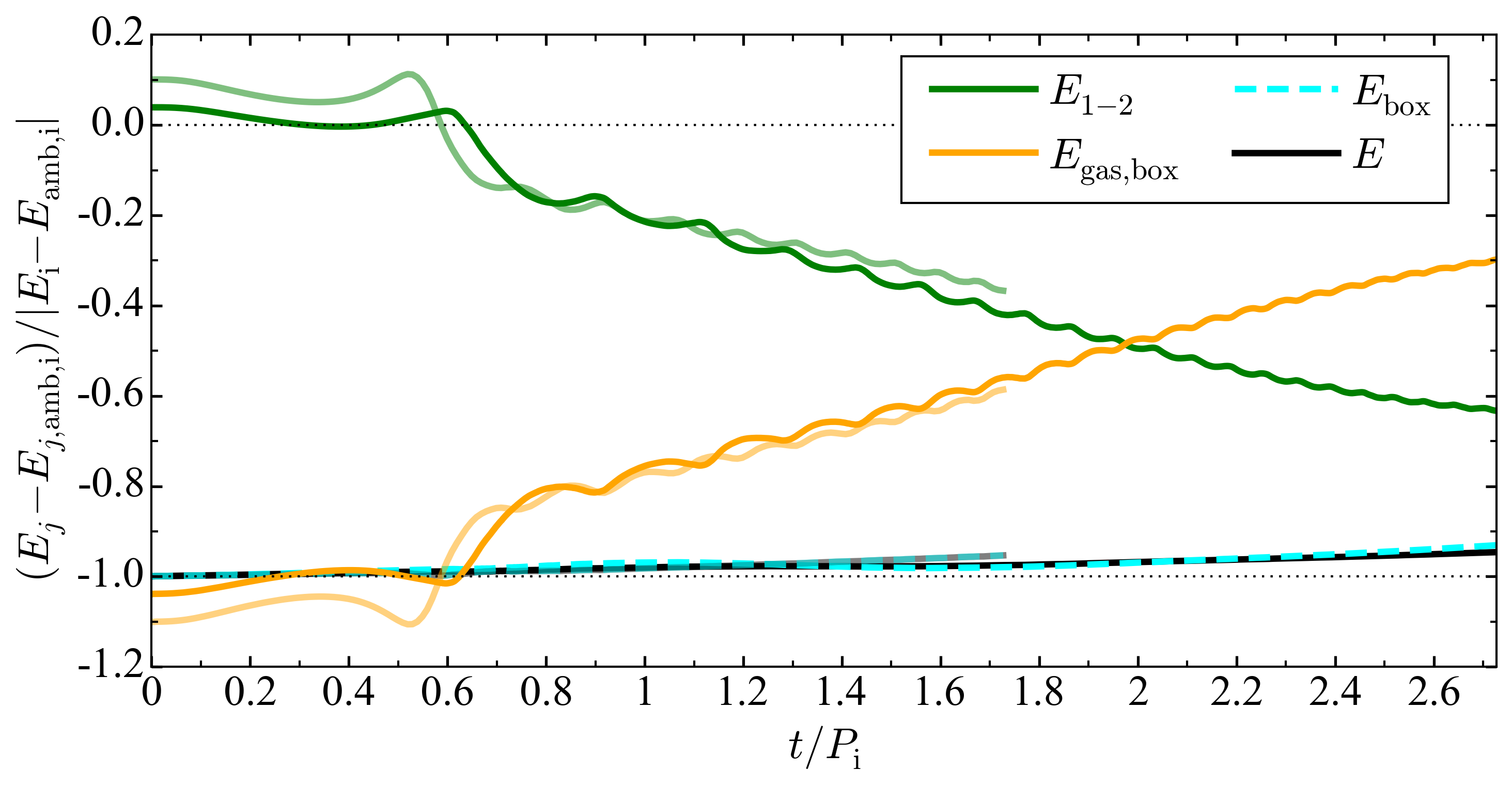}
  \vspace{-0.3cm}
  \caption{Energy terms (after subtracting initial ambient values).
           Both models are plotted using the same colours and line styles but the RGB run is plotted with paler shades,
           and the lines terminate just after $t/P\init=1.7$.
           Quantities are the total energy of terms involving gas in the simulation domain $E_\mathrm{gas,box}$,
           the total energy of terms involving particles only $E_\mathrm{1-2}$,
           the total energy in the simulation domain $E_\mathrm{box}$,
           and the total energy including the integrated flux through the boundaries, $E$.
           \label{fig:energy_terms_183_143_totals}
          }
\end{figure}

\begin{figure}
  %produced using anal_energy_terms_183_143.py
  \vspace{-0.3cm}
  \includegraphics[width=\columnwidth,clip=true,trim= 0 0 0 5]{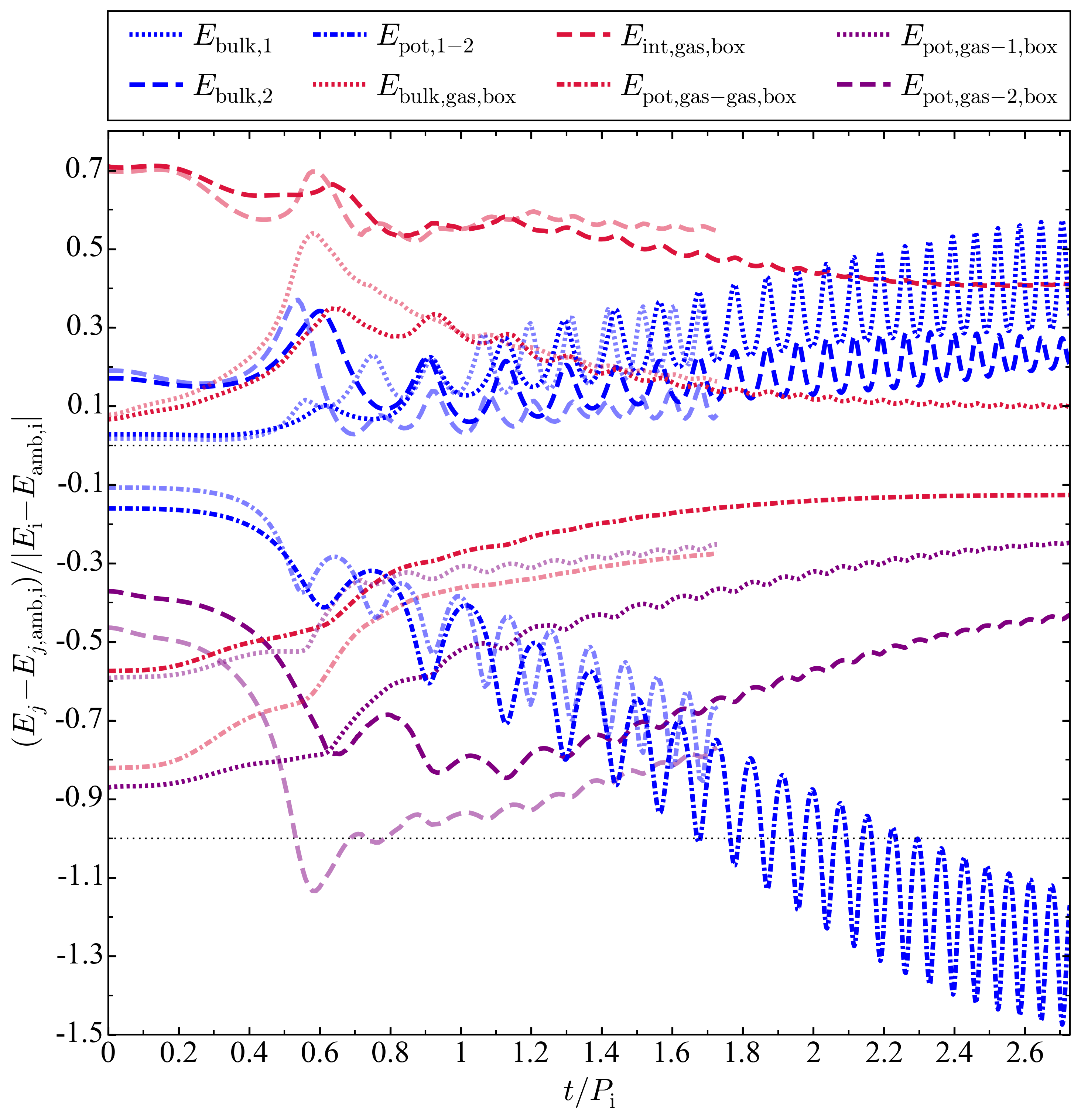}
  \vspace{-0.3cm}
  \caption{Energy terms in each simulation, plotted after subtracting initial ambient values
           and normalizing to initial total energy (minus ambient).
           Both models are plotted using the same colours and line styles but the RGB run is plotted with paler shades,
           and the lines terminate just after $t/P\init=1.7$.
           Terms are the bulk kinetic energy of the primary core particle and the secondary,
           $E_\mathrm{bulk,1}$ and $E_\mathrm{bulk,2}$, respectively,
           the potential energy due to the interaction between the particles, $E_\mathrm{pot,1-2}$, 
           the bulk kinetic, internal, and self-gravitational potential energies
           of gas inside the simulation domain $E_\mathrm{bulk,gas,box}$, $E_\mathrm{int,gas,box}$ and $E_\mathrm{pot,gas-gas,box}$,
           respectively, and the potential energy due to the interaction between the gas in the simulation domain and each particle, 
           $E_\mathrm{pot,gas-1,box}$ and $E_\mathrm{pot,gas-2,box}$. 
           \label{fig:energy_terms_183_143_individ}
          }
\end{figure}

We consider gas to be unbound if 
$\mathcal{E}'\gas\equiv$$\mathcal{E}_\mathrm{bulk,gas} +\mathcal{E}_\mathrm{int,gas}+\mathcal{E}_\mathrm{pot,gas-gas}
+2\mathcal{E}_\mathrm{pot,gas-1}+2\mathcal{E}_\mathrm{pot,gas-2}>0$,
where $\mathcal{E}$ denotes energy density, subscript  
$1$ refers to the primary core and $2$ refers to the secondary.
Here $\mathcal{E}_\mathrm{bulk,gas}=\tfrac{1}{2}\rho v^2$, where $v$ is the magnitude of the bulk velocity,
$\mathcal{E}_\mathrm{int,gas}=P/(\gamma-1)=\tfrac{3}{2}P$,
$\mathcal{E}_\mathrm{pot,gas-gas}=\tfrac{1}{2}\rho\Phi\gas$ with $\Phi\gas$ the potential due to gas only, 
and $\mathcal{E}_{\mathrm{pot,gas-}i}=\tfrac{1}{2}\rho\Phi_i$, where $\Phi_i$ is the potential due to particle $i$.
Using $2\mathcal{E}_{\mathrm{pot,gas-}i}$ rather than $\mathcal{E}_{\mathrm{pot,gas-}i}$ is a conservative choice
which can be thought of as distributing the particles' share of the gas-particle potential energy proportionately over the gas
to ensure that this contribution to the binding of the system is fully accounted for.
There is currently a lack of consensus with respect to the condition used to designate gas as bound or unbound.

The mass of unbound gas is plotted in Figure~\ref{fig:mass_183_143} as a solid dark blue (red) line for the AGB (RGB) run.
Solid light blue (orange) lines show the sum of the initial AGB (RGB) envelope mass and any mass change $\rmD M$ during the simulation
(the latter is negligible  so the lines are horizontal). 
Dashed lines show the values for the same quantities inside the simulation box -- that is, neglecting fluxes through the domain boundaries.
The early evolution of the unbound mass in the AGB case is rather similar to that in the RGB case.
The rate of mass unbinding accelerates until the first periastron passage, when it peaks, and then  decreases more slowly.
The decrease is due to energy transfer from unbound envelope material to the ambient medium \citep{Chamandy+19a}.
The fractional unbound mass of the envelope $\rmD M\unb/M_\mathrm{env,i}$ for the two runs at the peak is very similar, 
namely $11\%$ for the AGB run and $14\%$ for the RGB run.

However, in the AGB run, the mass of unbound gas  rises again after $t\sim125\da$.
That this upturn happens at about $4$ orbits and has not (yet) happened by the end of the RGB run after $10$ orbits
can partly be explained by the lower density (factor of $\approx7$) and pressure (factor of $\approx10$) of ambient gas in the AGB run.
The AGB run also exhibits significant gas outflow through the domain boundaries,
after which this gas cannot lose energy to ambient material.
The rate of unbinding between $t=125\da$ and the end of the AGB simulation is remarkably constant with mean value $\dot{M}\unb=0.17\Msunyr$
and standard deviation $0.03\Msunyr$.
At this rate, the envelope would completely unbind in $7\yr$.
Envelope ejection times of order $10\yr$ are comfortably shorter than estimates 
of the ages of PNe ($\sim10^3$--$10^4\yr$) and even pre-PNe ($\sim10^2$--$10^3\yr$; \citealt{Bujarrabal+01}).

Orbital plane slices, like those of the bottom row of Figure~\ref{fig:density_faceon_183}, 
are plotted for the local unbinding in Figure~\ref{fig:energy_faceon_183}.  
We normalize the quantity $\mathcal{E}'\gas$
with respect to either the sum of the positive contributions $\mathcal{E}_\mathrm{bulk,gas} +\mathcal{E}_\mathrm{int,gas}$,
or the modulus of the sum of the negative contributions
$|\mathcal{E}_\mathrm{pot,gas-gas}+2\mathcal{E}_\mathrm{pot,gas-1}+2\mathcal{E}_\mathrm{pot,gas-2}|$, whichever of the two is greater,
and plot this normalized quantity $\widetilde{\mathcal{E}}\gas$.
White represents marginally bound or unbound gas, blue (red) represents bound (unbound) gas, 
with darker shades for gas which is more bound (unbound).
Energy is transferred from the particles to the gas as the particles lose orbital energy (see also Section~\ref{sec:energy}).
Much of this liberated energy propagates outward within spiral density waves,
unbinding some gas that had been marginally bound \citep{Chamandy+19a}.
This process is visible in Figure~\ref{fig:energy_faceon_183}, 
where outward moving wave crests gradually turn the  outer-envelope from blue to red whilst expanding.
Moreover, the blue shade of most envelope gas whitens,
as it becomes less strongly bound.

To compare envelope unbinding for the AGB and RGB runs,
we turn to Figure~\ref{fig:comparison_143_183}.
Here the second and fourth columns show $\widetilde{\mathcal{E}}\gas$
in horizontal and vertical slices through the particles,
after 10 orbits (at $193\da$ for the AGB run and $40\da$ for the RGB run).
The AGB run is plotted in the top row and the RGB run in the bottom row.
The much paler  blue  in the top panels compared to the bottom panels
tell us that most of the envelope is less strongly bound for the AGB run.
Moreover, in the right column we see that gas along the orbital axis above and below the orbital plane 
is partially unbound in the AGB run, but not in the RGB run.
In short, outward transfer of energy is reduced in the RGB run at this stage  compared to the AGB run,
likely due to a much higher gas density  surrounding the particles in the RGB case.

\subsubsection{Influence of Ambient Gas}
Could diffusive mixing of bound envelope gas with the hot ambient medium evolved to $t=125\da$, rather than inspiral,
explain the change in $\rmD M\unb$ between $t=125\da$ and $t=263\da$ in the AGB?
The diffusivity at the interface between bound and unbound gas can be estimated as $\eta\sim\tfrac{1}{3}\delta_0 c\sound$,
where the sound speed is typically $c\sound\approx40\kms$ and the base numerical resolution is $\delta_0=2.25\Rsun$.
The diffusion length is given by $l\diff\sim (\eta \rmD t)^{1/2}$,
and adopting $\rmD t\approx138\da$ we obtain $l\diff\sim23\Rsun$.
The total mass that can be unbound by diffusion can be estimated 
as $M\diff \sim 4\uppi R^2 l\diff \rho \chi$.
Here $R\approx300\Rsun$ is the radius of the surface demarcating the bound envelope from unbound gas
and $\chi=\mathcal{E}'_\mathrm{gas,unb}/|\mathcal{E}'_\mathrm{gas,bou}|$, 
where `unb' and `bou' refer to unbound and bound gas on either side of the interface (up to a depth $\sim l\diff$).
Examination of 2D slices of $\mathcal{E}'\gas$ reveals that on average $\chi\sim\tfrac{1}{3}$. 
Near the interface, the density of bound material is $\rho\approx1\times10^{-8}\gcmcmcm$.
Thus, we estimate the mass unbound due to diffusive mixing with already-present hot gas as $M\diff \sim0.004\Msun$.
This is small compared with the change in $\rmD M\unb$ of $0.067\Msun$ between $t=125\da$ and $t=263\da$.
The surface area could be larger than $4\uppi R^2$ since some of the bound material is located within intermediate-scale structures
produced by prior mixing, which would increase the estimate of $M\diff$.
On the other hand, our estimate assumes, very conservatively, that {\it all} of the available energy in the surrounding unbound gas 
is transferred to the bound gas, and is distributed optimally such that previously bound material is unbound with $\mathcal{E}'\gas=0$.
Factoring in the inefficiency of this process would thus decrease the estimate of $M\diff$.
Also, much of the  unbound gas within a diffusion length could have originated in the envelope rather than sourced by the initial ambient medium,
so our estimate of $M\diff$ is likely an upper limit on the contribution to unbinding from the ambient gas, which is thus overall likely to be a subdominant effect.

Thus, while it seems likely that the orbital energy released by the inspiral is primarily responsible for the unbinding
between $t=125\da$ and $t=263\da$, the ambient medium might be playing some role.
Our choice of ambient medium parameter values was constrained by the need 
to keep the ambient pressure similar to that at the stellar surface 
and the ambient temperature small enough to avoid miniscule time steps.
Achieving reduced ambient density and temperature in future CE simulations is a priority.

\subsection{Overall Energy Budget}\label{sec:energy}
Here we describe the evolution of the various energy contributions, integrated over the simulation domain.
Figure~\ref{fig:energy_terms_183_143_totals} shows the evolution 
of the particle energy $E_\mathrm{1-2}=E_\mathrm{bulk,1}+E_\mathrm{bulk,2}+E_\mathrm{pot,1-2}$,
the gas energy in the domain $E_\mathrm{gas,box}$, equal to the integral over gas of $\mathcal{E}'\gas$, defined above,
and the total energy in the domain $E\bo=E_\mathrm{1-2}+E_\mathrm{gas,box}$, 
as well as the total energy accounting for fluxes across the domain boundaries, $E$.
The paler shaded curve of a given colour (extending to $t/P\init\sim1.7$) shows the RGB run, while the darker shade of the same colour shows the AGB run.
The initial energy of the ambient gas has been subtracted from the curves showing $E_\mathrm{gas,box}$, $E\bo$, and $E$,
and the curves have been normalized by the initial value of $E=E\bo$ (minus the ambient energy).

In both runs, the dashed cyan line does not deviate very much from the black/grey line, 
which implies that the flux of gas energy through the boundaries is small.
We also see that in both runs the total energy is reasonably well conserved 
but that there is a $\sim5\%$ energy gain by the end of each run owing to numerical effects (see Appendix~\ref{sec:energy_conservation}), but  evolution of the normalized gas (orange) and particle (green) energies 
are remarkably similar for the two runs.

Looking at the evolution of the individual energy terms reveals significant differences between the two runs,  
in addition to the similarities.
All of these terms are plotted in Figure~\ref{fig:energy_terms_183_143_individ},
again subtracting the respective initial value for the ambient medium (for terms involving gas)
and normalizing by the initial energy of the envelope-particle system with ambient energy subtracted.
The evolution curves of the terms $E_\mathrm{int,gas,box}$, $E_\mathrm{bulk,1}$, $E_\mathrm{bulk,2}$, and $E_\mathrm{pot,1-2}$
are  very similar between runs, so we focus on the other terms.
The term $E_\mathrm{pot,gas-gas,box}$ (dash-dotted red) is relatively more important in the RGB run.
This term scales roughly as $M\oneenv^2$, so the larger envelope mass of the RGB star makes more of a difference than for other terms.
Likewise, the larger envelope mass makes the term $E_\mathrm{pot,gas-2,box}$ (dashed purple)  relatively more important for the RGB run.
At around the first periastron passage, $|E_\mathrm{pot,gas-2,box}|$ peaks more strongly in the RGB run than in the AGB run,
and there is a corresponding increase in the bulk kinetic energy of gas $E_\mathrm{bulk,gas,box}$ (dotted red),
and reduction in the magnitude of $E_\mathrm{pot,gas-gas,box}$.
These features are consistent with the relatively deeper plunge of the secondary in the RGB run (Figure~\ref{fig:separation}),
and the associated violent ejection and expansion of envelope material \citep{Chamandy+19a}.
On the other hand, the term $E_\mathrm{pot,gas-1,box}$ (dotted purple) gives a relatively larger contribution in the AGB run
because of the larger value of $M\core$ and the more centrally 
concentrated density profile in the AGB case (Section~\ref{sec:morphology} and Appendix~\ref{sec:ic}).

\section{Summary and conclusions}
\label{sec:conclusions}
We carried out a high-resolution AMR hydrodynamic simulation of CE evolution involving a ZAMS $2\Msun$ AGB primary 
(modeled as a central point-particle  and extended envelope) and $1\Msun$ secondary (modeled as a point particle).
In the latter half of the simulation, the envelope steadily unbinds at the rate of $\sim 0.17\Msunyr$.
Were this to continue  until the envelope is completely unbound, the CE phase would last  $\sim 7\yr$. 
This is short compared to age estimates of PNe containing post-CE binary central stars.
At the end of the run, the mean inter-particle separation continues to decrease but is still $\sim 1.6$ times too large 
to place an upper limit on the commonly used theoretical parameter $\alpha\CE$.
Due to imperfect energy conservation, we stopped the simulation beyond about $20$ orbits (at $\sim5\%$ energy gain).
Energy conservation is a ubiquitous  problem in mesh-based CE simulations, and addressing it should be prioritized to enable longer runs.

We compared the results of this AGB run to one with the same secondary but a ZAMS $2\Msun$ RGB primary 
\citep{Chamandy+18,Chamandy+19a,Chamandy+19b}
by scaling our present results to the same relative initial binary separation $a\init$ (about 2\% larger than the primary radius for both runs), orbital period $P\init$, and initial energy of the binary. 
In these scaled units of time and distance, the separation-time curves for the particles are similar between runs,
but the first periastron passage occurs at somewhat larger values of $a/a\init$ and $t/P\init$ in the AGB case.
If instead we use orbital revolutions as the unit of time, then $a/a\init$ decreases faster for the AGB run than the RGB run 
between the first periastron passage and tenth apastron passage,
at which point we are closer to placing an upper limit on $\alpha\CE$ than for the RGB run.
Hence, while AGB CE simulations are more numerically demanding than their RGB counterparts,
they may offer certain strategic advantages. 

We  compared  the evolution of the drag force on the secondary in the non-inertial rest frame of the primary core particle.
Though an order of magnitude smaller in the AGB run, the drag force evolution in the two runs is very similar.
The discrepancy between the measured force and that crudely estimated
using Bondi-Hoyle-Lyttleton theory 
is approximately equal between the two runs.
In the AGB run, which lasts for twice as many orbital revolutions as the RGB run, 
the drag force becomes even smaller, 
and the discrepancy even greater, as the simulation progresses.

We also explored the evolution of gas surrounding the particles.
We found that gas stripping from that bound to the primary core 
near the time of first periastron passage and mass pileup near the secondary core 
are  less significant in the AGB run compared to the RGB run. 
This is likely  because the secondary does not get as close to the AGB core (particle+gas), 
and because the AGB core  is more tightly bound compared to the RGB core.  
Assuming gas to be unbound when the local quantity $\mathcal{E}'\gas$, 
defined in Section~\ref{sec:mass}, is positive,
the  maximum in the unbound mass occurs just after the first periastron passage in both cases 
with a very similar  peak value: 
$14\%$ for the RGB run and $11\%$ for the AGB run.
More simulations are needed to further explore the parameter dependence of this relative insensitivity to giant phase.

In the second, longer unbinding event in the AGB run, which was still ongoing at the end of the simulation, 
the outward transport of liberated orbital energy transforms gas 
from bound to unbound in the outermost envelope,
whilst gas in the bulk of the envelope becomes less strongly bound with time.
Simulations with smaller ambient temperature and density would further confirm that the unbinding seen 
is dominated by extraction of energy from the inspiral.
We compared the spatial distribution of $\widetilde{\mathcal{E}}\gas$,  which is like $\mathcal{E}'\gas$ but normalized 
such that $-1$ corresponds to maximally bound and $1$ to maximally unbound, in the two runs.
After $10$ orbits, the value of this quantity in the inner envelope is comparable between the two runs,
but is significantly larger in the outer envelope for the AGB case, 
apparently because outward energy transport by spiral shocks is impeded by a dense inner envelope in the RGB case.

For the volume-integrated energy terms (normalized by the total initial energy),
there was a high level of agreement between the two runs, but with a few notable differences. 
The potential energy term involving the primary core and envelope is relatively more important in the AGB case
(higher particle mass, more centrally condensed envelope),
while the potential energy terms involving the secondary and envelope and self-gravity of the envelope
are relatively more important in the RGB case (larger envelope mass).

Our setup can  eventually be improved by using a synchronously rotating primary initialized at the Roche limit separation 
\citep{Macleod+18a,Reichardt+19}, although this comes at a much higher computational cost.
Radiative transfer and a more realistic non-ideal equation of state are needed.  
Including these ingredients will allow us to model ionization and recombination, radiative cooling, and convection,
all of which are likely important for the budget and redistribution of energy in the envelope, 
and more accurate modeling of envelope unbinding.

Despite these caveats, the projection of $\sim7\yr$ to unbind the envelope 
suggests that the  universal failure in previous simulations to eject it without recombination 
is unlikely a consequence of insufficient  physics, 
but the result of a combination of numerical limitations and insufficient run time.
Nevertheless, the need  to improve both the physics and the numerical capabilities of simulations remains,
while also expanding into  new regions of phenomenologically relevant parameter space.

\section*{Acknowledgements}
The authors thank Jason Nordhaus, Orsola De~Marco, Thomas Reichardt and Javier S\'{a}nchez-Salcedo for discussions.
We thank Noam Soker for helpful comments.
This work used the computational and visualization resources in the Center for Integrated Research Computing (CIRC) at the University of Rochester and the computational resources of the Texas Advanced Computing Center (TACC) at The University of Texas at Austin, provided through allocation TG-AST120060 from the Extreme Science and Engineering Discovery Environment (XSEDE) \citep{xsede}, which is supported by National Science Foundation grant number ACI-1548562. Financial support for this project was provided by the Department of Energy grant DE-SC0001063, the National Science Foundation grants AST-1515648 and AST-181329, and the Space Telescope Science Institute grant HST-AR-12832.01-A.

\footnotesize{
\noindent
\bibliographystyle{mnras}
\bibliography{refs}

\begin{thebibliography}{}
\makeatletter
\relax
\def\mn@urlcharsother{\let\do\@makeother \do\$\do\&\do\#\do\^\do\_\do\%\do\~}
\def\mn@doi{\begingroup\mn@urlcharsother \@ifnextchar [ {\mn@doi@}
  {\mn@doi@[]}}
\def\mn@doi@[#1]#2{\def\@tempa{#1}\ifx\@tempa\@empty \href
  {http://dx.doi.org/#2} {doi:#2}\else \href {http://dx.doi.org/#2} {#1}\fi
  \endgroup}
\def\mn@eprint#1#2{\mn@eprint@#1:#2::\@nil}
\def\mn@eprint@arXiv#1{\href {http://arxiv.org/abs/#1} {{\tt arXiv:#1}}}
\def\mn@eprint@dblp#1{\href {http://dblp.uni-trier.de/rec/bibtex/#1.xml}
  {dblp:#1}}
\def\mn@eprint@#1:#2:#3:#4\@nil{\def\@tempa {#1}\def\@tempb {#2}\def\@tempc
  {#3}\ifx \@tempc \@empty \let \@tempc \@tempb \let \@tempb \@tempa \fi \ifx
  \@tempb \@empty \def\@tempb {arXiv}\fi \@ifundefined
  {mn@eprint@\@tempb}{\@tempb:\@tempc}{\expandafter \expandafter \csname
  mn@eprint@\@tempb\endcsname \expandafter{\@tempc}}}

\bibitem[\protect\citeauthoryear{{Bondi} \& {Hoyle}}{{Bondi} \&
  {Hoyle}}{1944}]{Bondi+Hoyle44}
{Bondi} H.,  {Hoyle} F.,  1944, \mn@doi [\mnras] {10.1093/mnras/104.5.273},
  \href {http://adsabs.harvard.edu/abs/1944MNRAS.104..273B} {104, 273}

\bibitem[\protect\citeauthoryear{{Briggs}, {Ferrario}, {Tout}  \&
  {Wickramasinghe}}{{Briggs} et~al.}{2018}]{Briggs+18}
{Briggs} G.~P.,  {Ferrario} L.,  {Tout} C.~A.,   {Wickramasinghe} D.~T.,  2018,
  \mn@doi [\mnras] {10.1093/mnras/sty2481}, \href
  {http://adsabs.harvard.edu/abs/2018MNRAS.481.3604B} {481, 3604}

\bibitem[\protect\citeauthoryear{{Bujarrabal}, {Castro-Carrizo}, {Alcolea}  \&
  {S{\'a}nchez Contreras}}{{Bujarrabal} et~al.}{2001}]{Bujarrabal+01}
{Bujarrabal} V.,  {Castro-Carrizo} A.,  {Alcolea} J.,   {S{\'a}nchez Contreras}
  C.,  2001, \mn@doi [\aap] {10.1051/0004-6361:20011090}, \href
  {http://adsabs.harvard.edu/abs/2001A%26A...377..868B} {377, 868}

\bibitem[\protect\citeauthoryear{{Carroll-Nellenback}, {Shroyer}, {Frank}  \&
  {Ding}}{{Carroll-Nellenback} et~al.}{2013}]{Carroll-Nellenback+13}
{Carroll-Nellenback} J.~J.,  {Shroyer} B.,  {Frank} A.,   {Ding} C.,  2013,
  \mn@doi [Journal of Computational Physics] {10.1016/j.jcp.2012.10.004}, \href
  {http://adsabs.harvard.edu/abs/2013JCoPh.236..461C} {236, 461}

\bibitem[\protect\citeauthoryear{{Chamandy} et~al.,}{{Chamandy}
  et~al.}{2018}]{Chamandy+18}
{Chamandy} L.,  et~al., 2018, \mn@doi [\mnras] {10.1093/mnras/sty1950}, \href
  {http://adsabs.harvard.edu/abs/2018MNRAS.480.1898C} {480, 1898}

\bibitem[\protect\citeauthoryear{{Chamandy}, {Tu}, {Blackman},
  {Carroll-Nellenback}, {Frank}, {Liu}  \& {Nordhaus}}{{Chamandy}
  et~al.}{2019a}]{Chamandy+19a}
{Chamandy} L.,  {Tu} Y.,  {Blackman} E.~G.,  {Carroll-Nellenback} J.,  {Frank}
  A.,  {Liu} B.,   {Nordhaus} J.,  2019a, \mn@doi [\mnras]
  {10.1093/mnras/stz887}, \href
  {https://ui.adsabs.harvard.edu/abs/2019MNRAS.486.1070C} {486, 1070}

\bibitem[\protect\citeauthoryear{{Chamandy}, {Blackman}, {Frank},
  {Carroll-Nellenback}, {Zou}  \& {Tu}}{{Chamandy}
  et~al.}{2019b}]{Chamandy+19b}
{Chamandy} L.,  {Blackman} E.~G.,  {Frank} A.,  {Carroll-Nellenback} J.,  {Zou}
  Y.,   {Tu} Y.,  2019b, \mn@doi [\mnras] {10.1093/mnras/stz2813}, \href
  {https://ui.adsabs.harvard.edu/abs/2019MNRAS.490.3727C} {490, 3727}

\bibitem[\protect\citeauthoryear{{Cojocaru}, {Rebassa-Mansergas}, {Torres}  \&
  {Garc{\'{\i}}a-Berro}}{{Cojocaru} et~al.}{2017}]{Cojocaru+17}
{Cojocaru} R.,  {Rebassa-Mansergas} A.,  {Torres} S.,   {Garc{\'{\i}}a-Berro}
  E.,  2017, \mn@doi [\mnras] {10.1093/mnras/stx1326}, \href
  {http://adsabs.harvard.edu/abs/2017MNRAS.470.1442C} {470, 1442}

\bibitem[\protect\citeauthoryear{Colella}{Colella}{1990}]{COLELLA1990171}
Colella P.,  1990, \mn@doi [Journal of Computational Physics]
  {https://doi.org/10.1016/0021-9991(90)90233-Q}, 87, 171

\bibitem[\protect\citeauthoryear{{Cunningham}, {Frank}, {Varni{\`e}re},
  {Mitran}  \& {Jones}}{{Cunningham} et~al.}{2009}]{Cunningham+09}
{Cunningham} A.~J.,  {Frank} A.,  {Varni{\`e}re} P.,  {Mitran} S.,   {Jones}
  T.~W.,  2009, \mn@doi [\apjs] {10.1088/0067-0049/182/2/519}, \href
  {http://adsabs.harvard.edu/abs/2009ApJS..182..519C} {182, 519}

\bibitem[\protect\citeauthoryear{{Davis}, {Kolb}  \& {Willems}}{{Davis}
  et~al.}{2010}]{Davis+10}
{Davis} P.~J.,  {Kolb} U.,   {Willems} B.,  2010, \mn@doi [\mnras]
  {10.1111/j.1365-2966.2009.16138.x}, \href
  {http://adsabs.harvard.edu/abs/2010MNRAS.403..179D} {403, 179}

\bibitem[\protect\citeauthoryear{{Dodd} \& {McCrea}}{{Dodd} \&
  {McCrea}}{1952}]{Dodd+Mccrea52}
{Dodd} K.~N.,  {McCrea} W.~J.,  1952, \mn@doi [\mnras]
  {10.1093/mnras/112.2.205}, \href
  {https://ui.adsabs.harvard.edu/abs/1952MNRAS.112..205D} {112, 205}

\bibitem[\protect\citeauthoryear{{Eggleton}}{{Eggleton}}{1983}]{Eggleton83}
{Eggleton} P.~P.,  1983, \mn@doi [\apj] {10.1086/160960}, \href
  {http://adsabs.harvard.edu/abs/1983ApJ...268..368E} {268, 368}

\bibitem[\protect\citeauthoryear{{Escala}, {Larson}, {Coppi}  \&
  {Mardones}}{{Escala} et~al.}{2004}]{Escala+04}
{Escala} A.,  {Larson} R.~B.,  {Coppi} P.~S.,   {Mardones} D.,  2004, \mn@doi
  [\apj] {10.1086/386278}, \href
  {https://ui.adsabs.harvard.edu/abs/2004ApJ...607..765E} {607, 765}

\bibitem[\protect\citeauthoryear{{Glanz} \& {Perets}}{{Glanz} \&
  {Perets}}{2018}]{Glanz+Perets18}
{Glanz} H.,  {Perets} H.~B.,  2018, preprint, \href
  {http://adsabs.harvard.edu/abs/2018arXiv180108130G} {} (\mn@eprint {arXiv}
  {1801.08130})

\bibitem[\protect\citeauthoryear{{Grichener}, {Sabach}  \& {Soker}}{{Grichener}
  et~al.}{2018}]{Grichener+18}
{Grichener} A.,  {Sabach} E.,   {Soker} N.,  2018, \mn@doi [\mnras]
  {10.1093/mnras/sty1178}, \href
  {https://ui.adsabs.harvard.edu/abs/2018MNRAS.478.1818G} {478, 1818}

\bibitem[\protect\citeauthoryear{{Hoyle} \& {Lyttleton}}{{Hoyle} \&
  {Lyttleton}}{1939}]{Hoyle+Lyttleton39}
{Hoyle} F.,  {Lyttleton} R.~A.,  1939, \mn@doi [Proceedings of the Cambridge
  Philosophical Society] {10.1017/S0305004100021150}, \href
  {http://adsabs.harvard.edu/abs/1939PCPS...35..405H} {35, 405}

\bibitem[\protect\citeauthoryear{{Iaconi} \& {De Marco}}{{Iaconi} \& {De
  Marco}}{2019}]{Iaconi+Demarco19}
{Iaconi} R.,  {De Marco} O.,  2019, \mn@doi [\mnras] {10.1093/mnras/stz2756},
  \href {https://ui.adsabs.harvard.edu/abs/2019MNRAS.490.2550I} {490, 2550}

\bibitem[\protect\citeauthoryear{{Iaconi}, {Reichardt}, {Staff}, {De Marco},
  {Passy}, {Price}, {Wurster}  \& {Herwig}}{{Iaconi} et~al.}{2017}]{Iaconi+17}
{Iaconi} R.,  {Reichardt} T.,  {Staff} J.,  {De Marco} O.,  {Passy} J.-C.,
  {Price} D.,  {Wurster} J.,   {Herwig} F.,  2017, \mn@doi [\mnras]
  {10.1093/mnras/stw2377}, \href
  {http://adsabs.harvard.edu/abs/2017MNRAS.464.4028I} {464, 4028}

\bibitem[\protect\citeauthoryear{{Iaconi}, {De Marco}, {Passy}  \&
  {Staff}}{{Iaconi} et~al.}{2018}]{Iaconi+18}
{Iaconi} R.,  {De Marco} O.,  {Passy} J.-C.,   {Staff} J.,  2018, \mn@doi
  [\mnras] {10.1093/mnras/sty794}, \href
  {http://adsabs.harvard.edu/abs/2018MNRAS.477.2349I} {477, 2349}

\bibitem[\protect\citeauthoryear{{Iaconi}, {Maeda}, {Nozawa}, {De Marco}  \&
  {Reichardt}}{{Iaconi} et~al.}{2020}]{Iaconi+20}
{Iaconi} R.,  {Maeda} K.,  {Nozawa} T.,  {De Marco} O.,   {Reichardt} T.,
  2020, arXiv e-prints, \href
  {https://ui.adsabs.harvard.edu/abs/2020arXiv200306151I} {p. arXiv:2003.06151}

\bibitem[\protect\citeauthoryear{{Iben} \& {Tutukov}}{{Iben} \&
  {Tutukov}}{1984}]{Iben+Tutukov84}
{Iben} I. J.,  {Tutukov} A.~V.,  1984, \mn@doi [\apjs] {10.1086/190932}, \href
  {https://ui.adsabs.harvard.edu/abs/1984ApJS...54..335I} {54, 335}

\bibitem[\protect\citeauthoryear{{Ivanova}}{{Ivanova}}{2018}]{Ivanova18}
{Ivanova} N.,  2018, \mn@doi [\apjl] {10.3847/2041-8213/aac101}, \href
  {http://adsabs.harvard.edu/abs/2018ApJ...858L..24I} {858, L24}

\bibitem[\protect\citeauthoryear{{Ivanova} et~al.,}{{Ivanova}
  et~al.}{2013}]{Ivanova+13a}
{Ivanova} N.,  et~al., 2013, \mn@doi [\aapr] {10.1007/s00159-013-0059-2}, \href
  {http://adsabs.harvard.edu/abs/2013A%26ARv..21...59I} {21, 59}

\bibitem[\protect\citeauthoryear{{Jones}}{{Jones}}{2020}]{Jones20a}
{Jones} D.,  2020, arXiv e-prints, \href
  {https://ui.adsabs.harvard.edu/abs/2020arXiv200103337J} {p. arXiv:2001.03337}

\bibitem[\protect\citeauthoryear{{Kashi} \& {Soker}}{{Kashi} \&
  {Soker}}{2011}]{Kashi+Soker11}
{Kashi} A.,  {Soker} N.,  2011, \mn@doi [\mnras]
  {10.1111/j.1365-2966.2011.19361.x}, \href
  {https://ui.adsabs.harvard.edu/abs/2011MNRAS.417.1466K} {417, 1466}

\bibitem[\protect\citeauthoryear{{Kim}}{{Kim}}{2010}]{Kim10}
{Kim} W.-T.,  2010, \mn@doi [\apj] {10.1088/0004-637X/725/1/1069}, \href
  {https://ui.adsabs.harvard.edu/abs/2010ApJ...725.1069K} {725, 1069}

\bibitem[\protect\citeauthoryear{{Kim} \& {Kim}}{{Kim} \&
  {Kim}}{2007}]{Kim+Kim07}
{Kim} H.,  {Kim} W.-T.,  2007, \mn@doi [\apj] {10.1086/519302}, \href
  {https://ui.adsabs.harvard.edu/abs/2007ApJ...665..432K} {665, 432}

\bibitem[\protect\citeauthoryear{{Kim}, {Kim}  \& {S{\'a}nchez-Salcedo}}{{Kim}
  et~al.}{2008}]{Kim+08}
{Kim} H.,  {Kim} W.-T.,   {S{\'a}nchez-Salcedo} F.~J.,  2008, \mn@doi [\apjl]
  {10.1086/589149}, \href
  {https://ui.adsabs.harvard.edu/abs/2008ApJ...679L..33K} {679, L33}

\bibitem[\protect\citeauthoryear{{L{\'o}pez-C{\'a}mara}, {De Colle}  \& {Moreno
  M{\'e}ndez}}{{L{\'o}pez-C{\'a}mara} et~al.}{2019}]{Lopez-camara+19}
{L{\'o}pez-C{\'a}mara} D.,  {De Colle} F.,   {Moreno M{\'e}ndez} E.,  2019,
  \mn@doi [\mnras] {10.1093/mnras/sty2959}, \href
  {https://ui.adsabs.harvard.edu/abs/2019MNRAS.482.3646L} {482, 3646}

\bibitem[\protect\citeauthoryear{{MacLeod}, {Antoni}, {Murguia-Berthier},
  {Macias}  \& {Ramirez-Ruiz}}{{MacLeod} et~al.}{2017}]{Macleod+17}
{MacLeod} M.,  {Antoni} A.,  {Murguia-Berthier} A.,  {Macias} P.,
  {Ramirez-Ruiz} E.,  2017, \mn@doi [\apj] {10.3847/1538-4357/aa6117}, \href
  {http://adsabs.harvard.edu/abs/2017ApJ...838...56M} {838, 56}

\bibitem[\protect\citeauthoryear{{MacLeod}, {Ostriker}  \& {Stone}}{{MacLeod}
  et~al.}{2018}]{Macleod+18a}
{MacLeod} M.,  {Ostriker} E.~C.,   {Stone} J.~M.,  2018, \mn@doi [\apj]
  {10.3847/1538-4357/aacf08}, \href
  {http://adsabs.harvard.edu/abs/2018ApJ...863....5M} {863, 5}

\bibitem[\protect\citeauthoryear{{Moreno M{\'e}ndez}, {L{\'o}pez-C{\'a}mara}
  \& {De Colle}}{{Moreno M{\'e}ndez} et~al.}{2017}]{Morenomendez+17}
{Moreno M{\'e}ndez} E.,  {L{\'o}pez-C{\'a}mara} D.,   {De Colle} F.,  2017,
  \mn@doi [\mnras] {10.1093/mnras/stx1385}, \href
  {http://adsabs.harvard.edu/abs/2017MNRAS.470.2929M} {470, 2929}

\bibitem[\protect\citeauthoryear{{Nandez} \& {Ivanova}}{{Nandez} \&
  {Ivanova}}{2016}]{Nandez+Ivanova16}
{Nandez} J.~L.~A.,  {Ivanova} N.,  2016, \mn@doi [\mnras]
  {10.1093/mnras/stw1266}, \href
  {http://adsabs.harvard.edu/abs/2016MNRAS.460.3992N} {460, 3992}

\bibitem[\protect\citeauthoryear{{Nandez}, {Ivanova}  \& {Lombardi}}{{Nandez}
  et~al.}{2015}]{Nandez+15}
{Nandez} J.~L.~A.,  {Ivanova} N.,   {Lombardi} J.~C.,  2015, \mn@doi [\mnras]
  {10.1093/mnrasl/slv043}, \href
  {http://adsabs.harvard.edu/abs/2015MNRAS.450L..39N} {450, L39}

\bibitem[\protect\citeauthoryear{{Ohlmann}}{{Ohlmann}}{2016}]{Ohlmann16}
{Ohlmann} S.~T.,  2016, PhD thesis, -

\bibitem[\protect\citeauthoryear{{Ohlmann}, {R{\"o}pke}, {Pakmor}  \&
  {Springel}}{{Ohlmann} et~al.}{2016}]{Ohlmann+16a}
{Ohlmann} S.~T.,  {R{\"o}pke} F.~K.,  {Pakmor} R.,   {Springel} V.,  2016,
  \mn@doi [\apjl] {10.3847/2041-8205/816/1/L9}, \href
  {http://adsabs.harvard.edu/abs/2016ApJ...816L...9O} {816, L9}

\bibitem[\protect\citeauthoryear{{Ohlmann}, {R{\"o}pke}, {Pakmor}  \&
  {Springel}}{{Ohlmann} et~al.}{2017}]{Ohlmann+17}
{Ohlmann} S.~T.,  {R{\"o}pke} F.~K.,  {Pakmor} R.,   {Springel} V.,  2017,
  \mn@doi [\aap] {10.1051/0004-6361/201629692}, \href
  {http://adsabs.harvard.edu/abs/2017A%26A...599A...5O} {599, A5}

\bibitem[\protect\citeauthoryear{{Ostriker}}{{Ostriker}}{1999}]{Ostriker99}
{Ostriker} E.~C.,  1999, \mn@doi [\apj] {10.1086/306858}, \href
  {http://adsabs.harvard.edu/abs/1999ApJ...513..252O} {513, 252}

\bibitem[\protect\citeauthoryear{{Paczynski}}{{Paczynski}}{1976}]{Paczynski76}
{Paczynski} B.,  1976, in {Eggleton} P.,  {Mitton} S.,   {Whelan} J.,  eds,
  IAU Symposium Vol. 73, Structure and Evolution of Close Binary Systems. p.~75

\bibitem[\protect\citeauthoryear{{Paxton}, {Bildsten}, {Dotter}, {Herwig},
  {Lesaffre}  \& {Timmes}}{{Paxton} et~al.}{2011}]{Paxton+11}
{Paxton} B.,  {Bildsten} L.,  {Dotter} A.,  {Herwig} F.,  {Lesaffre} P.,
  {Timmes} F.,  2011, \mn@doi [\apjs] {10.1088/0067-0049/192/1/3}, \href
  {http://adsabs.harvard.edu/abs/2011ApJS..192....3P} {192, 3}

\bibitem[\protect\citeauthoryear{{Paxton} et~al.,}{{Paxton}
  et~al.}{2013}]{Paxton+13}
{Paxton} B.,  et~al., 2013, \mn@doi [\apjs] {10.1088/0067-0049/208/1/4}, \href
  {http://adsabs.harvard.edu/abs/2013ApJS..208....4P} {208, 4}

\bibitem[\protect\citeauthoryear{{Paxton} et~al.,}{{Paxton}
  et~al.}{2015}]{Paxton+15}
{Paxton} B.,  et~al., 2015, \mn@doi [\apjs] {10.1088/0067-0049/220/1/15}, \href
  {http://adsabs.harvard.edu/abs/2015ApJS..220...15P} {220, 15}

\bibitem[\protect\citeauthoryear{{Prust} \& {Chang}}{{Prust} \&
  {Chang}}{2019}]{Prust+Chang19}
{Prust} L.~J.,  {Chang} P.,  2019, \mn@doi [\mnras] {10.1093/mnras/stz1219},
  \href {https://ui.adsabs.harvard.edu/abs/2019MNRAS.486.5809P} {486, 5809}

\bibitem[\protect\citeauthoryear{{Reichardt}, {De Marco}, {Iaconi}, {Tout}  \&
  {Price}}{{Reichardt} et~al.}{2019}]{Reichardt+19}
{Reichardt} T.~A.,  {De Marco} O.,  {Iaconi} R.,  {Tout} C.~A.,   {Price}
  D.~J.,  2019, \mn@doi [\mnras] {10.1093/mnras/sty3485}, \href
  {http://adsabs.harvard.edu/abs/2019MNRAS.484..631R} {484, 631}

\bibitem[\protect\citeauthoryear{{Reichardt}, {De Marco}, {Iaconi}, {Chamandy}
  \& {Price}}{{Reichardt} et~al.}{2020}]{Reichardt+20}
{Reichardt} T.~A.,  {De Marco} O.,  {Iaconi} R.,  {Chamandy} L.,   {Price}
  D.~J.,  2020, \mn@doi [\mnras] {10.1093/mnras/staa937}, \href
  {https://ui.adsabs.harvard.edu/abs/2020MNRAS.494.5333R} {494, 5333}

\bibitem[\protect\citeauthoryear{{Ruffert}}{{Ruffert}}{1993}]{Ruffert+93}
{Ruffert} M.,  1993, \aap, \href
  {http://adsabs.harvard.edu/abs/1993A%26A...280..141R} {280, 141}

\bibitem[\protect\citeauthoryear{{Sabach}, {Hillel}, {Schreier}  \&
  {Soker}}{{Sabach} et~al.}{2017}]{Sabach+17}
{Sabach} E.,  {Hillel} S.,  {Schreier} R.,   {Soker} N.,  2017, \mn@doi
  [\mnras] {10.1093/mnras/stx2272}, \href
  {http://adsabs.harvard.edu/abs/2017MNRAS.472.4361S} {472, 4361}

\bibitem[\protect\citeauthoryear{{S{\'a}nchez-Salcedo} \&
  {Brandenburg}}{{S{\'a}nchez-Salcedo} \&
  {Brandenburg}}{2001}]{Sanchez-salcedo+Brandenburg01}
{S{\'a}nchez-Salcedo} F.~J.,  {Brandenburg} A.,  2001, \mn@doi [\mnras]
  {10.1046/j.1365-8711.2001.04061.x}, \href
  {https://ui.adsabs.harvard.edu/abs/2001MNRAS.322...67S} {322, 67}

\bibitem[\protect\citeauthoryear{{S{\'a}nchez-Salcedo} \&
  {Chametla}}{{S{\'a}nchez-Salcedo} \&
  {Chametla}}{2014}]{Sanchez-salcedo+Chametla14}
{S{\'a}nchez-Salcedo} F.~J.,  {Chametla} R.~O.,  2014, \mn@doi [\apj]
  {10.1088/0004-637X/794/2/167}, \href
  {https://ui.adsabs.harvard.edu/abs/2014ApJ...794..167S} {794, 167}

\bibitem[\protect\citeauthoryear{{Sandquist}, {Taam}, {Chen}, {Bodenheimer}  \&
  {Burkert}}{{Sandquist} et~al.}{1998}]{Sandquist+98}
{Sandquist} E.~L.,  {Taam} R.~E.,  {Chen} X.,  {Bodenheimer} P.,   {Burkert}
  A.,  1998, \mn@doi [\apj] {10.1086/305778}, \href
  {http://adsabs.harvard.edu/abs/1998ApJ...500..909S} {500, 909}

\bibitem[\protect\citeauthoryear{{Shiber}, {Iaconi}, {De Marco}  \&
  {Soker}}{{Shiber} et~al.}{2019}]{Shiber+19}
{Shiber} S.,  {Iaconi} R.,  {De Marco} O.,   {Soker} N.,  2019, \mn@doi
  [\mnras] {10.1093/mnras/stz2013}, \href
  {https://ui.adsabs.harvard.edu/abs/2019MNRAS.488.5615S} {488, 5615}

\bibitem[\protect\citeauthoryear{{Soker}}{{Soker}}{1992}]{Soker92a}
{Soker} N.,  1992, \mn@doi [\apj] {10.1086/171004}, \href
  {https://ui.adsabs.harvard.edu/abs/1992ApJ...386..190S} {386, 190}

\bibitem[\protect\citeauthoryear{{Soker}}{{Soker}}{2004}]{Soker04}
{Soker} N.,  2004, \mn@doi [\na] {10.1016/j.newast.2004.01.004}, \href
  {http://adsabs.harvard.edu/abs/2004NewA....9..399S} {9, 399}

\bibitem[\protect\citeauthoryear{{Soker}}{{Soker}}{2017}]{Soker17b}
{Soker} N.,  2017, \mn@doi [\mnras] {10.1093/mnras/stx1978}, \href
  {https://ui.adsabs.harvard.edu/abs/2017MNRAS.471.4839S} {471, 4839}

\bibitem[\protect\citeauthoryear{{Staff}, {De Marco}, {Macdonald}, {Galaviz},
  {Passy}, {Iaconi}  \& {Low}}{{Staff} et~al.}{2016}]{Staff+16a}
{Staff} J.~E.,  {De Marco} O.,  {Macdonald} D.,  {Galaviz} P.,  {Passy} J.-C.,
  {Iaconi} R.,   {Low} M.-M.~M.,  2016, \mn@doi [\mnras]
  {10.1093/mnras/stv2548}, \href
  {http://adsabs.harvard.edu/abs/2016MNRAS.455.3511S} {455, 3511}

\bibitem[\protect\citeauthoryear{{Towns}, {Cockerill}, {Dahan}  \&
  {Foster}}{{Towns} et~al.}{2014}]{xsede}
{Towns} J.,  {Cockerill} T.,  {Dahan} M.,   {Foster} I.,  2014, \mn@doi
  [Computing in Science and Engineering] {10.1109/MCSE.2014.80}, 16, 62

\bibitem[\protect\citeauthoryear{{Webbink}}{{Webbink}}{1984}]{Webbink84}
{Webbink} R.~F.,  1984, \mn@doi [\apj] {10.1086/161701}, \href
  {https://ui.adsabs.harvard.edu/abs/1984ApJ...277..355W} {277, 355}

\bibitem[\protect\citeauthoryear{{Wilson} \& {Nordhaus}}{{Wilson} \&
  {Nordhaus}}{2019}]{Wilson+Nordhaus18}
{Wilson} E.~C.,  {Nordhaus} J.,  2019, \mn@doi [\mnras] {10.1093/mnras/stz601},
  \href {https://ui.adsabs.harvard.edu/abs/2019MNRAS.485.4492W} {485, 4492}

\bibitem[\protect\citeauthoryear{{Zorotovic}, {Schreiber}, {G{\"a}nsicke}  \&
  {Nebot G{\'o}mez-Mor{\'a}n}}{{Zorotovic} et~al.}{2010}]{Zorotovic+10}
{Zorotovic} M.,  {Schreiber} M.~R.,  {G{\"a}nsicke} B.~T.,   {Nebot
  G{\'o}mez-Mor{\'a}n} A.,  2010, \mn@doi [\aap] {10.1051/0004-6361/200913658},
  \href {http://adsabs.harvard.edu/abs/2010A%26A...520A..86Z} {520, A86}

\makeatother
\end{thebibliography}

\appendix

\section{Initial Conditions}
\label{sec:ic}
Initial profiles for the AGB primary star used are presented in Figure~\ref{fig:profile_183}.
We refer the reader to \citet{Chamandy+19a} for the same profiles for the RGB star used.

\begin{figure}
  \includegraphics[width=\columnwidth,clip=true,trim= 0 0 0 0]{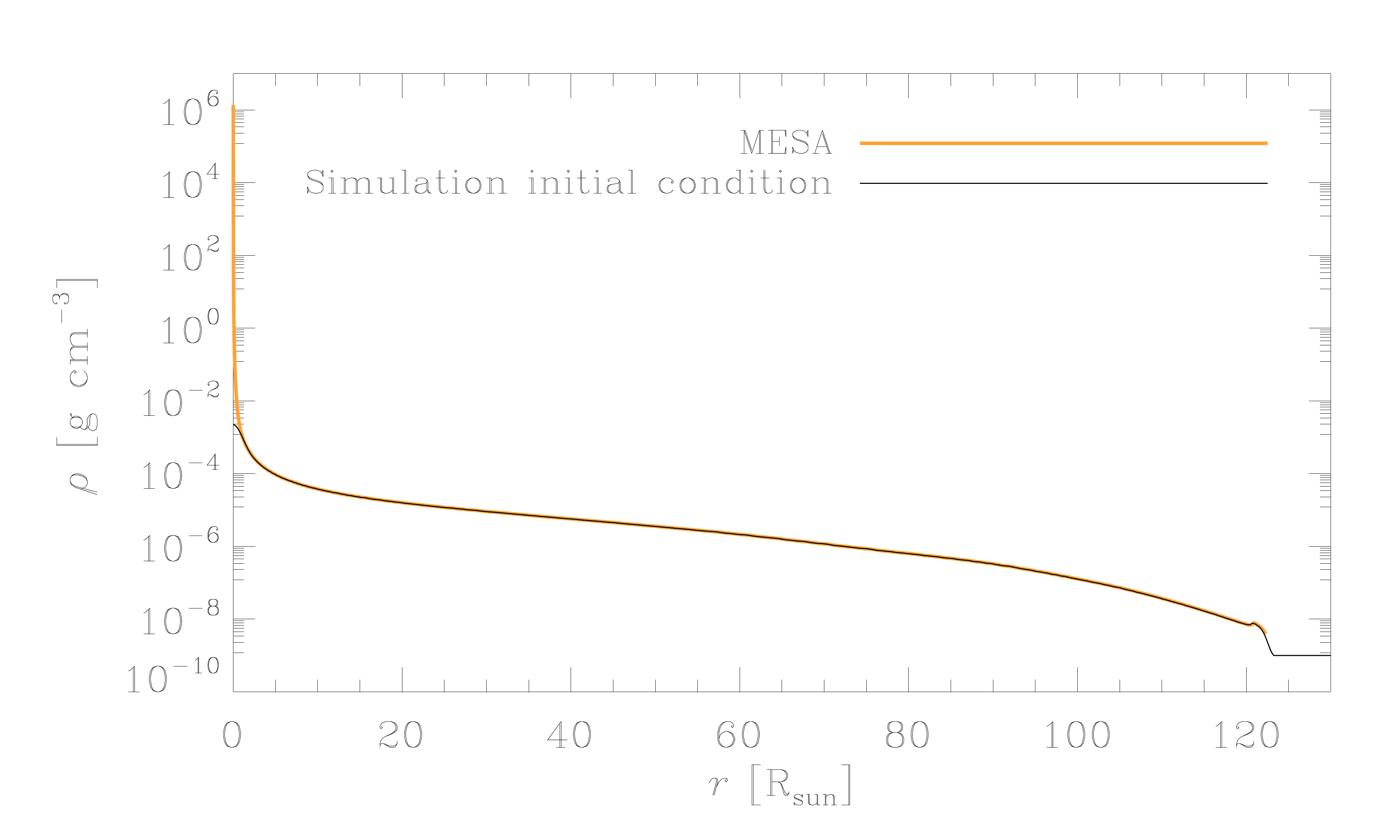}\\
  \includegraphics[width=\columnwidth,clip=true,trim= 0 0 0 0]{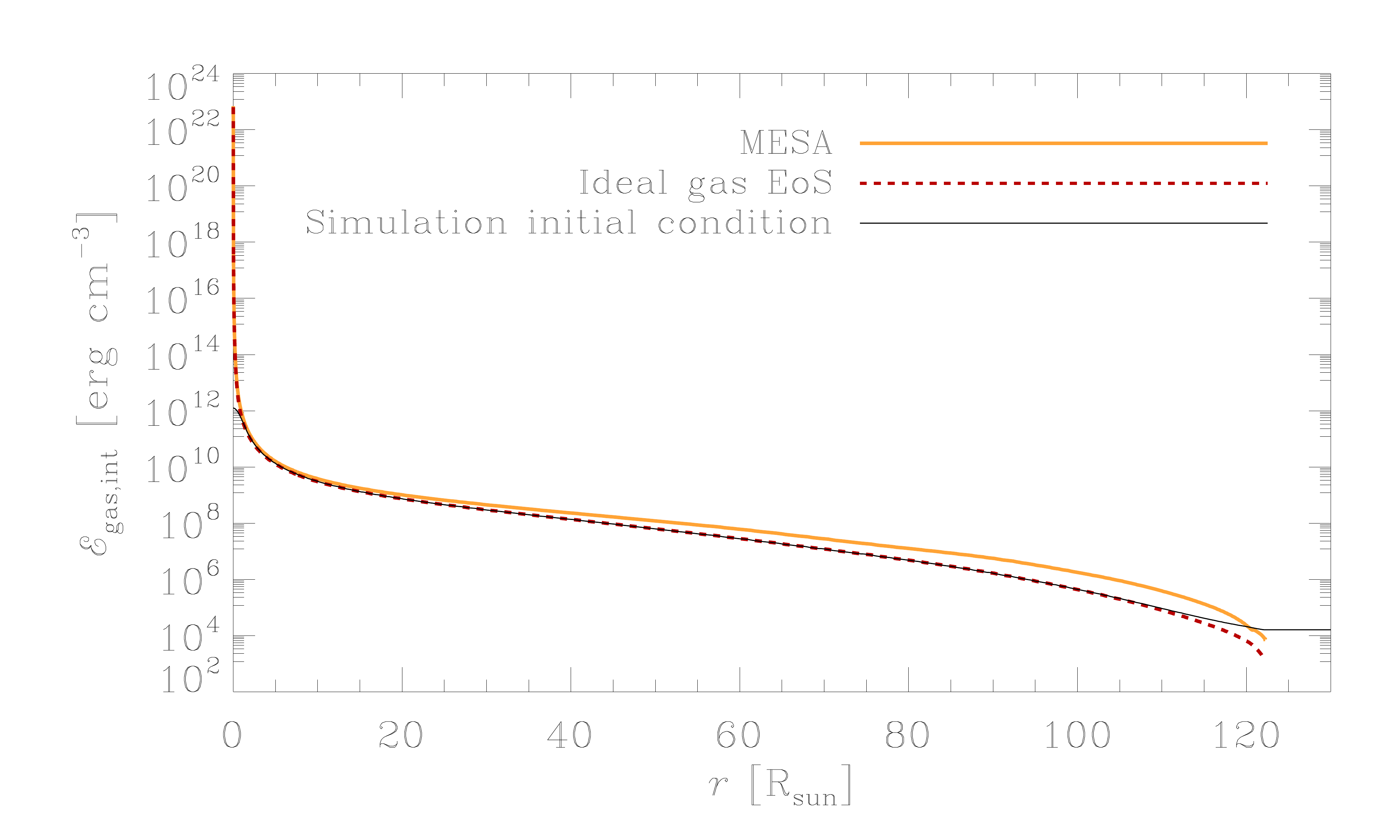}\\
  \includegraphics[width=\columnwidth,clip=true,trim= 0 0 0 0]{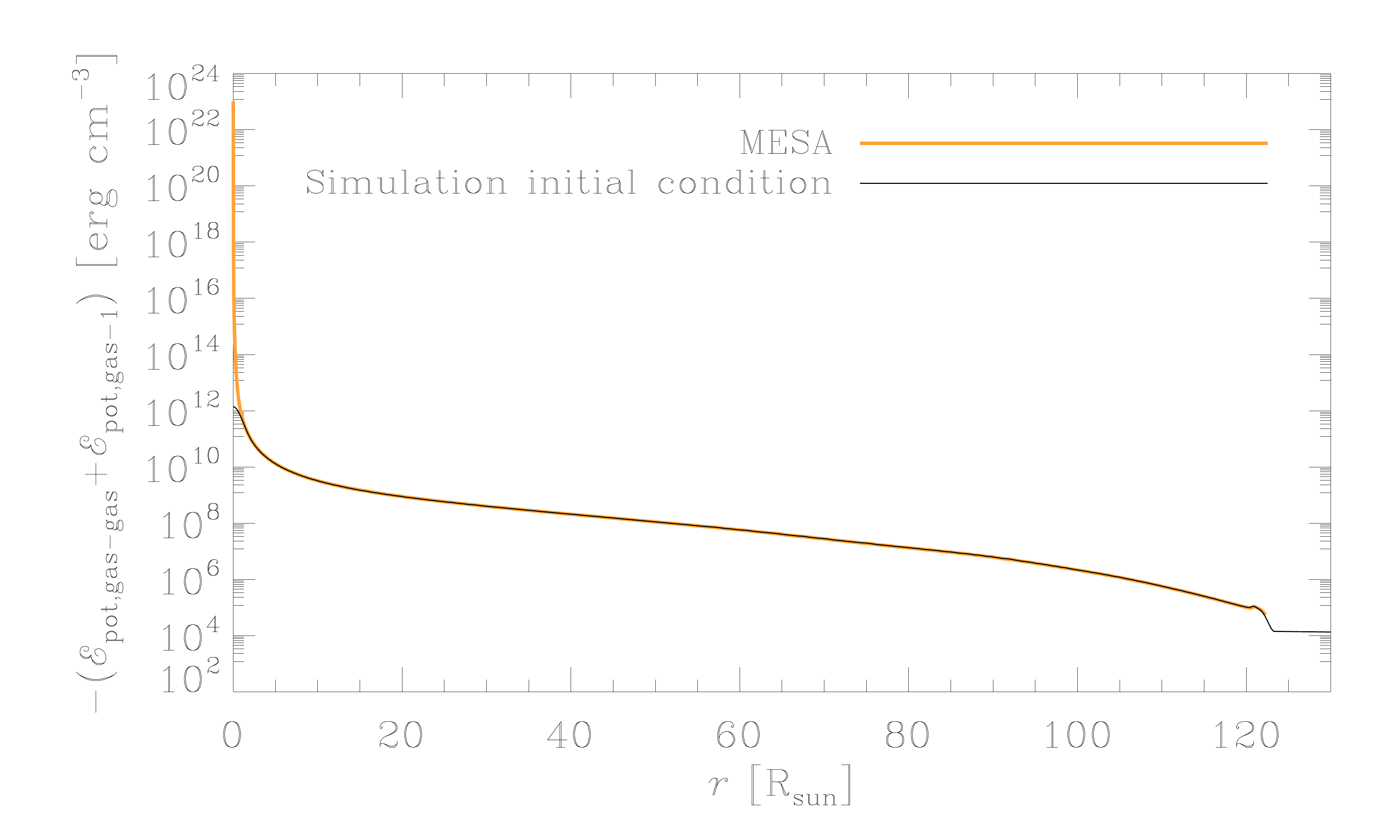}\\
 \caption{Top panel: Radial profile of gas density for the AGB star modeled using MESA (thick orange)
           and the envelope of our 3D AGB star in the simulation at $t=0$ (thin black).
           Middle panel: Comparison of the internal energy density profiles in the MESA model, the
           simulation initial envelope, and the MESA model with the equation of state replaced with an ideal gas equation of state,
           as in the simulation (dashed red).
           Bottom panel: Comparison of (negative of) potential energy density profiles 
           in the MESA model and the simulation initial envelope.
           \label{fig:profile_183}
          }            
\end{figure}

\section{Energy Non-Conservation}
\label{sec:energy_conservation}
The degree to which energy conservation is satisfied for our AGB run (denoted Model~A) and two lower resolution AGB runs (Models~B and C)
is shown in Figure~\ref{fig:energy_183_177_164}, where the fractional change in the total energy 
(accounting for flux through the domain boundaries) is plotted.
The level of adherence to energy conservation is sensitive to the resolution around the primary core particle,
and, after the first periastron passage, also to the resolution around the secondary.
The inter-particle separation is also plotted as a dashed line for each run both to illustrate the dependence of the total energy variation
on the orbital evolution, and to give a sense of how sensitive is the separation curve to small changes in resolution.

\begin{figure}
  \includegraphics[width=\columnwidth,clip=true,trim= 0 0 0 0]{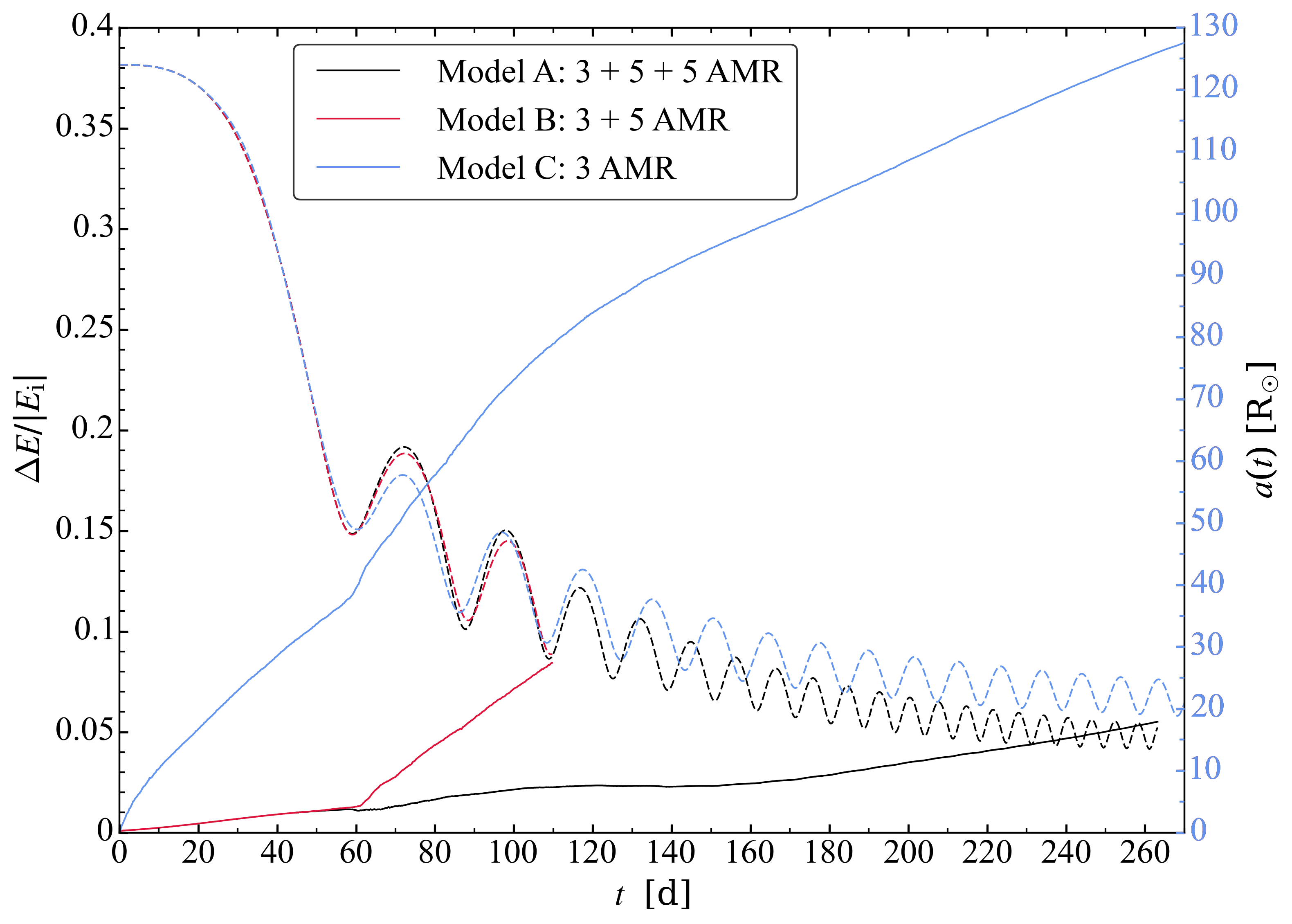}\\
  \caption{Evolution of the total energy $E$, as a fraction of the initial energy $|E_\mathrm{i}|$,
           for three different runs with varying resolution.
           This accounts for the energy within the simulation domain
           as well as that which has entered or exited through the boundaries.
           Model~A, which employs high resolution at AMR level 5 around both particles, 
           is the fiducial AGB run discussed in this paper.
           Model~A restarts from Model~B, for which high resolution is used near the primary core particle only.
           Model~C resolves the envelope at AMR level 3 like the other runs, 
           but does not employ higher resolution around either particle.
           \label{fig:energy_183_177_164}
          }            
\end{figure}

%-------------------------------------------------------------------------------------------
\end{document}